\documentclass[%
 aip, amsmath,amssymb, reprint,]{revtex4-1}
\pdfoutput =1
\usepackage{graphicx}
\usepackage{dcolumn}
\usepackage{bm}
\usepackage{color}
\usepackage{xcolor}
\usepackage[utf8]{inputenc}
\usepackage[T1]{fontenc}
\usepackage{mathptmx}
\usepackage{acronym}
\usepackage{amssymb}
\usepackage{physics}
\usepackage{amssymb}
\usepackage{amsmath}
\usepackage{bbold}
\begin{document}


\title{Photonic Quantum Metrology}

\author{Emanuele Polino}
\thanks{These authors contributed equally to the manuscript preparation}
\affiliation{Dipartimento di Fisica, Sapienza Universit\`{a} di Roma, Piazzale Aldo Moro 5, I-00185 Roma, Italy}
\author{Mauro Valeri}
\thanks{These authors contributed equally to the manuscript preparation}
\affiliation{Dipartimento di Fisica, Sapienza Universit\`{a} di Roma, Piazzale Aldo Moro 5, I-00185 Roma, Italy}
\author{Nicol\`o Spagnolo}
\affiliation{Dipartimento di Fisica, Sapienza Universit\`{a} di Roma, Piazzale Aldo Moro 5, I-00185 Roma, Italy}
\author{Fabio Sciarrino}
\email{fabio.sciarrino@uniroma1.it}
\affiliation{Dipartimento di Fisica, Sapienza Universit\`{a} di Roma, Piazzale Aldo Moro 5, I-00185 Roma, Italy}
\affiliation{Consiglio Nazionale delle Ricerche, Istituto dei Sistemi Complessi (CNR-ISC), Via dei Taurini 19, 00185 Roma, Italy}

\date{\today}

\begin{abstract}
Quantum Metrology is one of the most promising application of quantum technologies. The aim of this research field is the estimation of unknown parameters exploiting quantum resources, whose application can lead to enhanced performances with respect to classical strategies. Several physical quantum systems can be employed to develop quantum sensors, and photonic systems represent ideal probes for a large number of metrological tasks. Here we review the basic concepts behind quantum metrology and then focus on the application of photonic technology for this task, with particular attention to phase estimation. We describe the current state of the art in the field in terms of platforms and quantum resources. Furthermore, we present the research area of multiparameter quantum metrology, where multiple parameters have to be estimated at the same time. We conclude by discussing the current experimental and theoretical challenges, and the open questions towards implementation of photonic quantum sensors with quantum-enhanced performances in the presence of noise.
\end{abstract}

\maketitle

\tableofcontents

\section{Introduction}

One of the basic pillars of physical science is the measurement process. The goal of a measure is to associate a value to a physical quantity, giving an estimate of it. Together with each experimental estimate, an uncertainty has to be provided, that is the \emph{"parameter, associated with the result of a measurement, that characterizes the dispersion of the values that could reasonably be attributed to the measured quantity"}  (definition from  ISO \cite{ISO}). 
The statistical error which affects the measurement result, can have different natures: technical or fundamental. The technical one is mostly represented by the accidental error, due to out-of-control imperfections in the measurement process. Conversely, there are fundamental limits on uncertainty, such as those due to Heisenberg relations, that are imposed by physical laws.
Quantum mechanics is the most successful, predictive and fundamental theory describing small scales phenomena. For this reason, a study of the measurement process and the ultimate achievable precision bounds has to be done under the light of such theory \cite{helstrom1976quantum,holevo2011probabilistic}. On one hand, quantum mechanics imposes fundamental limits on the estimate precision. On the other hand, in order to achieve such limits, quantum resources have to be employed. 

Remarkably, the exploitation of quantum systems to estimate unknown parameters overcomes the precision limits that can be in principle obtained by using only classical resources. This idea is at the basis of the continuously growing research area of Quantum Metrology, that aims at reaching the ultimate fundamental bounds on estimation precision by exploiting quantum probes  \cite{caves1981quantum,braunstein1992quantum,braunstein1994statistical,lee2002quantum,giovannetti2004quantum,giovannetti2006quantum,paris2009quantum,giovannetti2011advances,simon2017quantum,demkowicz2015quantum,jacobs2014quantum,braun2018quantum}. Given an unknown parameter to be estimated and $m$ classical probes (with $m\gg 1$), each interacting a single time with the system under study, the estimation error will scale at best as $\sim m^{-1/2}$. This classical limit is consequence of central limit theorem, and is called \emph{Standard Quantum Limit} (SQL). The term "classical" stands for probes that are at most classically correlated. If quantum probes are allowed, the SQL can be surpassed so that the uncertainty of the estimator reaches the more fundamental scaling $\sim m^{-1}$, improving the precision by a factor $\sqrt{m}$ with respect to the SQL. Such new scaling represents the ultimate limit on estimation precision and is called \emph{Heisenberg Limit} (HL).

A key role of quantum enhancement in estimation theory is played by entanglement. 
This quantum phenomena is one of the most striking aspects of quantum mechanics and radically differs from classical view of the world, as recognized by Einstein, Podolsky and Rosen \cite{epr}. The  relevance of entanglement in the fundamental divergence between classical and quantum predictions, was theoretically demonstrated by Bell \cite{bell1964einstein}. Realizing his famous test, recent experiments showed, in a nearly unambiguous way, that the observed quantum correlations arising from distant entangled systems cannot be explained by any classical theory obeying local causality \cite{Hensen, Giustina, Shalm,rosenfeld2017event}.   Quantum correlations represent also an extraordinary resource allowing to enhance  performances in informational tasks \cite{Horodecki,Nielsen,Brunner,Adesso}. Quantum advantage with respect to the best classical strategies lies at the basis of the research area of quantum technologies,  comprising branches like quantum  computation, communication,  simulation and precisely metrology  \cite{Nielsen}.

Besides the fundamental interest about ultimate  precision limits, Quantum Metrology presents different applications. Indeed, quantum-enhanced sensitivity can benefit different research branches, such as:  measurement on biological systems \cite{cox2012optical,taylor2016quantum,mauranyapin2017evanescent,schirhagl2014nitrogen}, gravitational waves detection \cite{abadie2011gravitational}, atomic clocks \cite{borregaard2013near,ludlow2015optical, bollinger1996optimal,katori2011optical}, interferometry with atomic and molecular matter waves \cite{dowling1998correlated,cronin2009optics,che2019multiqubit,tsarev2018quantum}, plasmonic sensing \cite{lee2018quantum,xu2018quantum,dowran2018quantum}, magnetometry \cite{jones2009magnetic,wolfgramm2010squeezed,simmons2010magnetic, kucsko2013nanometre,toyli2013fluorescence,urizar2013macroscopic,brask2015improved,wasilewski2010quantum,altenburg2017estimation,danilin2018quantum,tanaka2015proposed,razzoli2019lattice,dinani2019bayesian,troiani2018universal,li2018quantum,evrard2019enhanced,albarelli2017ultimate,apellaniz2018precision}, spectroscopy and frequency measurements \cite{polzik1992spectroscopy,wineland1994squeezed,wineland1992spin,leibfried2004toward,dorfman2016nonlinear,udem2002optical,huelga1997improvement,naghiloo2017achieving,albarelli2020quantum},  lithography \cite{boto2000quantum,fonseca1999measurement,kawabe2007quantum,sciarrino2008experimental,d2001two,boyd2012quantum}, microscopy and imaging \cite{kolobov1999spatial,lugiato2002quantum,brambilla2008high,treps2002surpassing,shih2007quantum,giovannetti2009sub,genovese2016real, moreau2019imaging,brida2010experimental,guerrieri2010sub,tsang2016quantum,meda2017photon,rehacek2017optimal,zhou2019modern,schwartz2013superresolution,cui2013quantum,ram2006beyond,ono2013entanglement,tsang2017subdiffraction,lupo2016ultimate,yang2016far,paur2016achieving,edgar2012imaging,nasr2003demonstration,tsang2015quantum,monticone2014beating,juffmann2016multi,israel2017quantum,tham2017beating,classen2017superresolution,tenne2019super,bisketzi2019quantum,samantaray2017realization,tsang2019semiparametric,tsang2019resolving,lupo2019quantum,berchera2018quantum,prasad2019quantum,rossman2019rapid,toninelli2019resolution,sabines2019twin,aidukas2019phase,gregory2020imaging},  localization of incoherent point sources \cite{tsang2015quantum,nair2016far, nair2016interferometric,lupo2020subwavelength}, Hamiltonian estimation \cite{shabani2011estimation,cole2006identifying,granade2012robust,zhang2014quantumh,wiebe2014hamiltonian,wang2017quantum,wang2017experimental}, fundamental physics effects  \cite{aspachs2010optimal,yao2014quantum,bruschi2014quantum,ahmadi2014relativistic,hogan2012interferometers,berchera2013quantum,ruo2015one,downes2017quantum,tian2017entanglement,chou2017mhz,romano2017detection,howl2018gravity,branford2019quantum,kohlrus2019quantum}, coordinates transfer, synchronization and navigation \cite{gisin1999spin,chiribella2004efficient,bagan2001aligning, fink2019entanglement,valencia2004distant,de2005quantum,lamine2008quantum,quan2016demonstration,lee2019symmetrical}, absorption measurements \cite{xiao1988detection,tapster1991sub,moreau2017demonstrating,li2018enhanced,knyazev2019overcoming}, thermometry \cite{stace2010quantum,correa2015individual,mehboudi2015thermometry,hofer2017quantum,campbell2018precision,cavina2018bridging,mehboudi2019thermometry,razavian2019quantum,potts2019fundamental,fujiwara2020real} and general sensing technologies \cite{degen2017quantum,gefen2019overcoming}.  

Different physical systems can be employed to realize Quantum Metrology tasks \cite{pezze2018quantum,degen2017quantum}.  The most appropriate quantum systems for several scenarios are photons, due to their properties like high mobility, low interaction with the environment, together with the available technology for their generation, manipulation and detection \cite{Flamini19rev,slussarenko2019photonic,pirandola2018advances}. Hence, it is of significant importance the development of platforms and techniques to generate suitable quantum photonic states, able to provide quantum-enhancement in different  metrology tasks. In particular, since many physical problems can be mapped into phase estimation processes, interferometry represents one of the most relevant scenarios \cite{giovannetti2004quantum,giovannetti2006quantum,paris2009quantum,giovannetti2011advances}. 

Furthermore, the unknown parameters to be estimated in a physical problem can be, in general, more than one. Also in this case quantum resources can enhance the simultaneous estimation of all parameters \cite{szczykulska2016multi,albarelli2020perspective}. This represents a relatively new research area with different experimental and theoretical open questions, such as the capability of reaching the quantum ultimate bounds in the simultaneous estimation of all parameters.

\subsection{Outline}
In this review we present an overview of quantum metrology through photonic platforms, by analyzing recent advances and discussing open problems.
In Sec. II we will briefly review the definitions of the basic quantities, such as Fisher Information and Cram\'er-Rao bound, that characterize a general estimation process, together with the most relevant theorems. In particular, the treatment is focused on single parameter estimation.  
In Sec. III we describe photonic platforms for the generation of different quantum states, with the corresponding applications in quantum metrology and in particular on phase estimation problems.
Sec. IV is devoted to adaptive protocols able to enhance the estimation processes.
In Sec. V we study the generalization of single parameter to multiparameter quantum metrology. We introduce the generalized theoretical framework and then describe the state-of-the-art of experimental photonic realizations of simultaneous multiparameter estimation.  In Sec. VI we conclude by describing the challenges towards genuine quantum enhanced metrology and providing some perspectives on future developments.

\section{Quantum Metrology: Fundamentals}

\subsection{\label{sec:level2} Estimation Process} 

\begin{figure*}[ht!]
\centering
\includegraphics[width=1.\textwidth]{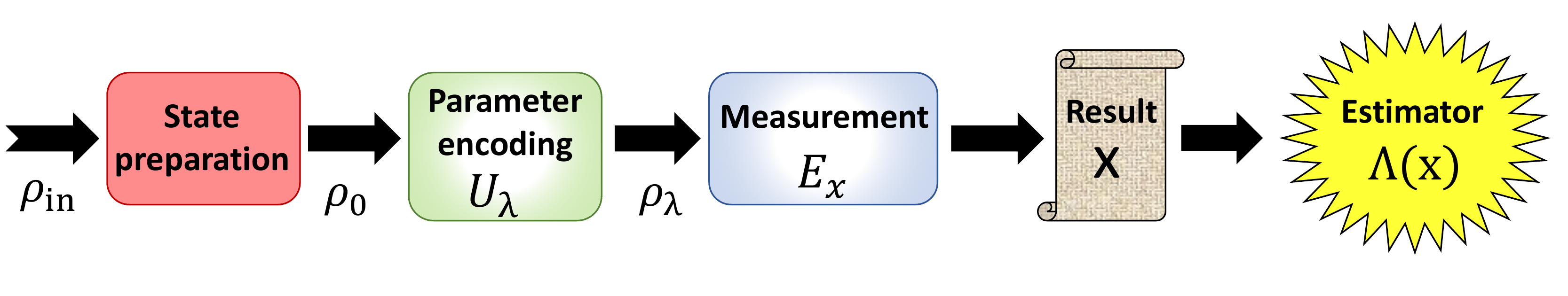}
\caption{{\bf Conceptual scheme of a parameter  estimation.} An initial probe is prepared (red box) in a state $\rho_0$ (eventually, from an initial state $\rho_{\mathrm{in}}$). Then, it interacts with the unknown parameter $\lambda$ through an evolution $U_\lambda$ (green box). The state $\rho_\lambda$ encoding the information on $\lambda$ is measured by a POVM $E_x$ (blue box) generating outcome $x$. Based on the outcomes $x$, a suitable estimator provides an estimate $\Lambda(x)$ of the parameter $\lambda$.}
\label{fig:figure1}
\end{figure*}

An estimation process aims at measuring an unknown parameter $\lambda$ embedded in a physical system. In this section we describe the case of single parameter estimation, while the problem of estimating more than one parameter is described in Sec. \ref{Sec:multi}. Through the interaction between a probe and the system,  information about the parameter is encoded in the probe state. The goal is to extract the information so that the estimation converges to the real value of the parameter.
The general scheme can be described through four steps (Fig. \ref{fig:figure1}):

\textit{(i)} preparation of a probe state $\rho_0$, such that it is sensitive to variations of the unknown parameter $\lambda$;

\textit{(ii)} interaction of the probe with the system through a unitary evolution $U_{\lambda}$ depending on $\lambda$ (for simplicity we consider only unitary evolution but this can be extended to non-unitary maps). From such interaction the evolved state $\rho_{\lambda}$ encodes the information on the unknown parameter: $\rho_{\lambda}=U_{\lambda} \rho_0 U_{\lambda}^\dagger$ ;

\textit{(iii)} the information is extracted by means of a suitable positive operator valued measure (POVM) $E_x$;

\textit{(iv)} finally, a suitable estimator, based on  measurement results $x$, provides an estimate $\Lambda(x)$ of the unknown parameter.  

Repeating this process $\nu$ independent times, the final estimator $\Lambda(\bm{x})$ in general depends on the complete sequence of measurement results $\bm{x}=(x_1,...,x_\nu)$.

A \emph{consistent} estimator asymptotically converges to the real value of the parameter. An estimator is said \emph{unbiased} if its mean value coincides with the unknown parameter: 

\begin{equation}
\bar{\Lambda} = \sum_{\bm{x}} P(\bm{x}|\lambda) \Lambda(\bm{x})=\lambda\;\;\;\quad   \forall \lambda,
\end{equation}
where $P(\bm{x}|\lambda)$ represents the conditional output probability of obtaining a sequence of measurement result $\bm{x}$, given a certain value of the parameter $\lambda$. Such probability is also called \emph{likelihood} and is given by the Born rule: \begin{equation} \label{quantumlikelihood}
P(x_i|\lambda)=\mathrm{Tr}(E_{x_i} \rho_\lambda),
\end{equation} 
in the case of single measurement result $x_i$. For $\nu$ independent measurements,  $P(\bm{x}|\lambda)=\prod_{i=1}^{\nu}P(x_i|\lambda)$. Furthermore, a \emph{locally unbiased} estimator is an unbiased estimator only for certain range of parameter's values, so that satisfies the relation: $\partial \bar{\Lambda}/\partial \lambda=1$. Finally, an estimator is \emph{asymptotically unbiased} when it converges to the real value in the limit of infinite number of probes: $ \lim_{m\to \infty} \bar{\Lambda}=\lambda \quad   \forall \lambda$.

In order to quantify the accuracy of an estimation process, the \emph{Mean Square Error} (MSE) can be defined: 
\begin{equation}\label{mse}
\mathrm{MSE}(\lambda)= \sum_{\bm{x}} (\Lambda(\bm{x})-\lambda)^2 P(\bm{x}|\lambda).
\end{equation}
For unbiased estimators, the MSE is equal to the variance on the estimate:  
\begin{equation}
\Delta \lambda ^2 \equiv \sum_{\bm{x}}(\Lambda(\bm{x})-\bar{\Lambda})^2P(\bm{x}|\lambda).
\end{equation}

In general, an estimator $\Lambda(\bm{x})$ can be non deterministic. In this case, it is possible to take into account  the probability $P^{\mathrm{exp}}(\Lambda|\bm{x})$ to generate an estimate $\Lambda$, given the experimental outcomes $\bm{x}$. More specifically, the following probability distribution has to be considered: 
\begin{equation}
P^{\mathrm{est}}(\Lambda|\lambda)=\sum_{\bm{x}}  P^{\mathrm{exp}}(\Lambda|\bm{x}) P(\bm{x}|\lambda).
\end{equation}
This expression provides the probability to obtain an estimate $\Lambda$ given the parameter $\lambda$.
In this case the MSE of the estimation becomes:
\begin{equation}
\mathrm{MSE}(\lambda)=\sum_{\Lambda} P^{\mathrm{est}}(\Lambda|\lambda) (\Lambda-\lambda)^2. 
\end{equation}

Quantum resources can be used both for the estimation of continuous unknown parameters and for parameters assuming discrete values, where the goal is to distinguish between them. This latter case corresponds to the quantum channel discrimination problem \cite{chefles2000quantum,pirandola2008computable,takeoka2008discrimination,barnett2009quantum,pirandola2017ultimate,zhuang2017optimum,pirandola2018advances,pirandola2019fundamental}.

\subsubsection{Estimators} \label{sec:estimators}

Different approaches exist to post-process experimental data and provide optimal estimation of the unknown parameter \cite{gianani2019assessing}.

One of the most widely adopted estimator is the \emph{maximum likelihood estimator} (MLE) \cite{fisher1922mathematical}. It is the value of the parameter that, given a list of experimental results $\bm{x}$, maximizes the likelihood probability:
\begin{equation}
\label{mle}
\Lambda^{\text{MLE}}(\bm{x})=\arg\left[\max_{\{\lambda\}}P(\bm{x}|\lambda)\right]\;.
\end{equation}
In the asymptotic limit the MLE is unbiased, consistent and saturates the Cramer Rao bound (see Sec.\ref{seccr}). Other estimators are \emph{Bayesian estimator} or the \emph{Method of Moments}, the latter not requiring full knowledge of the likelihood function \cite{pezze2014}.

The Bayesian approach is a natural framework to devise estimation protocols \cite{kay1993fundamentals,hradil1996quantum,box2011bayesian,berry2000optimal}. In this approach, the unknown parameter $\lambda$ and the experimental result $x$ are treated as random variables. Here the relevant quantity  is the degree of ignorance (or knowledge, equivalently) about the parameter. During a Bayesian estimation such knowledge, that can be regarded as subjective (degree of belief), is updated according to the measurement results. 

The starting point of the process is the \emph{prior distribution} $P(\lambda)$ that quantifies the initial ignorance on the unknown parameter. The experimental setup probing the system is described by the likelihood function $P(x|\lambda)$ [Eq.\eqref{quantumlikelihood}]. Once a measurement result $x$ is obtained, the degree of knowledge, described by the \emph{posterior probability} $P(\lambda|x)$, is updated by the Bayes' rule: \begin{equation}\label{bayes}
 P(\lambda|x)=\frac{P(\lambda)\;P(x|\lambda)}{\int \differential \lambda \; P(\lambda)\;P(x|\lambda)}\;,
\end{equation}
where the integral in the normalization term has to be replaced by a sum when the unknown parameter $\lambda$ assumes discrete values. The posterior in Eq.\eqref{bayes} contains the updated information from which interesting quantities can be calculated.

For instance, the mean square error (MSE) [Eq. \eqref{mse}], of an estimator $\Lambda(x)$, averaged over the parameter $\lambda$ is obtained as:
\begin{equation}\label{msebayes}
\langle \Delta \Lambda^2 \rangle=\int \differential \lambda \differential x \; P(\lambda) P(x|\lambda) \; (\Lambda(x)-\lambda)^2\;.
\end{equation}
By minimizing Eq. \eqref{msebayes}, the optimal Bayesian estimator $\Lambda^{\mathrm{opt}}(x)$ is calculated: 

\begin{equation}\label{meanbayes}
\Lambda^{\mathrm{opt}}(x)= \int \differential \lambda \; \lambda\, P(\lambda|x),
\end{equation}
that corresponds to the mean value of the parameter over the posterior distribution. Also other moments, such as the third moment, of such distribution can be informative on the estimation, especially to detect possible biases \cite{cimini2020diagnosing}. 

The phase shift $\phi$ estimated inside an interferometer is a circular parameter, where $\phi=\phi+2 k \pi$ with $k\in \mathbb{Z}$. 
For such parameter a circular mean, calculated over the posterior distribution, can be defined: 
\begin{equation}
   \langle \phi\rangle^{\mathrm{circ}}=\arg \left[\int\differential \phi e^{i \phi}\, P(\phi|x) \right].
\end{equation}
 
The standard variance with circular variables is no more adequate and the \emph{Holevo variance} $V^{\mathrm{H}}$ can be defined, as function of a quantity $S$ called \emph{Sharpness} \cite{holevo1984covariant}:
\begin{equation} \label{eq:sharpness}
  V^{\mathrm{H}}=S^{-2}-1\qquad \qquad S = |\langle e^{i \phi}\rangle|,
\end{equation}
where the mean $\langle \cdot \rangle$ is calculated over the probability distribution of the estimation process under study. The Holevo variance can describe the variance of unbiased phase estimators, $V^{\mathrm{H}}_{\mathrm{unbias}}=\left\lvert \langle e^{i \Phi}\rangle\right\rvert ^{-2}-1$, and coincides with the standard variance for sufficiently sharply picked distribution. With biased estimators, the variance is: $V^{\mathrm{H}}_{\mathrm{bias}}=\left\lvert \langle\cos (\Phi-\phi)\rangle\right\rvert ^{-2}-1$ .

A fundamental feature of a Bayesian approach is its direct application to adaptive protocols, described in Sec. \ref{qmsec5}. Note that, since a Bayesian approach allows to exploit prior knowledge on the parameters, the sensitivity bounds can be different from those relative to the frequentist approach \cite{van2007bayesian,li2018frequentist}.

\subsection{\label{seccr} Fisher Information and Cramer Rao bound}

Let us consider the estimation scenario where probes and measurements are fixed. A fundamental tool allowing to study the achievable bounds on estimation uncertainties is the \textit{Fisher Information} ($F$). It is a quantity able to catch the amount of information encoded in output probabilities of the estimation process, and is defined as \cite{fisher1922mathematical}:
\begin{equation}
F(\lambda)=\sum_x P(x|\lambda) \left( \frac{\partial \text{log} (P(x|\lambda)}{\partial \lambda}\right)^2 =\sum_x \frac{1}{ P(x|\lambda)} \left( \frac{\partial P(x|\lambda)}{\partial\lambda}\right)^2 .
\end{equation}

Note that all the previous definitions, written for the case of discrete values of measurement outcomes $x$, can be extended to the continuous case in which the sums are replaced by integrals over $x$. 

We introduce a useful quantity called \emph{symmetric logarithmic derivative} (SLD), $L_{\lambda}$, defined as the selfadjoint operator satisfying:

\begin{equation}
\frac{\partial \rho_{\lambda}}{\partial  \lambda}= \frac{(\rho_{\lambda}L_{\lambda}+L_{\lambda}\rho_{\lambda})}{2}.
\end{equation}

The Fisher Information is related to the SLD operator through the following relation: 

\begin{equation}
F(\lambda)=\sum_{x}\frac{\mathrm{Re}[\text{Tr}[\rho_{\lambda}E_x L_{\lambda}]]^2}{\text{Tr}[E_x \rho_{\lambda}]},
\end{equation}
where $E_x$ is a chosen POVM measurement. $F$ has the two following properties \cite{cohen1968fisher,pezze2014,fujiwara2001quantum}:

\textit{(i)} \emph{Convexity}: given a general mixed state $\rho=\sum_j c_j \rho_j$ with $\sum_j c_j=1$ and a fixed measurement, then: $F(\lambda) \le \sum_j c_j F^j(\lambda)$, where  $F(\lambda)$ is the  Fisher Information of the state while $F^j(\lambda)$ is the Fisher Information calculated for the single state $\rho_j$ of the mixture;

\textit{(ii)} \emph{Additivity}: given $\nu$ independent probes measured independently, the $F$ of the total ensemble will be: $F^{\mathrm{tot}}(\lambda)=\sum_i F^i(\lambda)$, where $F^i(\lambda)$ is the Fisher Information relative to the $i$-th probe together with its measurement.

Intuitively, being $F$ proportional to the derivative with respect to the parameter of the output probabilities, it allows to quantify the sensitivity of the system to a change of $\lambda$. More specifically, a larger amount of information are associated to larger variations of the output probabilities. This intuition was formalized with a fundamental result, called \emph{Cram\'er Rao bound} (CRB). It links $F$ to the ultimate bound achievable by the variance of \emph{any} arbitrary estimator, with fixed $\nu$ identical and independent probes and measurements \cite{cramer1946,rao1945information}: 

\begin{equation}
\label{CRBgen}
\Delta\Lambda^2 = \sum_{\bm{x}}(\Lambda(\bm{x})-\bar{\Lambda})^2P(\bm{x}|\lambda) \ge \frac{(\partial \bar{\Lambda}/\partial \lambda)}{ \nu\, F(\lambda)}.
\end{equation}

In the presence of an asymptotically locally unbiased estimator ($\partial \bar{\Lambda}/\partial \lambda=1$), the CRB becomes:

\begin{equation}
\label{CRB}
\Delta\lambda^2  \ge \frac{1}{\nu\, F(\lambda)}.
\end{equation}

An estimator that is able to saturate the inequality \eqref{CRB} is said to be \emph{efficient}. 

In the limit of large number of measurements, the maximum likelihood estimator is efficient, since its distribution normally converges to the real value with a variance that saturates the CRB \cite{lehmann2006theory}.  Also, Bayesian estimator is asymptotically efficient in the limit of large number of probes \cite{lehmann2006theory,van2007bayesian,le2012asymptotic,pezze2014}.  Nevertheless, in the case of limited measurements and data, the saturation of the bound is no more guaranteed\cite{rubio2019non}.

\subsection{Quantum Fisher Information and Quantum Cramer Rao Bound}
\label{subsecQCRB}

In the previous sections we have considered the scenario in which both probes and measurements are fixed, and we have discussed the ultimate limits optimizing for the best possible estimator. In this section, we review the ultimate precision limits obtained by optimizing over all possible measurements.

Given a state $\rho_{\lambda}$ encoding the information on the parameter $\lambda$, one can maximize the Fisher information $F$ over all possible POVMs $E_x$. This defines the \textit{Quantum Fisher Information} ($F_{\mathrm{Q}}$):

\begin{equation}
\label{QFI}
F_{\mathrm{Q}}(\rho_{\lambda})= \max_{\{E_x\}}F(\lambda).
\end{equation}

By definition $F_{\mathrm{Q}}(\lambda)\ge F(\lambda)$ and the CRB can be extended to the \textit{Quantum-CRB} (QCRB). For asymptotically locally unbiased estimators such bound reads:
\begin{equation}
\label{QCRB}
\Delta\lambda^2 \ge \frac{1}{\nu\, F(\lambda)} \ge \frac{1}{\nu\, F_{\mathrm{Q}}(\lambda)}.
\end{equation}

In other words, having a fixed probe state, the right hand side of \eqref{QCRB} represents the ultimate achievable precision bound regardless of the measurement. The only dependence is on the probe state $\rho_{\lambda}$.

$F_{\mathrm{Q}}$ is related to the symmetric logarithmic derivative operator $L_\lambda$. Indeed, it can be demonstrated that the following relation holds \cite{helstrom1976quantum,holevo2011probabilistic,braunstein1994statistical}: 
\begin{equation}
F_{\mathrm{Q}}(\rho_{\lambda})=(\Delta L_\lambda)^2=\text{Tr}[\rho_\lambda L_\lambda^2],
\end{equation}
where $(\Delta L_\lambda)^2$ is the variance of $L_\lambda$ over the state $\rho_{\lambda}$. If we write the probe state in the basis of its eigenstates $\rho_\lambda=\sum_n a_n \ket{\Psi_{n}}\bra{\Psi_{n}}$, then $F_{\mathrm{Q}}$ can be explicitly written as \cite{paris2009quantum}:
\begin{equation}
F_{\mathrm{Q}}(\rho_{\lambda})=\sum_n \frac{(\partial_\lambda a_n)^2}{a_n}+2\sum_{i\ne j}\frac{(a_i-a_j)^2}{a_i+a_j}|\braket{\Psi_i}{\partial_\lambda \Psi_j}|^2\;.
\end{equation}

Analogously to $F$, $F_{\mathrm{Q}}$ satisfies the properties of convexity and additivity:

\begin{eqnarray}
\label{convx}
F_{\mathrm{Q}}(\sum_j c_j \; \rho_\lambda^j) &\le& \sum_j c_j  \; F_{\mathrm{Q}}(\rho_\lambda^j),\\
\label{addit}
F_{\mathrm{Q}}(\bigotimes_i \rho_\lambda^i )&=&\sum_i F_{\mathrm{Q}}(\rho_\lambda^i).
\end{eqnarray}

If the evolution of the interaction is unitary, $\rho_{\lambda}=e^{i \lambda H}\rho_0 e^{-i \lambda H} $, or equivalently $\partial_{\lambda} \rho_{\lambda}=i[\rho_{\lambda}, H]$, where $H$ is an  Hermitian operator, $F_{\mathrm{Q}}$ does not depend on the unknown parameter. Then in this case $F_{\mathrm{Q}}$  is  function only  of the initial state $\rho_0$ and of the generator $H$.  Given $\rho_0=\sum_n b_n \ket{\Phi_{n}}\bra{\Phi_{n}}$, $F_{\mathrm{Q}}$'s explicit expression is \cite{paris2009quantum}:

\begin{equation}
F_{\mathrm{Q}}= 2\sum_{i\ne j}\frac{(b_i-b_j)^2}{b_i+b_j}|\bra{\Phi_i}H\ket{\Phi_j}|^2\;.
\end{equation}

Furthermore, for pure initial states $\rho_{0}=\ket{\Psi_{0}}\bra{\Psi_{0}}$, evolving in unitary evolution, a simple expression for $F_{\mathrm{Q}}$ can be found \cite{holevo2011probabilistic,helstrom1976quantum,braunstein1996generalized,giovannetti2011advances,paris2009quantum}:
\begin{equation}
\label{QFIsimple}
  F_{\mathrm{Q}}(\rho_{\lambda})=4 (\Delta H)^2 \ge (h_{\mathrm{max}}-h_{\mathrm{min}})^2\;,
\end{equation}
where the variance $(\Delta H)^2\equiv \langle (H-\langle H \rangle)^2 \rangle$ is calculated on the initial state $\rho_0$, and $h_{\mathrm{max}}$ and $h_{\mathrm{min}}$ are the maximum and minimum eigenvalues of $H$, respectively. For the more general case of mixed probe states the relation becomes:  $F_{\mathrm{Q}}(\rho_{\lambda})\le4 (\Delta H)^2$, then a mixed probe state cannot perform better than a pure one (see Sec. \ref{qqm} for discussion on optimal probes).

The maximum sensitivity of a quantum state for a parameter estimation is intimately related with the metric of the state \cite{wootters1981statistical,braunstein1994statistical,sidhu2020geometric}. In particular the distinguishability of the probe state for small variation of the parameters is directly linked to $F_{\mathrm{Q}}$.  The distinguishability between two states, $\rho_1$ and $\rho_2$ can be quantified by the normalized Bures distance: $\tilde{D}_{\mathrm{B}}(\rho_1,\rho_2)=\sqrt{1-\tilde{F}(\rho_1,\rho_2)}$, where $\tilde{F}(\rho_1,\rho_2)=\mathrm{Tr}\left[ \sqrt{\sqrt{\rho_1}\rho_2\sqrt{\rho_1}}\right]^2$ is the standard fidelity. Given the state $\rho_\lambda$ and an infinitesimal change $\delta \lambda$ of the parameter, the normalized distance squared between $\rho_\lambda$ and $\rho_{\lambda+\delta \lambda}$ is proportional to $F_{\mathrm{Q}}(\rho_\lambda)$ \cite{sommers2003bures,paris2009quantum}: 

\begin{equation}
\label{qfibures}
  \tilde{D}_\mathrm{B}(\rho_\lambda,\rho_{\lambda+\delta \lambda})^2 = \frac{1}{8}F_{\mathrm{Q}}(\rho_{\lambda})\;( \delta \lambda)^2.
\end{equation}

From this expression it is clear that the more $\rho_\lambda$ and $\rho_{\lambda+\delta \lambda}$ are "distant" (distinguishable) the greater is $F_{\mathrm{Q}} (\rho_{\lambda})$ and thus the sensibility of the state to $\lambda$.

One of the goals of quantum metrology is to find measurements that are able, given a probe state, to reach the ultimate precision and then to saturate the QCRB in Eq. \eqref{QCRB}. This task is equivalent to find the POVM such that the Fisher information $F$ associated to the process becomes equal to the corresponding $F_{\mathrm{Q}}$ associated to the probe state. The aim is then to find the measurement such that $F=F_{\mathrm{Q}}$. Indeed, if a large number of probes is available, the estimators to asymptotically saturate QCRB are known, such as maximum likelihood and Bayesian estimators. 
In the single parameter case it is always possible to saturate the QCRB through suitable measurements \cite{braunstein1994statistical}.
Because of $F_{\mathrm{Q}}$'s additivity property [Eq. \eqref{addit}] it is possible to saturate QCRB using local adaptive measurements  for each probe without entangling measurements  \cite{gill2000state,fujiwara2006strong,hayashi2017quantum,hayashi2008asymptotic,giovannetti2006quantum,nagaoka2005parameter}. Then quantum resources in the measurement stage do not enhance the estimation process \cite{giovannetti2004quantum,giovannetti2006quantum}.
In particular, an optimal choice of POVMs is realized by the projectors over the eigenstates of the symmetric logarithmic derivative $L_\lambda$. Furthermore an explicit expression for the optimal estimator $O_\lambda$ is \cite{paris2009quantum}: 

\begin{equation}
O_\lambda=\lambda \mathbb{1}+\frac{L_\lambda}{F_{\mathrm{Q}}(\lambda)}.
\end{equation}

Since in general the optimal POVMs can depend on its value, it may be necessary to have a priori knowledge on the parameter. This difficulty can be overcome through adaptive estimation protocols (see Sec.\ref{qmsec5}).

\subsection{\label{qqm}  Standard Quantum Limit, Entanglement and Heisenberg Limit}

In the last Section we described the optimization of the estimation precision over all possible quantum measurements. The last step, in order to find the ultimate fundamental bounds, is the optimization over all possible input states. This task can be done by optimizing $F_{\mathrm{Q}}$ over the initial probes.

\begin{figure}[ht!]
\includegraphics[width=.5\textwidth]{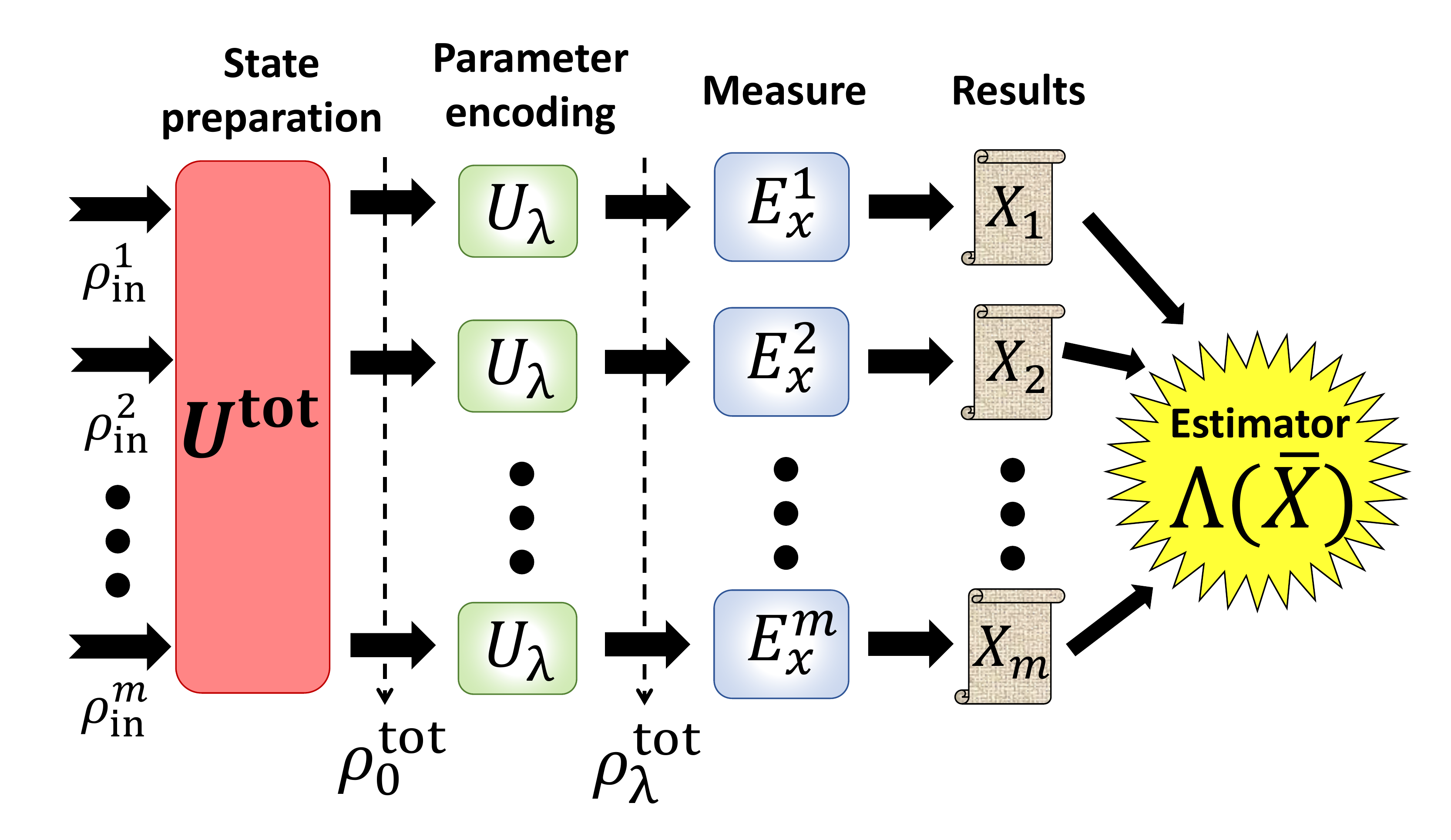}
\caption{{\bf Conceptual scheme of a parallel parameter  estimation}. The measurements here considered are separable. Indeed, empolying entanglement in the measurement process does not allow to obtain better performances than the optimal separable strategy. Conversely, state preparation can lead to quantum enhancement by exploiting entanglement between probes \cite{giovannetti2006quantum}.}
\label{parallel}
\end{figure}

For this analysis we consider the parallel strategy depicted in Fig.\ref{parallel}. Here, $m$ probes interact with the system, independently, with a separable linear unitary $U^{\mathrm{tot}}=\bigotimes_{i=1}^{m}U^{i}_\lambda$, with $U^i_\lambda$ acting only on the $i$-th probe and such that $U^i_\lambda=U_\lambda \; \forall \,i$.

The first property of optimal probes can be derived from the convexity \eqref{convx} of $F_{\mathrm{Q}}$: the maximum of $F_{\mathrm{Q}}$ is always achieved by pure states.

Now we initially focus on $m$ probes that are classically correlated, that is, non entangled. The total state can be then written as $\rho^{\mathrm{tot}}=\rho_1\otimes \rho_2  \cdot\cdot\cdot \otimes\rho_m$. The value of $F_{\mathrm{Q}}$ for this state is:

\begin{equation}
F_{\mathrm{Q}}(\rho_1\otimes \rho_2  \cdot\cdot\cdot \otimes \rho_m)=\sum_i^m F_{\mathrm{Q}}(\rho_i)\le m\;\;F_{\mathrm{Q}}^{\mathrm{max}},
\end{equation}
where for the first equality the additivity of $F_{\mathrm{Q}}$ has been exploited, and $F_{\mathrm{Q}}^{\mathrm{max}}$ represents the maximum of $F_{\mathrm{Q}}$ over the states $\rho_m$. Then, in presence of $\nu$ independent packets of $m$ classical correlated probes, from Eq.\eqref{QCRB} the minimum uncertainty $\Delta\lambda$ scales as \cite{giovannetti2006quantum, pezze2009entanglement}:  

\begin{equation}
\label{SQL}
\Delta\lambda\ge \frac{1}{\sqrt{\nu \;m\; F_{\mathrm{Q}}^{max}}}.
\end{equation}

Since $F_{\mathrm{Q}}^{\mathrm{max}}$ is a constant factor, the error scaling with the number of the probes $m$ is $\Delta\lambda\propto 1/\sqrt{m}$.  Namely this statistical bound is called \textit{Standard Quantum Limit}. Such bound corresponds to the QCRB optimized over any arbitrary classically correlated probe state and can be seen as a consequence of the central limit theorem.

In previous sections we have seen that quantum resources in measurement stage are not necessary to reach the QCRB.
Conversely, quantum resources employed for the preparation of probe states can enhance the sensitivity with respect to classical approaches, beating the SQL \cite{caves1981quantum,braunstein1992quantum, wineland1992spin,lee2002quantum,giovannetti2004quantum,giovannetti2006quantum,giovannetti2011advances}.

In the case of pure states and unitary evolution $e^{-i \lambda H}$, $F_{\mathrm{Q}}$ assumes the form of Eq. \eqref{QFIsimple}. The bound is saturated by states of the form:
\begin{equation}
\label{optstate}
\frac{\ket{h_{\mathrm{max}}}+e^{i\gamma} \ket{h_{\mathrm{min}}}}{\sqrt{2}},
\end{equation} 
where $\ket{h_{\mathrm{max}}}$ and $\ket{h_{\mathrm{min}}}$ are the eigenvectors corresponding to the maximum and minimum eigenvalues $h_{\mathrm{max}}$ and $h_{\mathrm{min}}$, respectively. 
Let us define $\ket{h_{\mathrm{S}}}$ and $\ket{h_{\mathrm{s}}}$ to be the single probe eigenstate of the generator $H$ relative to the maximum and minimum eigenvalues $h_{\mathrm{S}}$ and $h_{\mathrm{s}}$, respectively. Then, if we have $m$ probes, the optimal state in Eq. \eqref{optstate} is realized by 
\cite{giovannetti2006quantum}:
\begin{equation}
\label{optstatenoon}
\frac{\ket{h_{\mathrm{S}}}^{\otimes m}+e^{i\gamma}\ket{h_{\mathrm{s}}}^{\otimes m} }{\sqrt{2}}.
\end{equation}
which is a maximally entangled state.  For such state $(\Delta H)^2= m^2 (h_{\mathrm{S}}-h_{\mathrm{s}})^2/4$. Hence, by using Eq.\eqref{QFIsimple}, we find that $F= m^2 (h_{\mathrm{S}}-h_{\mathrm{s}})^2$. Then, for $\nu$ independent packets of states in Eq. \eqref{optstatenoon}, the QCRB becomes \cite{giovannetti2006quantum}:
\begin{equation}
\label{HL}
\Delta\lambda\ge \frac{1}{\sqrt{\nu} \;m\; (h_{\mathrm{S}}-h_{\mathrm{s}})}.
\end{equation}

The term $(h_{\mathrm{S}}-h_{\mathrm{s}})$ is a constant so, here, the error scaling with the number $m$ of probes is $\Delta\lambda \propto 1/m$, that corresponds to an improvement of the precision by a factor $\sqrt{m}$ with respect to SQL. This enhanced scaling is the ultimate limit on estimation precision and is called \emph{Heisenberg Limit}  (HL). Recently, in Ref. \onlinecite{gorecki2020pi} it was shown that the achievable HL has to be corrected by a constant factor $\pi$ when a finite amount of a-priori information, independent of $m$, is available.

\begin{table}[ht!]
\begin{center}
\begin{tabular}{|l|l|l|l|l|}
\hline
{\bf Quantity} & {\bf Probe} $\bm{\rho_0}$ & {\bf POVM} $\bm{E_x}$ & {\bf Estimator} ${\bm{\Lambda(x)}}$\\
\hline
$\mathrm{MSE}(\lambda)$&fixed&fixed&fixed\\
\hline
$F(\lambda)$&fixed&fixed&optimized\\
\hline
$F_{\mathrm{Q}}(\lambda)$&fixed&optimized&optimized\\
\hline
SQL&classically &optimized&optimized\\
& optimized&&\\
\hline
HL&quantum &optimized&optimized\\
& optimized&&\\
\hline
\end{tabular}
\end{center}
\caption{Table of the relevant metrology quantities, indicating which step of the estimation protocol is optimized.} \label{tabb3}
\end{table}

A key role to obtain quantum enhancement is played by entanglement. In particular, assuming unitary evolution, the relation: 
\begin{equation}
\label{usefent}
F_{\mathrm{Q}}(\rho_0,H)\ge m \; (h_{\mathrm{S}}-h_{\mathrm{s}}),
\end{equation}
is a sufficient condition for the presence of entanglement in the probe state $\rho_0$ \cite{pezze2009entanglement}. It turns out that entanglement, in the considered estimation scheme, is necessary in order to have an enhancement in estimation. Also, Eq.\eqref{usefent} can be exploited to detect entanglement: if a state shows a sensitivity greater than SQL, then it is entangled \cite{lucke2011twin,krischek2011useful,hyllus2012fisher,toth2014quantum}.
This captures the fundamental relation between entanglement and quantum enhanced metrology.
However not all entangled states are able to satisfy inequality \eqref{usefent}. Then such relation defines also the concept of \emph{useful entanglement} for quantum metrology \cite{pezze2009entanglement}. Note that the entanglement can be defined only after that the considered Hilbert space is divided in subsystems \cite{demkowicz2015quantum}. In this way, one can define mode- and particle-entanglement depending on which Hilbert spaces are considered \cite{killoran2014extracting}. For instance, two indistinguishable photons, $1$ and $2$, along two different modes, $a$ and $b$, are mode-separable, $\ket{1}_a \ket{1}_b$ but particle entangled, $\ket{a}_1 \ket{b}_2+\ket{a}_2 \ket{b}_1$. In particular, in Refs. \onlinecite{demkowicz2015quantum,gessner2018sensitivity} the authors investigate the role of mode- and particle-entanglement for quantum-enhanced performances in parameters estimation.

Up to now we have defined the HL scaling for the case of parallel estimation strategies and linear unitary evolutions, in which the Hamiltonian does not generate correlation between different probes. If  we consider schemes with non-linear interactions between probes and system, the scaling can be different \cite{luis2004nonlinear,boixo2007generalized,choi2008bose,zwierz2012ultimate}.
Furthermore, if one exploits resources other than the number of particles, the SQL can also be beaten with non-entangled probes \cite{giovannetti2006quantum}. This is obtained for instance through  multiround protocols \cite{van2007optimal,higgins2007entanglement,resch2007time,berry2009perform,higgins2009demonstrating,afek2010classical}, in which the additional employed resource is the running time of the estimation process.

\section{Photonic Quantum Metrology: Schemes and Platforms}

Quantum metrology and in general quantum information tasks can be realized through different physical systems. Among them, photons possess different properties which render them fundamental in different scenarios \cite{o2009photonic,walmsley2015quantum,pirandola2018advances,pan2012multiphoton,dowling2015quantum,Flamini19rev,slussarenko2019photonic,moreau2019imaging}. 
Indeed, photons present high mobility  and, at the same time, possess very low interaction with the environment (then low decoherence). This makes them the optimal choice for tasks like quantum communication \cite{gisin2007quantum} or for long distance metrological problems such as coordinate transfer \cite{gisin1999spin,chiribella2004efficient,bagan2001aligning} or large interferometers \cite{abadie2011gravitational}.
Furthermore, different degrees of freedom of light can be exploited to encode and extract information. Photonics accomplishes suitable  technologies for the generation, manipulation and detection of quantum states encoded in the various degrees of freedom of light \cite{o2009photonic,Flamini19rev,slussarenko2019photonic}.

The quantized electromagnetic field has an associated  Hamiltonian of a quantum harmonic oscillator, with total energy:

\begin{equation}\label{quantumoscill}
H_{\mathrm{em}}= \sum_{\bm{k}} \hbar \omega_{\vec{k}} (a^\dagger_{\bm{k}}a_{\bm{k}}+\frac{1}{2}),
\end{equation}
where $\vec{k}$ is the wave vector, while $\bm{k}$ represents the electromagnetic mode, that comprise wave vector, polarization, frequency, time bin, and in general any degree of freedom of the field. The operators $a_{\bm{k}}$ and $a^\dagger_{\bm{k}}$ represent, respectively, the annihilation and creation operators of  photons with  energy $\hbar \omega_{\vec{k}}$. Such  operators obey the following bosonic commutation rules: 
\begin{equation}\label{commann}
    [a_{\bm{k_i}},a_{\bm{k_j}}]=[a_{\bm{k_i}}^\dagger,a_{\bm{k_j}}^\dagger]=0 \qquad \qquad 
[a_{\bm{k_i}},a_{\bm{k_j}}^\dagger]=\delta_{ij},
\end{equation}
where $\bm{k_i}$ and $\bm{k_j}$ are two modes of the field. The \emph{number operator} $n_{\bm{k}}$ along mode $\bm{k}$ is represented by: $n_{\bm{k}}=a^\dagger_{\bm{k}}a_{\bm{k}}$, and the energy can be written as: $H_{\mathrm{em}}= \sum_{\bm{k}} \hbar \omega_{\vec{k}} (n_{\bm{k}} +\frac{1}{2})$. 

The eigenstates of the Hamiltonian along mode $\bm{k}$ are the \emph{Fock states}, $\ket{N_{\bm{k}}}$, having fixed photon-number $N_{\bm{k}}$, and corresponding energy $E_{N_{\bm{k}}}= \hbar \omega_{\vec{k}} (N_{\bm{k}}+\frac{1}{2})$. The action of annihilation (creation) operators on Fock states is to destroy (create) a photon along mode $\bm{k}$, according to the relations:
\begin{equation}
    a_{\bm{k}}\ket{N_{\bm{k}}}= \sqrt{N_{\bm{k}}}\ket{N_{\bm{k}}-1}\qquad
 a_{\bm{k}}^\dagger\ket{N_{\bm{k}}}= \sqrt{N_{\bm{k}}+1}\ket{N_{\bm{k}}+1}.
\end{equation}
The number of photons excited in a particular mode is given by the photon-number operator $n_{\bm{k}}$:
\begin{equation}
n_{\bm{k}}\ket{N_{\bm{k}}}=a^\dagger_{\bm{k}}a_{\bm{k}}\ket{N_{\bm{k}}}=N_{\bm{k}}|N_{\bm{k}}\rangle.
\end{equation}

Since the photon-number operators corresponding to different modes are commuting observables [see relations \eqref{commann}], and each acts only on the corresponding mode, it is possible to completely describe the whole radiation field, at fixed number of photons along $d$ modes, by taking the tensor product of the individual states:
\begin{equation}\label{focktot}
|\{N_{\bm{k}}\}\rangle=\prod_{\bm{k}}|N_{\bm{k}}\rangle=|N_{\textbf{k}_1}\rangle|N_{\textbf{k}_2}\rangle\ldots|N_{\textbf{k}_d}\rangle\equiv|N_{\textbf{k}_1}N_{\textbf{k}_2}\ldots N_{\textbf{k}_d}\rangle.
\end{equation}
Note that, since in this notation a mode comprises all degrees of freedom,  the photons along each single mode $\bm{k_i}$  ($i=1,...,d$) in Eq.\eqref{focktot},  are \emph{indistinguishable}.

The state in which the occupation numbers of all modes are $0$ is called  \textit{vacuum} state $|\{0\}\rangle\equiv\ket{0}$, defined as the state such that $a_{\bm{k}}\ket{0}=0\;\;\forall \bm{k}$. We can then generate any Fock state from  vacuum by iteratively applying creation operators on the modes:
\begin{equation}\label{Fockannihi}
|N_{\bm{k}}\rangle=\frac{a^{\dagger\, N_{\bm{k}}}_{\bm{k}}}{\sqrt{N_{\bm{k}}!}}|0\rangle.
\end{equation}

In second quantization the dimensionless position-like and momentum-like operators $x_{\bm{k}}$ and  $p_{\bm{k}}$, also called \emph{quadratures}, can be defined and expressed as a function of annihilation and creation operators:
\begin{equation}\label{quadratures}
x_{\bm{k}} =a_{\bm{k}}+a_{\bm{k}}^\dagger\;\quad\qquad p_{\bm{k}} =-i(a_{\bm{k}}-a_{\bm{k}}^\dagger).
\end{equation}
Note that $x_{\bm{k}}$ and $p_{\bm{k}}$ are Hermitian operators, and therefore represent field observables. The commutation relations follow from \eqref{quadratures} and \eqref{commann}:
\begin{equation}
\label{commrelquadratures}
    [x_{\bm{k_l}},x_{\bm{k_j}}]=[p_{\bm{k_l}}^\dagger,p_{\bm{k_j}}^\dagger]=0 \quad \qquad 
[x_{\bm{k_l}},p_{\bm{k_j}}^\dagger]=2i \delta_{lj}.
\end{equation}
Equivalently to Eq.\eqref{quantumoscill}, we can write the energy as function of  $x_{\bm{k}}$ and $p_{\bm{k}}$: 
\begin{equation}\label{quantumoscill2}
H_{\mathrm{em}}= \sum_{\bm{k}} \frac{\hbar \omega_{\vec{k}}}{4} (p_{\bm{k}}^2+x_{\bm{k}}^2).
\end{equation}
Quadratures are useful to describe photonic  states  in the \emph{phase space}, in particular this formalism provides insights for the study of continuous variable states (see Sec. \ref{cvsqueezed}). General quadratures rotated by an angle $\theta$ are defined as: 
\begin{equation}\begin{split}\label{rotatedquadratures}
x_{\bm{k}}(\theta) =e^{-i\theta}a_{\bm{k}}+e^{i\theta}a_{\bm{k}}^\dagger\;\qquad\; p_{\bm{k}}(\theta)& =-i(e^{-i\theta}a_{\bm{k}}-e^{i\theta}a_{\bm{k}}^\dagger)
\\
x_{\bm{k}}(\theta)=\cos \theta\, x_{\bm{k}}+\sin\theta\, p_{\bm{k}}.
\end{split}
\end{equation}

The electromagnetic field can undergo different evolutions. A common and important class of transformations is represented by the linear and bilinear ones that allow  mode operators to evolve through an arbitrary Bogoliubov transformation  \cite{weedbrook2012gaussian}:  $a_{\bm{k}}\rightarrow  \sum_{\bm{j}}(\eta_{\bm{k}\bm{j}}a_{\bm{j}}+\beta_{\bm{k}\bm{j}}a_{\bm{j}}^{\dagger})+\gamma_{\bm{k}}$, where the matrices $\eta_{\bm{j}}$, $\beta_{\bm{j}}$ are related by the so called Bloch-Messiah reduction for bosons \cite{braunstein2005squeezing}.
The Hamiltonian evolution of such processes, in a space of $d$ different modes, has the following form \cite{ferraro2005gaussian}: 
\begin{equation}\label{bilinearev}
H= \sum_{\bm{k}=1}^d g_{\bm{k}}^{(1)}a_{\bm{k}}^\dagger+\sum_{\bm{k}>{\bm{l}}=1}^d g_{\bm{kl}}^{(2)}a_{\bm{k}}^\dagger a_{\bm{l}}+\sum_{\bm{k}\bm{l}=1}^d  g_{\bm{kl}}^{(3)} a_{\bm{k}}^\dagger a_{\bm{l}}^\dagger+h.c.\;,
\end{equation}
where $g^{(i)}$ represent the coefficients of the corresponding evolution terms.

The second evolution term $\sum_{\bm{k}>{\bm{l}}}g_{\bm{kl}}^{(2)}a_{\bm{k}}^\dagger a_{\bm{l}}+h.c.$ conserves photon number and describes linear mixing between modes. Such operations can be implemented by passive optical elements such as beam splitter and phase shifters [Eq.\eqref{scw}].
Conversely, the first and third terms, $\sum_{\bm{k}}g_{\bm{k}}^{(1)}a_{\bm{k}}+h.c.$ and $\sum_{\bm{k}\bm{l}} g_{\bm{kl}}^{(3)} a_{\bm{k}}^\dagger a_{\bm{l}}^\dagger+h.c.$, describes transformations that do not conserve the total number of photons and are associated to displacement and squeezing operators, respectively (see Sec. \ref{cvsqueezed}).

Photons represent a fundamental probe for quantum metrology. In particular a paradigmatic scenario is phase estimation, in which the unknown parameter is a phase shift between different optical modes.

\subsection{Phase estimation problem}
\label{qmsec2}

One of the most relevant scenarios for quantum metrology is phase estimation \cite{d1998general,giovannetti2004quantum,pezze2014}. The problem consists in estimating an unknown phase shift $\phi$ between two different modes, such as polarization, OAM or different paths. A lot of physical problems can be cast in a general phase shift estimation, and different physical probes can be employed. Tasks such as measurements of atomic properties \cite{ekstrom1995measurement,bouchendira2011new}, atomic clocks \cite{diddams2004standards,ludlow2015optical}, measurements of forces \cite{peters1999measurement,gustavson1997precision,gietka2019supersolid}, require the use of atomic probes \cite{pezze2018quantum}.
Conversely, for tasks like the estimation of phase shifts produced by gravitational waves \cite{abadie2011gravitational},  lithography \cite{boto2000quantum,fonseca1999measurement,kawabe2007quantum,d2001two},  imaging \cite{kolobov1999spatial,lugiato2002quantum,brambilla2008high,treps2002surpassing,giovannetti2009sub,genovese2016real,moreau2019imaging,tsang2016quantum,meda2017photon,bisketzi2019quantum,samantaray2017realization,schwartz2013superresolution,ono2013entanglement,juffmann2016multi,tenne2019super,berchera2018quantum,yang2016far,li2018enhanced,bonsma2019realistic,sabines2017sub,sabines2019twin,gregory2020imaging}, sensing on biological systems \cite{taylor2016quantum}, quantum key distribution \cite{inoue2002differential}, measurements of velocity, displacements and lengths \cite{giovannetti2011advances}, photons are the most suitable systems. Besides the practical applications, phase estimation represents also a standard benchmark for general metrological protocols. 

Consider an estimation of a phase shift between two paths. The transition of a system through a phase shift along a mode, say mode 1, is described by the unitary evolution: 

\begin{equation}
\label{phaseshift}
U_{\mathrm{ps}}=e^{i \phi H_{\mathrm{ps}}}=e^{i \phi a_1^\dagger a_1 },
\end{equation}
where $a_1$ is the particle annihilation operator along  mode 1. The generator and conjugated operator \cite{shapiro1991quantum} of the phase shift is the number operator $n_1$ along the corresponding mode: 

\begin{equation}
\label{number}
H_{\mathrm{ps}}= a^\dagger_1 a_1 = n_1.
\end{equation}

For the number operator $n_1$, the difference of possible eigenvalues, with a single probe, is $h_{\mathrm{S}}-h_{\mathrm{s}}=1$. Then the SQL, Eq. \eqref{SQL}, for phase estimation reads: 

\begin{equation}
\label{pssql}
\Delta\phi_{\mathrm{SQL}}\ge \frac{1}{\sqrt{\nu \;m}}.
\end{equation}
corresponding to the standard quantum limit. Conversely, the HL then reads:
\begin{equation}
\label{pshl}
\Delta\phi_{\mathrm{HL}}\ge \frac{1}{\sqrt{\nu} \;m}.
\end{equation}
Since a general definition of a standard selfadjoint operator associated to phase shift measurement  is problematic, its direct sharp measurement is not possible \cite{lynch1995quantum,pegg1997quantum,barnett1986phase}. Nevertheless, a phase shift can be treated as an evolution parameter and estimated from other observables whose values depend on it.  
In particular in optical phase estimation, the phase shifts is a difference between optical paths that can be estimated  through interferometers. One of the most common and simple two-mode optical interferometers, suitable for phase estimation, is the \emph{Mach-Zehnder interferometer} (MZI) \cite{pezze2006phase,pezze2008mach}. 

The two key elements of a MZI are the \emph{phase shifter} (PS) and the \emph{beam splitter} (BS). The former adds a phase shift $\phi$ between two modes whose annihilation operators are $a_1$ and $a_2$. The beam splitter (BS) represents a basic optical element that allows mixing between two input electromagnetic modes. It can be realized with a partially reflective mirror that transmits or reflects the incoming light. In particular we consider here the balanced BS whose transmission and reflection probabilities are equal to $0.5$.

The action of these elements are described by:
\begin{equation}
\label{psbs}
\text{PS}(\phi)=\begin{pmatrix}
1&0\\
0&e^{i\phi}\\
\end{pmatrix}\quad\quad
\text{BS}_{\pm}=\frac{1}{\sqrt{2}}\begin{pmatrix}
1&\pm i\\
\pm i&1\\
\end{pmatrix},
\end{equation}
where BS$_{+}$ and BS$_{-}$ differ of an irrelevant, for our purpose, phase shift.
The mode operator $b_i^\dagger$ generated by  a unitary evolution $U$ on modes $a_k^\dagger$,  will be: $b_i^\dagger= \sum_{k}U_{ik}\,a_k^{\dagger}\, $. The cascaded combination of these two elements can  realize any unitary linear operation in arbitrary dimension \cite{reck1994experimental,kok2007linear,clements2016optimal}. Such decompositions represent the basis for the realization of universal linear optics circuits \cite{carolan2015universal}.

\begin{figure}[ht!]
\includegraphics[width=.5\textwidth]{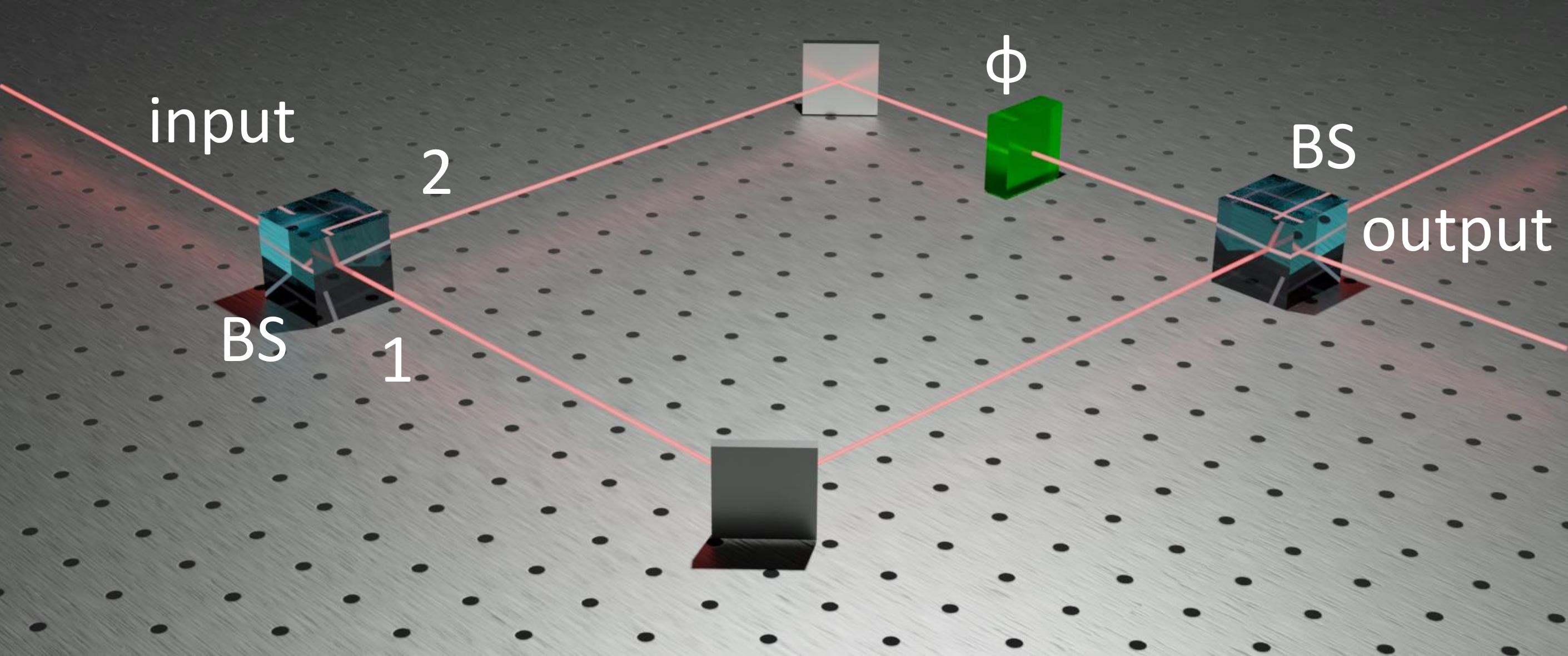}
\caption{{\bf Scheme of a Mach-Zehnder interferometer.} A MZI is composed of two beam splitters (BS) and a phase shift $\phi$ between the modes, 1 and 2, of the interferometer.}
\label{mziimm}
\end{figure}

A MZI interferometer is composed of cascaded two BS interspersed with a PS (Fig.\ref{mziimm}).
In the lossless scenario, up to a global phase, it is described by: 
\begin{equation}
\label{mzieq}
\text{MZI}(\phi)=\text{BS}_{+}\,\text{PS}(\phi)\,\text{BS}_{-}=\begin{pmatrix}
\cos(\frac{\phi}{2})&-\sin(\frac{\phi}{2})\\
\sin(\frac{\phi}{2})&\cos(\frac{\phi}{2})\\
\end{pmatrix}.
\end{equation}
The first BS can be seen as a preparation step of the estimation process, while the last one as part of the measurement step. In general the output probabilities of photons exiting from a MZI depend on the phase $\phi$. Since the Fisher Information depends on the derivatives of the output probabilities, the probe is more sensitive to a phase shift change for larger variation of the fringe pattern.

A convenient way to express the MZI operation on electromagnetic modes is based on Pauli matrices expressed through the annihilation operators for modes 1 and 2 ($a_1$ and $ a_2$): $  \sigma_x= a_1^\dagger a_2+ a_2^\dagger a_1$, $   \sigma_y=-i( a_1^\dagger a_2- a_2^\dagger a_1)$ and $  \sigma_z= a_1^\dagger a_1- a_2^\dagger a_2$. 
The following relations hold \cite{yurke19862, sanders1995optimal}:
\begin{equation} \begin{split}
\label{scw}
\mathrm{PS}(\phi)=e^{-\imath\phi \sigma_z/2}\quad\;
\mathrm{BS}_{\pm}=e^{\pm \imath\pi \sigma_x/4}\quad\; \\
\mathrm{MZI}(\phi)=\mathrm{BS}_{+}\,\mathrm{PS}(\phi)\,\mathrm{BS}_{-}=e^{- \imath \phi  \sigma_y/2}.
\end{split}
\end{equation}
Two cascaded independent PSs interspersed by a MZI can realize any unitary belonging to Lie SU(2) group. MZI transformation is used also as interferometer in other degrees of freedom like polarization, for which the BSs are replaced by HWPs rotated by $22.5^\circ$.

Besides the applications to quantum metrology and in general to quantum information tasks, a MZI can be also the testbed for foundational tests, like those exploring wave-particle duality of photons \cite{shadbolt2014testing,Ma,huang2019compatibility,chaves2018causal,polino,yu2019realization,Adil,wang2019quantum,qureshi2019coherence} or even quantum gravity phenomena when the probes are massive systems \cite{marletto2017gravitationally,bose2017spin,christodoulou2019possibility}.

\subsection{States and schemes}

We describe now some important states of light used for phase estimation schemes. We note that attention has to be paid to the accounting of the external phase reference in the study of  sensitivity in  optical interferometry \cite{jarzyna2012quantum}. A complete review on the limits achievable in quantum optical interferometry can be found in Ref. \onlinecite{demkowicz2009quantum}.

\subsubsection{Coherent states}
\label{coherent}

Among the most relevant classical states are \textit{coherent states}. A coherent state $|\alpha\rangle$ of a single radiation field mode is the eigenstate of the annihilation operator $a$:  
\begin{equation}
a|\alpha\rangle=\alpha|\alpha\rangle,
\end{equation}
where $\alpha \in \mathbb{C}$ is called \emph{displacement}. If we expand the state in terms of the complete basis of number states and substitute the expression into the eigenvalues equation, we find the general expression:
\begin{equation}
\label{coherentSTATE}
|\alpha\rangle=e^{-\frac{|\alpha|^2}{2}}\sum_{N=0}^{\infty}\frac{\alpha^N}{\sqrt{N!}}|N\rangle.
\end{equation}
The coherent state is not an eigenstate of the Hamiltonian in Eq. \eqref{quantumoscill}. Hence, the continuous eigenvalues $\alpha$ of $a$ are time-dependent, and their values change following the time evolution of the system $|\alpha(t)\rangle=e^{i\omega t}|\alpha\rangle$. Mean values of quadratures are related to $\alpha$ by: $\alpha=(\langle x \rangle + i \langle p \rangle)/2$ .
Since $|\alpha\rangle$ is a linear superposition of states with different number of photons, it does not possess a definite number of photons. However, the expectation value and variance of the photon number operator $n$ in such a state can be calculated as:
\begin{equation}
\langle n\rangle=\langle\alpha| a^{\dagger}a|\alpha\rangle=|\alpha|^2,
\end{equation}
\begin{equation}
\Delta n^2=\langle n^2\rangle-\langle n\rangle^2=\langle\alpha |a^{\dagger}a\;a^{\dagger}a|\alpha\rangle-|\alpha|^4=|\alpha|^2,
\end{equation}
Hence, that the root-mean-squared deviation is $\Delta n=\sqrt{\langle n\rangle}$. This is a typical property of the \textit{Poisson} distribution. Indeed, it can be shown from Eq. \eqref{coherentSTATE}, that the photon number probability distribution is
\begin{equation}
P(N)=|\langle N|\alpha\rangle|^2=e^{-{\langle n\rangle}}\sum_{N=0}^{\infty}\frac{(|\alpha|^2)^N}{N!}=e^{-{\langle n\rangle}}\sum_{N=0}^{\infty}\frac{\langle n\rangle^N}{N!},
\end{equation}
with mean value and variance equal to $\langle n\rangle=\Delta n^2=|\alpha|^2$. The relative value then decreases as $\Delta n/\langle n\rangle=1/\sqrt{\langle n\rangle}$. Analogously to the vacuum state, coherent states symmetrically saturate Heisenberg relation for quadratures. Hence, they are states of minimum uncertainties with $\Delta x= \Delta p=1$. 

Coherent states can be also expressed through the displacement operator $D(\alpha)$. Such operator is associated to the first term of the evoluion in Eq. \eqref{bilinearev}, being defined by the relation:
\begin{equation}\label{displacement}
D(\alpha)=e^{\alpha a^\dagger-\alpha^* a}\;.
\end{equation}
The annihilation and quadrature operators, along mode $\bm{k}$ and under displacement evolution, are transformed in the following way:
\begin{equation}\label{displacementEvolution}\begin{split}
&a_{\bm{k}}\xrightarrow[]{D(\alpha)} a_{\bm{k}}+\alpha \qquad \qquad   a_{\bm{k}}^\dagger \xrightarrow[]{D(\alpha)} a_{\bm{k}}^\dagger+\alpha^*
\\
&x_{\bm{k}}\xrightarrow[]{D(\alpha)} x_{\bm{k}}+\Re[\alpha] \qquad \qquad   p_{\bm{k}} \xrightarrow[]{D(\alpha)} p_{\bm{k}}+\Im[\alpha].
\end{split}
\end{equation}
Then $D(\alpha)$ produces a translation in phase space without changing the uncertainties.
A coherent states can be seen as the result of a displacement operation applied to vacuum state: 
\begin{equation}
\label{coherentdisplace}
|\alpha\rangle=D(\alpha)\ket{0}\;.
\end{equation}

An experimental approximation of the displacement operation on a target state can be obtained by injecting a coherent state 
$|\alpha\rangle$, used as ancilla, in an unbalanced beam splitter together with the target \cite{kim1998scheme}. 

Furthermore these states describe with good approximation the light emitted from lasers and said to be "semi-classical" for their statistical properties \cite{gerry2005introductory}. Therefore, in many contests, coherent states represent a suitable classical benchmark to be surpassed in order to certify quantum performances of other states. In this spirit, the statistics of a state is called non-classical if it cannot be simulated by proper mixtures of coherent ones \cite{shchukin2005nonclassicality}.

When applied to phase estimation problem, coherent states can reach the SQL but not the HL. The scaling of the error on a phase estimation with coherent states is $1/\sqrt{N}=1/|\alpha|$, consistent with the SQL. 

\subsubsection{N00N states} 

One of the most important and paradigmatic classes of quantum states with fixed number of particles, which enables quantum enhanced phase estimation, is represented by the so-called N00N states, that are maximally entangled multipartite states distributed along two modes \cite{lee2002quantum,bollinger1996optimal}: 
\begin{equation}
\label{NOON}
\ket{\Psi}_{\text{N00N}}\equiv \frac{\ket{N,0}+e^{i\gamma}\ket{0,N}}{\sqrt{2}},
\end{equation}
where $N$ is the number of particles of the state and $\gamma$ is the relative phase between the two components of the balanced superposition. The variance of the number operator calculated on $\ket{\Psi}_{\text{N00N}}$ is:  $(\Delta n)^2=N^2/4$, hence, from Eq. \eqref{QFIsimple}, the sensitivity achievable by N00N states is:
\begin{equation}
\Delta\phi_{\text{N00N}}\ge \frac{1}{N},
\end{equation}
that is, the Heisenberg limit in Eq. \eqref{pshl}. For this reason, N00N states play a key role in quantum metrology and in particular in phase estimation processes \cite{lee2002quantum,bollinger1996optimal,mitchell2005metrology}. In an ideal interferometer, with a relative phase shift $\phi$, a N00N state evolves as $(\ket{N,0}+\ket{0,N})/\sqrt{2} \xrightarrow{U_{\phi}}(\ket{N,0}+e^{iN\phi}\ket{0,N})/\sqrt{2}$, thus acquiring an amplified shift equal to $N\phi$. This faster change in the phase shift, proportional to the number of photons, is at the basis of the improved metrological performances of such class of states.

For the particular case of N00N states with $N=2$, there exists a deterministic generation recipe. This is possible by exploiting indistinguishability of photons which, entering along the inputs of a beam splitter, interfere through the \emph{Hong-Ou-Mandel effect} (HOM) \cite{hong1987measurement}. Consider two monochromatic photons which share the same degrees of freedom (frequency, polarization, and so on...) apart from spatial modes. Each photon is injected along each of two inputs of a beam splitter, respectively. If the beam splitter is symmetric, the probability to find the photons in different output modes is zero. This is a \emph{bunching} effect due to purely quantum interference. The final state at the output of symmetric BS will be:
\begin{equation}
\ket{\Psi}_{\mathrm{out}}= \frac{i}{\sqrt{2}}(\ket{2,0}+\ket{0,2}),
\end{equation}
that is, a N00N state with $N=2$. This kind of effect is due to bosonic quantum interference happening inside a BS. Indistinguishability of photons then leads to entanglement and a BS acts as an entangling gate, even if the two photons are independent and not properly interacting. In general, indistinguishability is considered a quantum resource \cite{tsujino2004distinguishing,killoran2014extracting,franco2016quantum,franco2018indistinguishability,castellini2019indistinguishability,morris2019entanglement}. Unfortunately no equivalent deterministic schemes exist to generate N00N states with $N>2$.

\begin{figure}[ht]
\centering
  \includegraphics[width=.5\textwidth]{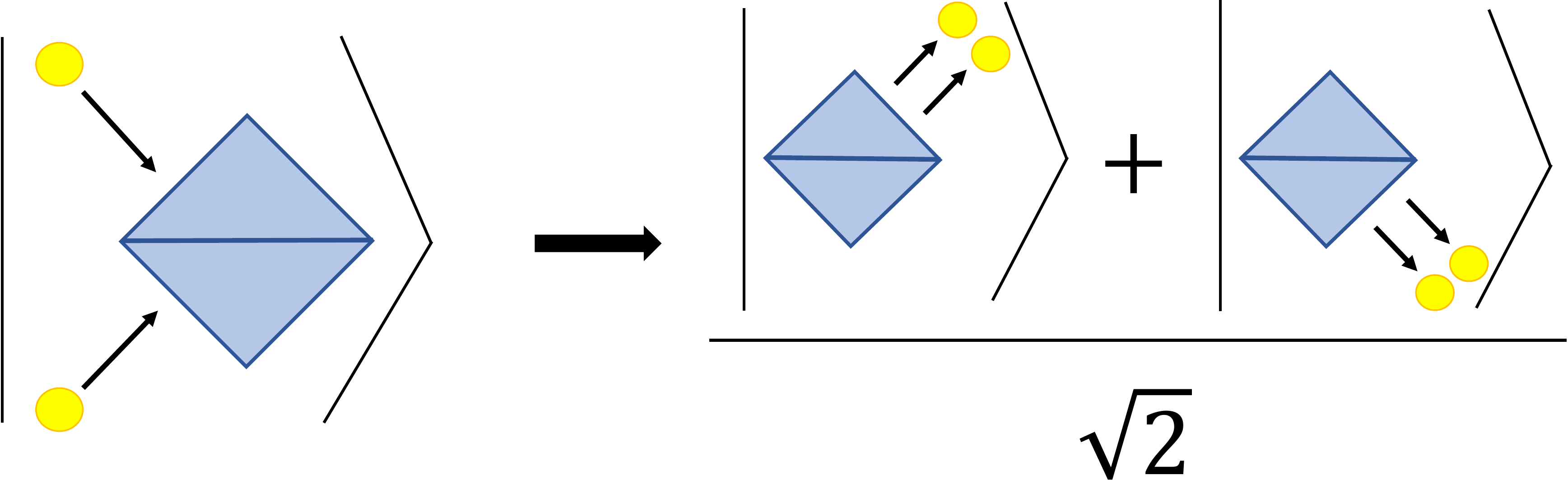}
 \caption{{\bf Scheme of HOM effect with symmetric beam splitter.} Two indistinguishable photons are injected along the two input of a symmetric beam splitter. The final state is a balanced superposition of states in which the two photons are along the same output mode.}
\label{ou}
  \end{figure}

N00N states are optimal for \emph{local estimation} when the unknown phase shift is small. However, such states cannot distinguish phase shifts that differ of $\pi/N$ without a prior knowledge on the unknown phase shift \cite{lee2002quantum,bollinger1996optimal,durkin2007local}. 

\subsubsection{Continuous variable states} \label{cvstates}
Fock states are expressed in the discrete variable (DV) formalism of mode operators $a_{\bm{k}}$. However, in order to describe some states and processes, it is useful to employ an equivalent formalism, based on continuous variables (CV) \cite{adesso2014continuous,braunstein2005quantum,ferraro2005gaussian,kwon2019nonclassicality,yadin2018operational}. A convenient choice of CV is represented by quadrature operators $x_{\bm{k}}$ and $p_{\bm{k}}$ in Eq.\eqref{quadratures}, defining the phase space. Such observable operators do not commute [Eq.\eqref{commrelquadratures}] and then have to satisfy the Heisenberg relation. The latter imposes the following constraint on uncertainties: 
\begin{equation}\label{heisenbquadratures}
\Delta x_{\bm{k}}\Delta p_{\bm{k}}\ge1.
\end{equation}

Contrary to photon number operator, the spectrum of the quadrature operators is continuous and their eigenstates form a complete orthogonal basis:   $\{\ket{X}\}$  of $x$ with eigenvalues $\{X\}$ and $\{\ket{P}\}$  of  $p$ with eigenvalues $\{P\}$. Such states satisfy the following conditions: $\langle X_1|X_2\rangle=\delta(X_1-X_2)$ and $\langle P_1|P_2\rangle=\delta(P_1-P_2)$. Since $x$ and $p$ are conjugate observables, their eigenstates $\ket{X}$ and $\ket{P}$ are related by a Fourier transformation.

In order to describe a general state $\rho$ in quantum phase space, a useful tool is provided by quadratures quasi-probability distributions  $P(\bm{X},\bm{P})$. Among the possible quasi-distributions  \cite{schleich2011quantum,cahill1969density} one of the most used is the \emph{Wigner function} \cite{wigner1932quantum,schleich2011quantum,braunstein2005quantum,gerry2005introductory,demkowicz2015quantum,tilma2016wigner,weinbub2018recent} that, for a $d$-mode state $\rho$, is defined as:
\begin{equation}\begin{split}
&W_\rho(\bm{X},\bm{P})= 
\\&=\frac{1}{(2\pi^2)^d}\int_{-\infty}^{+\infty}d^d\bm{X}'\; d^d\bm{P}'\;Tr\left[ \rho\; e^{i[\bm{P}'(\bm{x}-\bm{X})-\bm{X}'(\bm{p}-\bm{P})]} \right],
\end{split}
\end{equation}
where $\bm{X}=(X_1,...,X_d)$ and $\bm{P}=(P_1,...,P_d)$ are the values assumed by quadratures operators $\bm{x}=(x_1,...,x_d)$ and $\bm{p}=(p_1,...,p_d)$, respectively, along the $d$ modes. The Wigner function is normalized: $\int_{-\infty}^{+\infty}d^d\bm{X}\; d^d\bm{P}\;W_\rho(\bm{X},\bm{P})=1$. From $W_\rho(\bm{X},\bm{P})$ one can recover the real marginal quadratures distributions:
\begin{equation}\begin{split}
P(\bm{P})=\int_{-\infty}^{+\infty}d^d\bm{X}\; \;W_\rho(\bm{X},\bm{P}),\\
P(\bm{X})=\int_{-\infty}^{+\infty}d^d\bm{P}\; \;W_\rho(\bm{X},\bm{P}).
\end{split}
\end{equation}
However the global Wigner function $W_\rho(\bm{X},\bm{P})$ is not a proper distribution and can also assume negative values. Indeed the joint probability distribution $P(\bm{X},\bm{P})$ for two no- commuting quantum operators $\bm{x}$ and $\bm{p}$ cannot be properly defined because of Heisenberg uncertainty relations \eqref{heisenbquadratures}. In particular, the negativity of Wigner function can be used to certify the non-classical nature of quantum states.
Different experimental techniques are available to measure and reconstruct Wigner functions of photonic states \cite{banaszek1999direct,breitenbach1997measurement,lvovsky2001quantum,schleich2011quantum,ciampini2017wigner}.

An important class of CV states useful also for quantum metrology is represented by Gaussian states \cite{pinel2012ultimate,pinel2013quantum,weedbrook2012gaussian,braunstein2005quantum, ferraro2005gaussian,monras2006optimal}, that are described by a Wigner function corresponding to a multidimensional Gaussian distribution. Then, in order to characterize a Gaussian state, it is sufficient to acquire knowledge of the two moments of the associated distribution. The simplest Gaussian state is the vacuum state $\ket{0}$, with zero valued mean quadratures, $\langle x \rangle=\langle p \rangle=0$ and minimum symmetric uncertainties saturating relation \eqref{heisenbquadratures}: $\Delta x\; \Delta p=1$. Conversely, thermal states are Gaussian states with $\langle x \rangle=\langle p \rangle=0$ but their uncertainties do not saturate Eq.\eqref{heisenbquadratures}. Restricting on a single mode, they can be expressed in the photon number basis as an incoherent mixture:
\begin{equation}\label{thermal}
\rho^{\mathrm{th}}= \sum_{N=0}^\infty \frac{\langle n \rangle ^N}{(1+\langle n \rangle)^{1+N}}\ket{N}\bra{N},
\end{equation}
where $\langle n \rangle$ is the mean photon number. The photon-number probability $P(N)$ follows the distribution describing black-body radiation: $P(N)=\frac{\langle n \rangle ^N}{(1+\langle n \rangle)^{1+N}}$ in which $\langle n \rangle=1/(e^{\frac{\hbar}{K_B T}}-1)$, where $K_\mathrm{B}$ is the Boltzmann constant and $T$ is the temperature. The quadratures fluctuations of such states are \cite{weedbrook2012gaussian}: $\Delta x^2 =\Delta p^2=1+2\langle n \rangle$.

An intense research activity is devoted to study the interferometric properties of Gaussian states \cite{adesso2014gaussian,sparaciari2015bounds,monras2006optimal,pinel2013quantum,aspachs2009phase,vsafranek2015quantum,friis2015heisenberg,olivares2018high,oh2019optimal}. In the next three sections we provide a description of some of the most used Gaussian states for quantum metrology tasks. Different reviews providing more detailed description of Gaussian states are available, such as those in Refs. \onlinecite{ferraro2005gaussian,braunstein2005quantum,weedbrook2012gaussian,adesso2014continuous}.

\subsubsection{Squeezed states}  \label{cvsqueezed}

Squeezed light states are among the most used Gaussian states to enhance quantum metrology tasks   \cite{jaekel1990quantum,braunstein2005quantum, ferraro2005gaussian,pezze2013ultrasensitive,lvovsky2015squeezed,schnabel2017squeezed,xu2019sensing,maccone2019squeezing,lawrie2019quantum}. A quantum state is said to be squeezed when an observable on this state presents a fluctuation (second moment of the Wigner distribution) that is lower than that of vacuum state. Generally, the continuous squeezed variables for quantum metrology tasks are quadratures. Furthermore, squeezed states saturate the Heisenberg relation and are minimum uncertainty states. However, such saturation is asymmetric due to squeezing of one quadrature and the corresponding anti-squeezing of the conjugate one:
\begin{equation}
    \begin{cases}
    \Delta x<1\;\;\mathrm{and}\;\;\Delta p>1\\\qquad or\\\Delta x>1\;\;\mathrm{and}\;\;\Delta p<1
    \end{cases} \qquad \qquad  \Delta x\; \Delta p=1.
\end{equation}

\paragraph{Single mode squeezed states}

All single mode states saturating Heisenberg inequality are called \emph{single mode squeezed states} \cite{walls2007quantum}. Hence, coherent and vacuum states are a particular cases of such class of states. The operation of quadratures uncertainties squeezing is related to the third term in Eq.\eqref{bilinearev}, and can be described through the squeezing single mode operator $S(r)$:

\begin{equation}\label{squeezingOp}
S(r)=e^{\frac{1}{2}r\,a^{\dagger 2} -\frac{1}{2}r^*\,a^2},
\end{equation}
where $r=|r|e^{i\theta}$, with $\vert r \vert$ called  \emph{squeezing factor} and $\theta$ being the \emph{squeezing angle}. The annihilation and quadrature operators, under squeezing evolution along mode $\bm{k}$, are transformed following the relations:
\begin{equation}\label{squeezingEvolution}\begin{split}
&a_{\bm{k}}\xrightarrow[]{S(r)} \cosh |r| \;a_{\bm{k}}+e^{i \theta}\sinh |r| \;a_{\bm{k}}^\dagger \\
&a_{\bm{k}}^\dagger \xrightarrow[]{S(r)} \cosh |r| \;a_{\bm{k}}^\dagger+e^{-i \theta}\sinh |r| \;a_{\bm{k}}
\\
&x_{\bm{k}}\xrightarrow[]{S(r)} \cosh |r| \;x_{\bm{k}}+\cos \theta\;\sinh |r|\; x_{\bm{k}}+ \sin \theta\;\sinh |r|\; p_{\bm{k}} 
\\
&p_{\bm{k}}\xrightarrow[]{S(r)} \cosh |r| \;p_{\bm{k}}+\sin \theta\;\sinh |r|\; x_{\bm{k}}-\cos \theta\;\sinh |r|\; p_{\bm{k}}.
\end{split}
\end{equation}
In the case of $\theta=0$, the uncertainty on $p$ is squeezed by a factor $e^{|r|}$: $\Delta p ^2 \xrightarrow[]{S(|r|,\theta=0)} \Delta p^2 /e^{|r|}$
while that of $x$ is anti-squeezed by the same factor: $\Delta x ^2\xrightarrow[]{S(|r|,\theta=0)} e^{|r|} \Delta x^2$. Therefore, the squeezing factor can be estimated as the ratio between the uncertainty before and after squeezing, reducing exponentially one quadrature uncertainty at the expense of the other.  For this reason it represents the most relevant parameter in squeezing generation and it is commonly expressed in dB ($(\Delta p_{\mathrm{squeezed}}/\Delta p_{\mathrm{not}\,\mathrm{squeezed}})_{\text{dB}}  \propto \vert r \vert$).

The simplest single mode squeezed states are the \emph{squeezed vacuum states} generated by applying the squeezing operator in Eq.\eqref{squeezingOp} to the vacuum. Such class of states can also be written in Fock basis: 
\begin{equation}\label{vacuumsqueezedstate}
\ket{r}\equiv S(r)\ket{0}=\frac{1}{\sqrt{\cosh |r| }}\sum_{N=0}^\infty  \frac{ (e^{i  \theta}\,\tanh |r|)^N\sqrt{(2N)!}}{2^N N!} \ket{2N}.
\end{equation}
Such states have zero mean quadratures, are composed of a linear combination of only even photon number states and have a mean photon number $\langle n \rangle=\sinh^2 \vert r \vert$.

A general single mode squeezed state, with mean quadratures different from zero, can be generated by the application of the squeezing operator [Eq.\eqref{squeezingOp}] and of the displacement one [Eq.\eqref{displacement}] on a vacuum state:
\begin{equation}
\label{squeezedstate}
\ket{\Psi}_{\mathrm{squeezed}}^{\mathrm{sm}}\equiv D(\alpha)S(r)\ket{0}.
\end{equation}
 
Photonic single mode squeezing can be experimentally generated exploiting parametric
down conversion-based setups \cite{bachor2004guide}. Here a second order nonlinear crystal is pumped by a pump of frequency $2\nu$ with a phase able to create amplification (optical parametric amplification, OPA). In absence of other excitation field, a squeezed vacuum state at frequency $\nu$ is generated. The pumped cavity is maintained slightly below the oscillation threshold so that no bright light is generated. The squeezing factor is higher as the working point is closer to the oscillation threshold. In order to enhance the amplification efficiency of the OPA, the nonlinear crystal can be placed inside an optical cavity, realizing an optical parametric oscillator (OPO) \cite{schnabel2017squeezed}.

\paragraph{Two mode squeezed states}
\label{twomodesqueezedSec}

Squeezing along two modes is a fundamental resource for quantum metrology  \cite{jaekel1990quantum,braunstein2005quantum, ferraro2005gaussian,anisimov2010quantum,lvovsky2015squeezed,schnabel2017squeezed}. The evolution operator associated to two-mode squeezing is the contribution of the third term in Eq.\eqref{bilinearev} that, contrary to single mode operator in Eq.\eqref{squeezingOp}, involves two different modes $\bm{k_1}$ and $\bm{k_2}$:
\begin{equation}
S_2(r)=e^{r^*\,a_{\bm{k_1}} \,a_{\bm{k_2}} -r\,a_{\bm{k_1}}^\dagger \,a_{\bm{k_2}}^{\dagger}}\;\;,
\end{equation}
where again $r=|r|e^{i\theta}$. The modes operators evolve under $S_2(r)$ according to the following relations: 

\begin{equation}\label{twomodesqueezingEvolution}\begin{split}
&a_{\bm{k_1}}\xrightarrow[]{S_2(r)} \cosh |r| \;a_{\bm{k_1}}+e^{i \theta}\sinh |r| \;a_{\bm{k_1}}^\dagger \\ & a_{\bm{k_2}} \xrightarrow[]{S_2(r)} \cosh |r| \;a_{\bm{k_2}}-e^{-i \theta}\sinh |r| \;a_{\bm{k_2}}
\end{split}
\end{equation}

The simplest states, generated by the two-mode squeezing operator, are the \emph{two mode squeezed vacuum states} or \emph{twin-beam states}, which present non-classical correlation involving two modes (mode-entanglement) \cite{demkowicz2015quantum}.  They are obtained by the action of $S_2$ on the vacuum state, leading to the following output state in the Fock basis: 
\begin{equation}
\ket{r}_2\equiv S_2(r)\ket{0,0}=\frac{1}{\sqrt{\cosh |r| }}\sum_{N=0}^\infty (e^{i \theta} \tanh |r|)^N \ket{N,N}.
\end{equation}
For this state the total number of photons along the modes is even for each superposition component, and the mean values of quadratures are zero.  The  mean total photon number $\langle n \rangle=\langle n_{\bm{k_1}}+n_{\bm{k_2}}\rangle= 2 \sinh^2 |r|$. These states, in photon number basis, clearly show entanglement between the two modes. In particular, when the squeezing parameter $|r| \rightarrow \infty$ (large squeezing regime), this state corresponds to the EPR state \cite{banaszek1998nonlocality}. 

Since the variance of the number operator is $2(\langle n \rangle^2+\langle n \rangle)$, we find that its application to phase estimation in a MZI leads to:

\begin{equation}
\Delta\phi_{\mathrm{squeezed}}\ge \frac{1}{2\sqrt{2}}\frac{1}{\sqrt{\langle n \rangle^2+\langle n \rangle}},
\end{equation}
which for large $\langle n \rangle$ shows Heisenberg-limited scaling.

Two mode squeezed states can be realized combining in a beam splitter two squeezed states generated through a type I parametric down conversion process with an opportune phase shift, or also by dividing in a polarizing beam splitter a squeezed state generated by a type II OPA \cite{schnabel2017squeezed}.

Recently a more general concept of nonlinear squeezing, applicable to non-Gaussian states,  has been introduced \cite{gessner2019metrological}.

\paragraph{Homodyne measurements}
Since squeezed states are well described by CV quadratures, their detection mostly relies on measuring such variables. In order to measure quadratures, a suitable technique is represented by \emph{homodyne detection} \cite{breitenbach1997measurement,braunstein2005quantum,ferraro2005gaussian,kumar2012versatile,schnabel2017squeezed}. 
To experimentally implement a homodyne apparatus, the idea is to interfere in a beam-splitter the target optical signal together with an additional coherent state of the same frequency, that acts as additional phase reference. Such reference beam is called \emph{local oscillator} (LO)  \cite{ferraro2005gaussian,braunstein2005quantum,weedbrook2012gaussian}.   

A first scheme employs a balanced beam splitter and is called \emph{balanced homodyne detection}.
Consider a target state, with annihilation operator $a_{\mathrm{tgt}}$, whose quadratures have to be measured. Such state is injected in one input mode of a balanced beam splitter. At the same time, a coherent state $\ket{\alpha_{\mathrm{LO}}}$, with $\alpha_{\mathrm{LO}}=|\alpha_{\mathrm{LO}}|e^{\imath \theta}$ and associated photon annihilation operator $a_{\mathrm{LO}}$, is injected along the other input of beam splitter. The output annihilation operators are then composed by the superposition of the operators relative to the two input beams: $a_{\mathrm{out}_1}=\frac{1}{\sqrt{2}}(a_{\mathrm{tgt}}+a_{\mathrm{LO}})$ and $a_{\mathrm{out}_2}=\frac{1}{\sqrt{2}}(-a_{\mathrm{tgt}}+a_{\mathrm{LO}})$. The light exiting from BS outputs is then detected by photodiodes that reveal a current with intensity proportional to the number of photons: $I_{\mathrm{out}_1}\propto  n_{\mathrm{out}_1}= a^\dagger_{\mathrm{out}_1} a_{\mathrm{out}_1}$ and $I_{\mathrm{out}_2}\propto n_{\mathrm{out}_2}=a^\dagger_{\mathrm{out}_2} a_{\mathrm{out}_2}$. Then, the difference of the two output intensities for the above input state in the limit of $|\alpha_{\mathrm{LO}}|\gg 1$ ($a_{\mathrm{LO}}\sim \alpha_{\mathrm{LO}}$) is:
\begin{equation}
\Delta I=I_{\mathrm{out}_1}-I_{\mathrm{out}_2} \propto |\alpha_{\mathrm{LO}}|(e^{\imath \theta}a_{\mathrm{tgt}}^\dagger+e^{-\imath \theta}a_{\mathrm{tgt}}).
\end{equation}
This expression corresponds to the rotated quadrature $x_{\mathrm{tgt}}(\theta)$ (see Eq. \eqref{rotatedquadratures}) for the target state. Note that, by tuning the local oscillator phase $\theta$, one can measure all the quadrature space. For instance with $\theta=\frac{\pi}{2}$, quadrature $p_{\mathrm{tgt}}$ is measured since $x_{\mathrm{tgt}}(\theta=\frac{\pi}{2})=p_{\mathrm{tgt}}(\theta=0)$ [Eq. \eqref{rotatedquadratures}]. Through this kind of measurements it is then possible to perform quantum tomographies in the phase space \cite{lvovsky2009continuous,kumar2012versatile,d2003quantum,paris2004quantum}. Furthermore, homodyne detection can be performed with high quantum efficiency. 

When the beam splitter is unbalanced the previous scheme is called \emph{unbalanced homodyne detection} \cite{ferraro2005gaussian}. In this case only one BS output is measured and, tracing out the local oscillator, quadrature statistics can be measured in the limit $|\alpha_{\mathrm{LO}}|\gg 1$. 

Homodyne detection allows also to perform entangled measurements, such as Bell measurements, in quadratures space through multi-homodyne detector schemes. \cite{ferraro2005gaussian,braunstein2005quantum}. Alternatively such measurements can be realized by heterodyne detectors \cite{yuen1980optical}, in which the local oscillator has a different frequency with respect to the target state \cite{ferraro2005gaussian}. 

\subsubsection{Other states}

Holland-Burnett (HB) states \cite{holland1993interferometric,campos2003optical} have a similar form to the one of N00N states, with some advantages in non-ideal conditions \cite{thomas2011real,datta2011quantum}. The idea, for generating such entangled states, is to generalize the HOM effect, injecting two beams of indistinguishable $N/2$ photons (with $N$ even) in the two input ports of a beam splitter.  The output state, after a relative phase shift $\phi$ between the two output modes, can be written in the following form:

\begin{eqnarray}
\label{eq:hbstates}
    \ket{\Psi}_{\mathrm{HB}}&=&\sum_{n=0}^{N/2} C_n \ket{2n, \, N-2n}, \\
     C_n&=&e^{i2 n\phi}\,\,\frac{\sqrt{(2n)!\,(N-2n)!}}{2^{N/2} n! \,(N/2-n)!}.
\end{eqnarray}
When $N=2$, Eq.\eqref{eq:hbstates} coincides with a N00N state. Such states reach Heisenberg scaling up to a constant factor $\sqrt{2}$. The two beams injected along the input modes of the beam splitter can be generated with a SPDC process.

Another interesting class of states with fixed number of subsystems are Dicke states \cite{dicke1954coherence,lucke2014detecting}. A symmetric Dicke state $\ket{D^k_N}$,  composed of $N$ qubits (in the basis $\{ \ket{0}, \ket{1}\}$) with $k$ excitations, has the following form: 
\begin{equation} 
\label{eq:dicke}
\ket{D^k_N}=\begin{pmatrix} N\\k \end{pmatrix}^{-\frac{1}{2}}\sum_l P_l(\, \ket{1}^{\otimes k} \otimes \ket{0}^{\otimes (N-k)} \,),
\end{equation}
where the balanced superposition involves all the permutations $P_l$ of $k$ qubits in the excited state $\ket{1}$ and the other $N-k$ in the state $\ket{0}$ (note that in this case the notation $\ket{0}$ and $\ket{1}$ stands for logical values of the qubits and not for the number of photons).
For instance, the Dicke state with $N=4$ different qubits and $k=2$ is: $\ket{D^2_4}=(\,\ket{0,0,1,1}+\ket{0,1,0,1}+\ket{1,0,1,0}+\ket{1,1,0,0}+\ket{1,0,0,1}+\ket{0,1,1,0}\,)/\sqrt{6}$. In general, the case with $k=N/2$ for $N$ even, is called twin Fock state and has relevance in quantum metrology applications \cite{toth2014quantum, pezze2018quantum}, such as the generation of the previously described HB states. A more general class of states, robust against noises, are Dicke squeezed states \cite{zhang2014quantumm}. It has been recently demonstrated that the multimode photon states emitted by the phenomenon of collective Dicke superradiance  can be a resource for quantum metrology \cite{paulisch2019quantum}.

Other states, suitable for experimental generation and robust with respect to noises, are \emph{entangled coherent states} (ECS) \cite{sanders1992entangled,sanders2012review,zhang2013quantum,joo2011quantum,joo2012quantum} along two modes, that in Fock basis read:
\begin{equation}\label{eq:ecs}
    \ket{\mathrm{ECS}_{\alpha}}= \frac{1}{\sqrt{2(1+e^{-|\alpha|^2})}}(\ket{\alpha}_1\ket{0}_2+\ket{0}_1\ket{\alpha}_2),
\end{equation}
where $\ket{0}_i$ and $\ket{\alpha}_i$ represent, respectively, a vacuum and a coherent state along mode $i$. Such state can be realized by injecting a coherent state $\ket{\alpha}$ along an input of a beam splitter and the state $(\ket{\alpha}+\ket{-\alpha})/(2+2e^{-2|\alpha|^2})$, also called Schr\"{o}dinger cat state, along the other input \cite{luis2001equivalence,joo2011quantum}. 

Finally, also cluster states \cite{friis2017flexible} and random symmetric states \cite{oszmaniec2016random} are useful for quantum metrology tasks.

\subsection{Photonic platforms}

Photons represent the ideal probes for several interferometric tasks and quantum sensing \cite{pirandola2018advances,tan2019nonclassical}. Here, we briefly review some of the basic photonic platforms exploited for quantum information and in particular metrology tasks. Note that recent and more in-depth reviews on photonic technologies for quantum information can be found in Refs. \onlinecite{Flamini19rev, slussarenko2019photonic}.

\subsubsection{Photonic degrees of freedom}

Quantum information can be encoded in photonic states by exploiting different degrees of freedom. A first possibility is provided by the polarization, or spin angular momentum (SAM), that can encode 2-dimensional quantum states spanned by the orthonormal basis of horizontal and vertical polarizations  $\{|H\rangle , |V\rangle\}$ \cite{kok2010introduction}. The interaction between polarized light and matter allows the manipulation of the polarization through linear optical elements like waveplates. These are birefringent materials that introduce a phase delay between the two orthogonal polarizations. Any unitary transformation in the 2-dimensional Hilbert space of polarization can be realized by a suitable sequence of waveplates. Projection in polarization space can be realized through polarizing beam splitter (PBS), that spatially separates orthogonal polarizations \cite{kok2010introduction}. Polarization is then a largely used encoding for quantum information protocols and is also often coupled with other degrees of freedom \cite{Flamini19rev}. 

Besides SAM, angular momentum of light is also composed of another contribution carrying orbital angular momentum (OAM) that is related to light spatial distribution \cite{allen1992orbital,padgett2004light}. SAM and OAM can be considered separately in the paraxial limit and are exploited as independent carriers of quantum states. OAM can be described conveniently by Laguerre-Gauss (LG) modes, carrying, in single photon regime, a quantized amount of angular momentum $m \hbar$, where $m\in \mathbb{Z}$ indicates the azimuthal phase structure of the beam. Thanks to the unbounded Hilbert space in which OAM lives, which spanned by the basis of optical vortexes $\ket{LG_m}$,  states of arbitrary discrete dimensions (qudits) can be realized \cite{roadMap,giordani2019experimental,malik2016multi,erhard2019advances,cozzolino2019high,shen2019optical}. Generation and manipulation of OAM modes can be performed by means of different devices. Two of the most important techniques are: $q$-plate (QP) and spatial light modulator (SLM). 
A QP is an inhomogeneous anisotropic material that, based on light polarization, changes the OAM state: $\ket{L} \ket{LG_{m}} \stackrel{\text{QP}}{\longrightarrow} \ket{R} \ket{LG_{m+2q}}$ and $\ket{R} \ket{LG_{m}} \stackrel{\text{QP}}{\longrightarrow} \ket{L} \ket{LG_{m-2q}}$, where $\ket{R}$ ($\ket{L}$) is the right (left) circularly polarized state, and  $q\in\mathbb{Z}$ represents the topological charge of the QP \cite{marrucci2006optical,Cardano:12,d2012deterministic,rubano2019q}. Such devices, naturally entangle OAM and polarization degrees of freedom, thus are capable to generate vector-vortex beams \cite{erhard2018twisted,Miles,ndagano2017creation,liu2017generation,rosales2018review}. The latter are a class of states with several applications in quantum information, and very recently it has been also demonstrated the possibility of fiber propagation \cite{cozzolino2019air}.
Conversely, a SLM can induce directly phase and intensity changes on the optical beam in correspondence of each pixel \cite{heckenberg1992generation,gruneisen2008holographic}. An SLM, followed by coupling in single mode fibers, can be exploited to measure  different OAM modes \cite{mair2001entanglement,qassim2014limitations,bouchard2018measuring,bavaresco2018measurements, giordani2019experimental}. Other methods different from QP and SLM can be exploited to manipulate and measure OAM states, as shown in Refs. \onlinecite{leach2002measuring,malik2012measurement,kulkarni2017single,zhou2017orbital,yang2019manipulation,liu2019compact,liu2019superhigh}.

Time-bin \cite{marcikic2003long,schreiber20122d,lorz2019photonic} and time-frequency \cite{zhong2015photon,nunn2013large,steinlechner2017distribution,gianani2020measuring} degrees of freedom can be also exploited as quantum resources. For such encodings, photon manipulation can be done through interferometric schemes \cite{Flamini19rev}. In parallel, field quadratures, amplitude and phase-squeezing, can suitably encode continuous variables states \cite{ma2011quantum,adesso2014continuous,braunstein2012quantum}.

In the context of quantum metrology, one of the most widely adopted photonic degrees of freedom for quantum metrology is path encoding. The latter corresponds to a set of spatial modes occupied by the photons \cite{lee2002quantum,kok2010introduction,giovannetti2004quantum}. In this framework, the two elements that allow a complete manipulation of single photons in spatial modes are the beam splitter and the phase shifter  \cite{kok2010introduction}. Through these devices, interferometric setups, that are at the basis of most quantum metrology tasks, can be realized (see Sec. \ref{qmsec2}).

\subsubsection{Generation and detection of photons}

Different optical platforms are used to generate photonic quantum states, in particular single or entangled photons. Photon sources based on the spontaneous parametric down conversion (SPDC) process in non-linear $\chi^2$ materials are those more commonly employed \cite{boyd2003nonlinear,eisaman2011invited}. During this process, a pump photon (momentum $\vec{k}_p$ and  frequency $\omega_p$) passing through the crystal is annihilated, while a pair of photons is generated: idler (momentum  $\vec{k}_i$ and frequency $\omega_i$) and signal (momentum  $\vec{k}_s$ and frequency $\omega_s$). Given $\vert \alpha \rangle_{p}$ the pump coherent state, the process in the Fock basis representation reads $\vert \alpha \rangle_{p} | 0\rangle_{i}| 0\rangle_{s} \longrightarrow a_pa_i^{\dagger}a_s^{\dagger}|\alpha\rangle_{p} | 0\rangle_{i}| 0\rangle_{s} \approx |\alpha\rangle_{p}| 1\rangle_{i} | 1\rangle_{s}$, where $a_x$ is the particle annihilation operator of photon $x$ with $x=p,s,i$. SPDC processes must satisfy the following conditions:  \texttt{(i)} energy conservation $\omega_p=\omega_i+\omega_s$ and \texttt{(ii)} momentum conservation (phase matching condition) $\vec{k}_p=\vec{k}_i+\vec{k}_s$. 
SPDC sources permit to generate both heralded single photons and entangled states in the different degrees of freedom \cite{eisaman2011invited,kwiat1995new,ansari2018tailoring,mair2001entanglement,giovannetti2002generating,brendel1999pulsed,graffitti2018independent}.

Another process that can be exploited to generate single and entangled photons is spontaneous four-wave-mixing (SFWM) inside $\chi^{(3)}$ nonlinear waveguides \cite{fiorentino2002all,rarity2005photonic,fan2005efficient}. Inside the waveguide, two pump photons (with frequencies $\omega_{p_1}$ and $\omega_{p_2}$) interact  and generate two photons, signal and idler. Given two coherent pumps $\vert \alpha \rangle_{p_1}$ and $\vert \alpha \rangle_{p_2}$, the first order process generating a single  pair is:
$\vert \alpha \rangle_{p_1} \vert \alpha \rangle_{p_2} | 0\rangle_{i}| 0\rangle_{s} \longrightarrow a_{p_1}a_{p_2}a_i^{\dagger}a_s^{\dagger}|\alpha\rangle_{p_1} |\alpha\rangle_{p_2} | 0\rangle_{i}| 0\rangle_{s} \approx  |\alpha\rangle_{p_1} |\alpha\rangle_{p_2} | 1\rangle_{i} | 1\rangle_{s}$. The energy and momentum conservation relations are: $\omega_i+\omega_s=\omega_{p_1}+\omega_{p_2}$, and  $\vec{k}_{p_1}+\vec{k}_{p_2}=\vec{k}_i+\vec{k}_s$.

Both SPDC and SFWM are probabilistic processes, and the probabilities to generate photon pairs are typically low. In order to achieve a deterministic on-demand generation of single photons, other kinds of sources have to be employed, such as quantum dots \cite{lodahl2017quantum,senellart2017high,loredo2019generation,dusanowski2019near}, colour centers \cite{benedikter2017cavity} and others \cite{eisaman2011invited}.

At the end of a process, photons have to be measured in order to extract the encoded information. A lot of technologies exist to detect single photons \cite{buller2009single,migdall2013single,eisaman2011invited}. Depending on the photon wavelength, different platforms are employed for detector technology. In the visible range, Si-based avalanche photodiode detectors (APDs) are commonly used \cite{eisaman2011invited}, while for the telecom range (around $1310$ nm and $1550$ nm) superconductive nanowire single photon detectors (SNSPD) reach efficiencies above $0.95$ \cite{reddy2019exceeding,natarajan2012superconducting,renema2017probing}. Transition-edge sensors (TES) detectors are suitable for high efficiency photon number resolving detection, in both visible and telecom ranges \cite{cabrera1998detection,fukuda2011titanium}. Other detectors are those based on quantum dots \cite{li2007quantum} and up-conversion \cite{vandevender2004high}. In order to measure the spatial profile of the photons and realize imaging studies, single-photon sensitive cameras can be used \cite{jost1998spatial,morris2015imaging,aspden2013epr,lemos2014quantum}. 

\subsubsection{Integrated photonic circuits}

The realization of linear optical platforms for quantum applications requires a large number of components, especially when dealing with operations on the path degree of freedom. Bulk optical platforms possess limitations on the scalability on the experimental platforms. Complex schemes can require hundreds of optical elements, and thus the size of the bulk apparatus would be far greater than a standard optical table. In particular, the most severe limitation concerns the stability of the apparatus: without a strict control over temperature, vibrations and other environmental noise effects it can be impossible to reach a sufficient accuracy in the phase control.

Then, while simple interferometers involving a small number of modes can be implemented with bulk optics, other platforms are necessary for more complex interferometers. A possible solution to these issues is provided by integrated photonic circuits \cite{carolan2015universal,tanzilli2012genesis,Flamini19rev,slussarenko2019photonic,o2009photonic,wang2019integrated,paesani2019large}. Such circuits provide stability, scalability, miniaturization, flexibility, cost reduction, standardization, greater efficiency and precision for quantum applications with respect to a bulk approach.  

In integrated circuits, light is confined in waveguides fabricated inside or on the devices materials.  Interactions between optical modes are realized through directional couplers. The latter elements are composed of two waveguides brought close together, so that the evanescent fields inside the waveguides overlap and tunnelling between the two modes can happen. Also, tunable phase shifts between optical paths can be implemented by appropriate technologies depending on the integrated platform. Indeed, different platforms can be used for the integration of optical circuits each with its own advantages and drawbacks \cite{bogdanov2017material,tanzilli2012genesis,Flamini19rev,slussarenko2019photonic}. 

Common and powerful platforms for integrated components are Si-based technologies \cite{silverstone2016silicon,politi2008silica} such as Silica-on-Silicon   \cite{silverstone2016silicon,politi2008silica,politi2009integrated} and Silicon-on-Insulator platforms \cite{silverstone2016silicon,qiang2018large,zhang2019generation,paesani2019generation}. These devices allows high density circuits thanks to the strong difference of refraction indexes between substrate and surface, even if polarization qubits are not supported. Furthermore, several additional components, such as electronics elements and fast optical modulators, are compatible with this technology and can be exploited also for high rate quantum key distribution \cite{sibson2017integrated,bunandar2018metropolitan,paraiso2019modulator,pirandola2019advances}. Silicon waveguides are naturally suitable also for the generation of single and entangled photons through SFWM process \cite{silverstone2014chip,silverstone2015qubit}.

Femtosecond laser waveguide writing technique  \cite{della2008micromachining,marshall2009laser,ceccarelli2020low} exploits laser pulses to write waveguides inside glass (but also crystalline or polymeric) material substrates.  Such technique guarantees low cost, low loss ($\sim 0.1-0.3$ dB/cm), high speed fabrication and allows for 3-D circuit geometries \cite{nolte2003femtosecond,rodenas2019three}. Furthermore, thanks to the low birefringence of glass, polarization insensitive waveguides can be realized \cite{rojas2014analytical} and polarization qubits can be supported and manipulated \cite{sansoni2010polarization,corrielli2014rotated,fernandes2011femtosecond,crespi2011integrated,wang2019chip}.

Among other integrated platforms there are those based on III-V semiconductors \cite{wang2014gallium, xiong2011integrated,abellan2016quantum,orieux2013direct,dietrich2016gaas} and UV writing  \cite{smith2009phase, kundys2009use}.

The final goal of quantum integrated photonics is the complete simultaneous integration of all steps of a quantum information protocols in a single chip: generation, manipulation and detection of quantum photonic states. Furthermore, integrated circuits should possibly be able to support all the degrees of freedom of light. During the last years much progress, towards this goal, has been made. 

Integrated sources of single and entangled photons have been developed for different circuits and different generation processes \cite{sharping2006generation,davanco2012telecommunications,silverstone2014chip,silverstone2015qubit,zadeh2016deterministic,khasminskaya2016fully,caspani2017integrated,atzeni2018integrated,meyer2018high,spring2017chip}.  Efforts for integration of single photons detector have also been performed  \cite{ferrari2018waveguide,najafi2015chip,munzberg2018superconducting,calkins2013high}. Regarding integration of different degrees of freedom, path and polarization are within current state of the art as previously discussed. Furthermore, progress have been made towards integration of time \cite{xiong2015compact} and OAM \cite{oamchip,cai2012integrated,chiptofiber}. Finally, it has been demonstrated that quantum states can be converted between different degrees of freedom on a chip \cite{feng2016chip}.

\subsection{Platforms for the generation of   phase-sensitive quantum states}

The final target in quantum phase estimation is to reach a genuine enhancement with respect to the classical limit (SQL) when all the employed resources are carefully taken into account. More specifically, the relevant parameter is the total number of photons effectively employed throughout the experiment. In this way, we have to consider in the resource count also those photons that are lost and are not detected by the final measurement stage. Crucial parameters of the experimental setups to reach unconditional violation of SQL are thus the total transmission, the detection efficiency and the visibility of quantum interference \cite{slussarenko2017unconditional}. Such parameters depend on the technical details of the particular experimental implementation and of the employed optical elements. Furthermore, photons can be lost during the  post-selection process eventually needed for the implementation of the employed scheme. In this case, even with lossless optical elements, an enhanced sensitivity, with respect to classical resources could be impossible to reach \cite{combes2014quantum}.  

A less stringent requirement for quantum phase estimation is super-resolution:  the achievement of fringes of interference that oscillate faster than any classical state \cite{jacobson1995photonic}. The simple super-resolution of a phase is less demanding than the unconditional violation of SQL, and can be obtained also starting from classical coherent states by applying suitable filtering protocols \cite{resch2007time,datta2011quantum}.

Several photonic realizations of quantum phase estimation experiments demonstrated a super-resolution with schemes that, in principle, allow unconditional super-sensitivity, while experimental imperfections prevented to effectively reach it.

\subsubsection{Platforms for N00N-like states}

N00N states are definite photon-number (discrete variable) quantum states, which are typically measured via single-photon counting. Hence, the privileged basis is the field mode photon number. Such states can be generated in a heralded configurations reaching high quality levels. However, the drawback of most  protocols is the low efficiency of generation and detection that prevents scalability towards unconditional quantum advantages for large photon numbers $N$. Indeed, besides the case of $N=2$, no scheme is able to generate in a deterministic way a N00N state with arbitrary $N$.  Hence, most of the protocols, realizing higher photon-number N00N states, rely upon post-selection schemes \cite{kok2002creation,cable2007efficient,pryde2003creation,fiuravsek2002conditional,hofmann2004generation}, in the same spirit of linear optic quantum computation \cite{knill2001scheme,kok2007linear}. In this way, the desired states are successfully generated, conditioned on the occurrence of specific probabilistic events. 

N00N states with $N=2$ can be deterministically generated, through Hong-Ou-Mandel effect \cite{hong1987measurement}, starting from two indistinguishable photons injected each along a different input of a Mach-Zehnder interferometer. This protocol has been demonstrated with bulk optics configurations in path  \cite{rarity1990two, ou1990experiment,edamatsu2002measurement} and polarization \cite{kuzmich1998sub} degrees of freedom. Polarization two-photon N00N states were also generated using the nonlocal correlation of entangled photons \cite{eisenberg2005multiphoton}, and were exploited for imaging of samples that surpasses SQL, even if without unconditional violation, because of losses and technical imperfections \cite{ono2013entanglement}. Two-photon polarization N00N states were also used to probe, in a non-destructive way, an atomic spin ensemble exploiting Faraday rotation \cite{wolfgramm2013entanglement}.

For numbers of photons $N>2$, the generation of N00N states is not straightforward. Consider a general state, with fixed photon-number, along two modes:
\begin{equation}\label{eq:nphotstate}
    \sum_{k=0}^{N} c_k \ket{k,N-k}.
\end{equation}
The goal is to avoid all the non-N00N contributions $\ket{k,N-k}$ with $k\ne 0,N$. Such undesired terms could be suppressed deterministically through quantum interference, such as the terms $\ket{1,1}$ in the HOM effect, or could be discarded through post-selection schemes, thus introducing losses in the generation and/or measurement stages. Post selection schemes are generally needed to generate high number N00N states.

Conversely, another approach is to employ, as probes,  states that are much easier to generate compared to N00N states. For this purpose, even classical states can be employed, given their greater  robustness against losses.
From such states one can finally recover N00N typical super-resolution, through suitable state projections that discard all the undesired terms of Eq. \eqref{eq:nphotstate}, so that only a super-sensitive N00N contribution is observed   \cite{wang2005phase,liu2008demonstration,sun2006observation,pregnell2004retrodictive, resch2007time,nagata2007beating,okamoto2008beating}.

The first N00N state with $N>2$ was realized in the polarization degree of freedom of three photons ($N=3$) along a single spatial mode\cite{mitchell2004super,shalm2009squeezing}, in a post-selected configuration. States with $N=3$ were also generated through the photon subtraction technique applied on two pairs of photons \cite{kim2009three}. Another demonstration of N00N super-resolution at $N=3$ was  obtained  through state-projection measurements \cite{liu2008demonstration}.

The case of $N=4$ super-resolving photon states was demonstrated in  polarization  \cite{sun2006observation} and path \cite{walther2004broglie,nagata2007beating,okamoto2008beating}  degrees of freedom, in four\cite{walther2004broglie} and  two\cite{nagata2007beating,okamoto2008beating} modes bulk interferometers.  Some of such works \cite{sun2006observation,nagata2007beating,okamoto2008beating} studied interference fringes after state-projection measurements on contributions showing super-resolution. Super-resolution with N00N states of up to 4 photons was also obtained for quantum lithography application by using  optical centroid measurement \cite{rozema2014scalable} that allows to reach higher efficiencies. Indeed, such technique does not require  photons to be detected all at the same point \cite{tsang2009quantum, shin2011quantum}.

\begin{figure}[!ht]
\centering
\includegraphics[width=.5\textwidth]{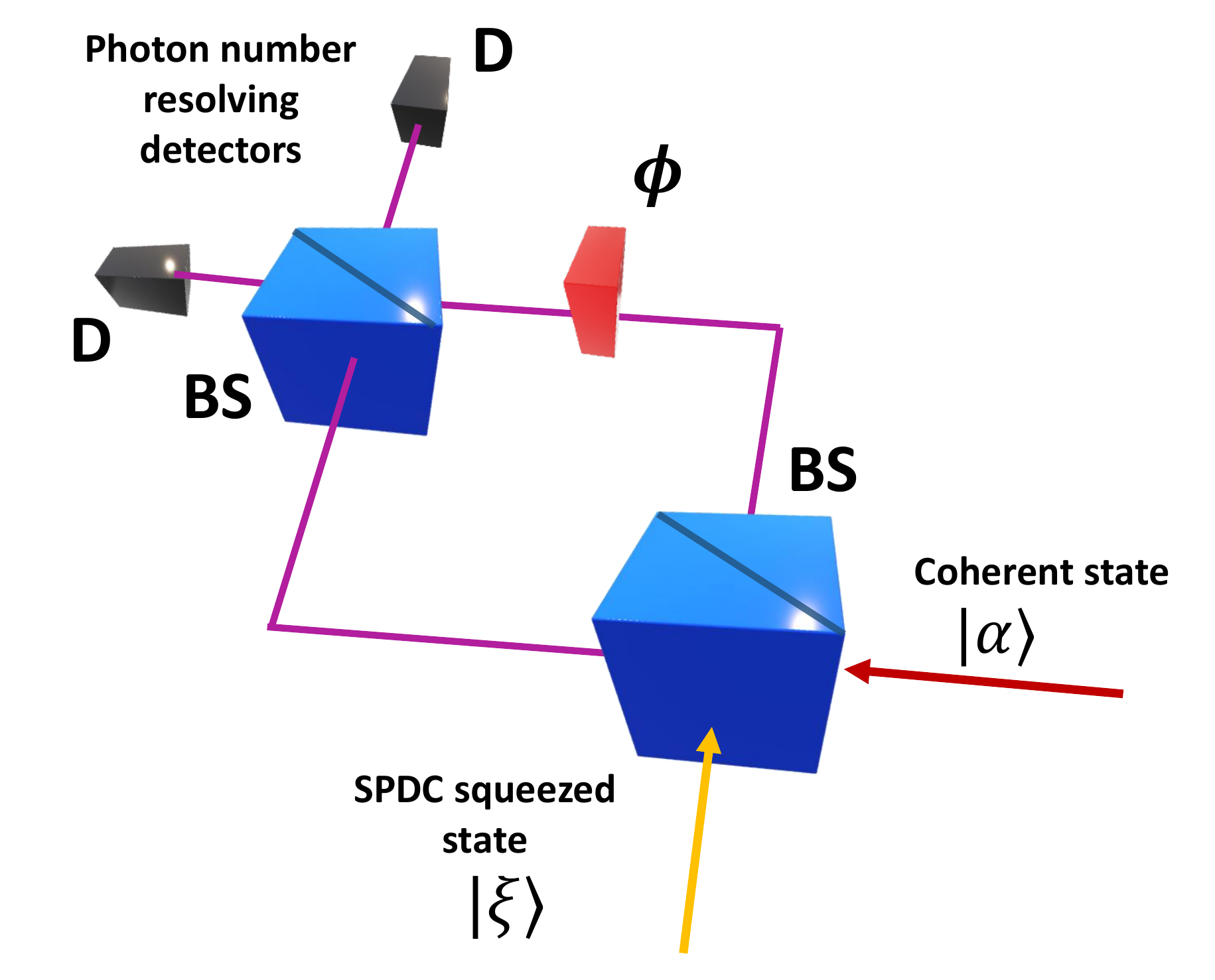}
\caption{{\bf High photon-number N00N state generation.} Conceptual scheme for the generation of N00N state in the path degree of freedom, with high photon-number $N$. A coherent state $\ket{\alpha}$ and the output state $\ket{\xi}$ of a SPDC state interfere in a beam splitter (BS) at the input of a Mach-Zehnder interferometer. Finally, after evolution the photon number of the output ports of the final BS is measured by detectors D.}
\label{fig:noonhigh}
\end{figure}

A clever way to generate N00N-like states is to let a coherent state interfere with the state generated by a SPDC process \cite{hofmann2007high,pezze2008mach} (Fig. \ref{fig:noonhigh}). Such scheme was experimentally realized, in the polarization degree of freedom, to perform imaging with $N=2,3$ \cite{ono2013entanglement} and allowed to generate (in a post-selected configuration) N00N states with up to $N=5$ photons \cite{rozema2014scalable,israel2012experimental,afek2010high}.

States showing super-resolution with photon number $N=6$, were demonstrated through a coherent probe state and suitable state-projection measurements along N00N states in a time-reversal configuration \cite{resch2007time}. The probe state interacting with the system and embedding the phase is a classical one. Hence, even with ideal optical elements, SQL cannot be unconditionally beaten. Such scheme was also used to perform imaging through a polarization state of $N=2,3$ photons \cite{israel2014supersensitive}.

A class of N00N-like states, where photons are distributed along different modes (for example $N$ photons along $N$ spatial modes maximally entangled in the polarization degree of freedom), are called Greenberger-Horne-Zeilinger (GHZ) states \cite{greenberger1990bell}. Such kind of states were realized with up to 10 spatially separated photons entangled in the polarization degree of freedom (with an experimentally observed super-resolution up to $N=8$) \cite{wang2016experimental}, with up to 10 qubits  encoded in two degrees of freedom (polarization and path) of 5 spatially separated photons (with a  super-resolution shown for $N=8$ qubits) \cite{gao2010experimental}, and with up to 18 qubits encoded in three degrees of freedom (polarization, path and OAM) of 6 spatially separated photons    \cite{wang201818}. GHZ states were also exploited for noisy phase estimations. In Ref. \onlinecite{proietti2019enhanced} each qubit of a four-photon polarization GHZ state is locally encoded in the diagonal basis, in order to improve the robustness of the phase estimation under dephasing noise along the computational basis \cite{chaves2013noisy}. Similarly, Ref. \onlinecite{zhang2019demonstrating} demonstrated the robustness to transversal noise using up to six-photon GHZ polarization states.

The first unconditional quantum violation of SQL, taking into account all the employed resources, was obtained in 2017 with a N00N state ($N=2$) in polarization degree of freedom evolving through a bulk Mach-Zehnder interferometer \cite{slussarenko2017unconditional}. The SQL, corrected by the efficiency $\eta$ and the visibility $V$, has the following form \cite{slussarenko2017unconditional}: 
\begin{equation}
    \eta^N V^2 N <1.
\end{equation}
The reported experimental violation was $ \eta ^N V^2 N \approx 1.23 $, thus showing genuine quantum enhancement. In order to reach this violation, high efficiency ($>0.95$) detectors, low losses circuits and high visibility fringes ($V\approx 0.98$) were obtained.

The generation of N00N-like states and their use for quantum phase estimation can be realized inside integrated circuits, that allow for high stability and fine tunability. Super-resolution with generated N00N states was achieved for $N=2$ in silica-on-silicon \cite{matthews2009manipulation,matthews2011heralding,li2018two}, UV-written \cite{smith2009phase}, silicon-on-insulator \cite{bonneau2012quantum,silverstone2014chip} and femtosecond laser written  \cite{crespi2012measuring,flamini2015thermally}  circuits. The case $N=4$ was obtained with state projections in silica-on-silicon circuits \cite{matthews2009manipulation}, also in a heralded configuration \cite{matthews2011heralding}. In these circuits, also states that resemble N00N ones and are ideal against symmetric photon-losses \cite{demkowicz2009quantum} were generated, by adopting a heralded configuration \cite{matthews2011heralding}. Such states contain $M$ photons along one mode and $L=N-M$ photons along the other: 
\begin{equation}
    \ket{M::L}\equiv \frac{1}{\sqrt{2}}(\ket{M,L}+e^{\imath \phi}\ket{L,M}).
\end{equation}
Such states correspond to N00N states when $L=0$. They are entangled when $M\ne L$, and show interference fringes whose frequency is enhanced by a factor $|M-L|$.

Note that photons can be generated outside and then injected inside the circuits, where they interfere to generated N00N states, or they can be directly generated on chip through nonlinear processes \cite{caspani2017integrated,solntsev2017path}. SFWM generation process is naturally exploited in Si-based circuits \cite{shadbolt2012generating,silverstone2014chip, silverstone2015qubit, wang2016chip}.
Also SPDC process can be realized in integrated waveguides to generate N00N states  \cite{setzpfandt2016tunable,kruse2015dual,jin2014chip,atzeni2018integrated}. Finally, quantum dots can be exploited to generate N00N states \cite{muller2017quantum,kamide2017method}.

The use of N00N states for practical quantum metrology is limited by two factors: \textit{(i)} it is hard to generate high photon-number states without relying on post-selection or filtering, and \textit{(ii)} in the noisy regime N00N states maintain their optimality only for small $N$. Indeed the quantum Fisher Information, in presence of symmetric losses $\eta$, is \cite{demkowicz2015quantum}: $F_{\mathrm{Q}}^{\text{N00N}}=\eta^N N^2$, and thus it decreases exponentially in $N$. For these reasons, for most of practical applications where high number of photons is required, other states such as squeezed ones are employed. 

\begin{table*}[t!]
\centering
\begin{tabular}{|l|l|l|l|l|}
\hline
{\bf Probe}&{\bf Refs.}&{\bf Platform} & {\bf Number of}\\
& && {\bf  photons} \\
\hline
N00N& [\onlinecite{ono2013entanglement}]&Bulk polarization& $N=2$ \\
 & [\onlinecite{israel2014supersensitive}]& 2-mode MZI microscope&$N=2,3$\\
\hline
N00N &[\onlinecite{afek2010high}] &Bulk polarization&$N=5$\\
  &&2-mode MZI&\\
\hline
N00N &[\onlinecite{slussarenko2017unconditional}] &Bulk polarization 2-mode MZI&$N=2$\\
  && unconditional enhancement&\\
\hline
 N00N &[\onlinecite{matthews2011heralding}] &Integrated silica-on-silicon &$N=2,4$\\
 &&heralded 2-mode-MZI&\\
 \hline
 N00N &[\onlinecite{matthews2009manipulation}] &Integrated silica-on-silicon &$N=2,4$\\
 && 2-mode-MZI&\\
 \hline
 N00N &[\onlinecite{bonneau2012quantum}] &Integrated silicon-on-insulator &$N=2$\\
 && 2-mode-MZI&\\
  \hline
 N00N &[\onlinecite{silverstone2014chip}] &Integrated silicon-on-insulator &$N=2$\\
 &&2-sources 2-mode-MZI&\\
  \hline
 N00N &[\onlinecite{wolfgramm2013entanglement}] &Bulk polarization &$N=2$ \\
 && 2-mode-MZI&\\
   \hline
 N00N &[\onlinecite{rozema2014scalable}] &Bulk  polarization to path 2-mode-MZI &$N=2,3,4$\\
 && with optical centroid measurement&\\
 \hline
 N00N &[\onlinecite{crespi2012measuring},\onlinecite{flamini2015thermally}] &Integrated FLW&$N=2$\\
 &&2-mode-MZI&\\
  \hline
  GHZ &[\onlinecite{wang2016experimental}] &Bulk polarization&$N=8$\\
  &&&\\
   \hline
  GHZ &[\onlinecite{proietti2019enhanced}] &Bulk polarization&$N=4$\\
  &[\onlinecite{zhang2019demonstrating}]&noisy estimation&$N=1,2,3,4,6$\\
   \hline
  GHZ &[\onlinecite{gao2010experimental}] &Bulk polarization-path&$N=4$ (8 qubits)\\
  &&hyper-entanglement&\\
     \hline
  GHZ &[\onlinecite{wang201818}] &Bulk polarization-path-OAM&$N=6$ (18 qubits)\\
  &&hyper-entanglement&\\
  \hline
 QFT & [\onlinecite{su2017multiphoton}] &Bulk path/polarization&$N=3$\\
  &&multi-mode MZI &$N=4$\\
\hline
HB &[\onlinecite{xiang2011entanglement}] &Bulk  polarization&$N=2,4$\\
 &[\onlinecite{xiang2013optimal,jin2016detection}]&2-mode-MZI&$N=2,4,6$\\
\hline
Dicke & [\onlinecite{wieczorek2009experimental,prevedel2009experimental}] &Bulk  polarization&$N=6$\\
 &&&\\
\hline
Dicke &[\onlinecite{chiuri2012experimental}] &Bulk  path-polarization&$N=2$ (4 qubits)\\
 &&hyper-entanglement&\\
 \hline
\end{tabular}
\caption{Table of some platforms used to measure phase shifts with fixed photon-number quantum probes, realized during the last 10 years.} \label{tabb1}
\end{table*}

\subsubsection{Platforms for squeezed states}

Several proof-of-principle experiments have demonstrated that squeezed states of light are fundamentals to beating shot-noise \cite{lugiato2002quantum, treps2003quantum} and improve interferometer sensibility \cite{grangier1987squeezed,xiao1987precision,mckenzie2002experimental,vahlbruch2005demonstration,goda2008quantum,ma2011quantum,andersen201630,schnabel2017squeezed}. Squeezed states and homodyne detection have been recently exploited to deterministically achieve simultaneously the super-sensitivity and the super-resolution conditions \cite{schafermeier2018deterministic}. First generation of squeezed light was realized by exploiting nonlinear processes as four-wave mixing (FWM) and parametric down conversion (PDC) inside sodium atoms and nonlinear materials, together with optical cavities and fibers \cite{slusher1985observation, shelby1986broad, wu1986generation}. The three main technologies that have been adopted to generate squeezed states of light are: atoms, nonlinear crystals and optomechanical systems \cite{chua2014quantum}. Squeezing by atoms exploit the third-order nonlinear susceptibility, through FWM process \cite{slusher1985observation,horrom2013all,corzo2013rotation}. Conversely, squeezed light from PDC process is generated by inserting nonlinear crystals inside optical cavities \cite{villar2006direct, jing2006experimental, grangier1987squeezed,ou1992realization,vahlbruch2008observation,eberle2010quantum,mehmet2011squeezed,stefszky2012balanced,vahlbruch2016detection}. Finally, coupling optical fields with mechanical modes of given structures, such as crystalline resonators and membranes \cite{purdy2013strong, safavi2013squeezed, brooks2012non,sudhir2017quantum,clark2017sideband,otterpohl2019squeezed}, allow squeezing generation in optomechanical systems.

Currently, gravitational wave detection represents the most direct and relevant metrological application \cite{barsotti2018squeezed}. For this purpose, squeezing factors above 10 dB have been experimental achieved: $10$ dB\cite{vahlbruch2008observation}, $12.7$ dB\cite{eberle2010quantum}, $12.3$ dB \cite{mehmet2011squeezed}, $11.6$ dB\cite{stefszky2012balanced}, $15$ dB\cite{vahlbruch2016detection}. All these realizations exploit cavity-enhanced optical-parametric amplification, working below its oscillation threshold and using a pumped type-I nonlinear crystal.

To date, photonic implementation using atoms, optomechanics and nonlinear crystals, have produced largest squeezing factors amount relatively of $14.9$ dB \cite{chua2014quantum}, $25$ dB \cite{purdy2013strong} and $19$ dB \cite{eberle2010quantum} respectively. 
Squeezed light have been generated at different wavelengths, such as 795 nm \cite{corzo2013rotation,purdy2013strong}, 860 nm \cite{serikawa2016creation}, 946 nm \cite{aoki2006squeezing}, 1064 nm \cite{stefszky2012balanced,eberle2010quantum}, 1540 nm \cite{safavi2013squeezed} and 1550 nm \cite{mehmet2011squeezed,schonbeck201813}. Squeezed vacuum states have been demonstrated only exploiting atoms \cite{chua2014quantum} and nonlinear crystals \cite{eberle2010quantum}. Although optomechanics demonstrated largest squeezing \cite{purdy2013strong}, in gravitational wave frequencies (audio-band regime) non linear crystals hold the record both in terms of generated and measured squeezing \cite{eberle2010quantum}. Two-mode squeezed states can potentially enhance interferometry \cite{berchera2013quantum,souza2015gaussian} and are fundamental for practical quantum metrology. Two mode squeezing was achieved on platforms able to reach squeezing factors even greater than $10$dB \cite{lawrie2016coherence, eberle2013stable, furst2011quantum, boyer2008generation, boyer2008entangled}. Furthermore, while most of the largest squeezing values were achieved in bulk optics, squeezed light generation and measurement have been also investigated in integrated platforms \cite{lenzini2018integrated,porto2018detection,kaiser2016fully,dutt2015chip,dutt2016tunable,masada2015continuous,safavi2013squeezed,kanter2002squeezing,raffaelli2018homodyne,vaidya2019broadband,zhao2020near}.
Even thermal mixtures of squeezed states can be exploited for quantum metrology \cite{kim1989properties,aspachs2009phase} (also combined with other states\cite{tan2014enhanced}) as experimentally analyzed using non-degenerate OPO  to realized thermal squeezed states with different purity and balanced homodyne detection  \cite{yu2020quantum}.
Finally, realization of squeezed states in the polarization degree of freedom have been reported \cite{grangier1987squeezed,bowen2002experimental,korolkova2002polarization,laurat2005effects}

As discussed above, one the most direct application of squeezed states relies in the enhancement of estimation of gravitational waves  \cite{hollenhorst1979quantum,caves1980measurement,abramovici1992ligo,adhikari2014gravitational,goda2008quantum,demkowicz2013fundamental,schnabel2010quantum, abbott2017exploring}. Starting from the recent observations\cite{abbott2016observation, abbott2016gw151226, aasi2015advanced}, investigation of gravitational waves is a very challenging research area and represent the first actual application of quantum metrology \cite{caves1981quantum, slusher1985observation, xiao1987precision,grangier1987squeezed, oelker2014squeezed,vahlbruch2018laser,chan2018binary,danilishin2019advanced}. The small amplitude ($\sim 10^{-22}$) of gravitational waves needs very long interferometer to be measured, together with very low overall noise \cite{schnabel2010quantum, abbott2017exploring}. Isolation from thermal contributions and radiation pressure \cite{caves1980quantum} over free-falling interferometer mirrors is fundamental to allow reaching sensitivity sufficient to detect such amplitudes. Furthermore, detectors are affected by photon-counting noise (or shot-noise), which follow the Poisson statistics. Therefore, improving signal-to-shot noise ratio (SNR) requires increasing the number of input photons. On the contrary, thermal mirrors displacement is proportional to the input laser power. Quantum metrological techniques, such as those adopting squeezing resource as input state in the interferometer \cite{lynch2015effect}, seem to be the only possible short-term solution, able to reduce the quantum noise contributes without increasing the laser power. In particular, an interferometer-based detection depends on the optical path difference between internal arms. Entanglement in probe state can correlate both noise and signal between internal arms in a way that noise is cancel out, thus enhancing the SNR. For example, if a coherent state is overlapped with a vacuum squeezed state into the input beam splitter of a Michelson interferometer, correlation after this interference permits squeezing in the output state and the improvement of the SNR \cite{schnabel2010quantum}. In particular, the higher the squeezing factor, the greater the SNR enhancement. Indeed, it has been demonstrated that strong squeezing could led to approximately 10-fold improvement in gravitational wave detection \cite{schnabel2010quantum}. Furthermore, the capability of reducing the quantum noise in gravitational wave interferometers depends on the frequency which has to be detected \cite{danilishin2012quantum}. Frequency-dependent squeezing schemes can be used to reduce quantum noise in the audio-band spectrum, and an EPR entangled squeezing-based setup has been proposed to achieve a broadband solution \cite{brown2017broadband,ma2017proposal}. One of this proposal has been recently realized in Refs. \onlinecite{sudbeck2020demonstration} and \onlinecite{yap2020generation}. In principle, N00N states can give access to better performances respect to squeezed ones. However, the difficulty in realizing these states with high number of photons makes N00N-based setup for these tasks impracticable. 

Starting from 2007, GEO600 gravitational wave detector has successful adopted squeezed light for its detection \cite{abadie2011gravitational,grote2013first}. Some successful experimental tests have been done on LIGO, injecting squeezing light into the interferometer  \cite{aasi2013enhanced}. Other GW detectors have almost achieved best technological performances in their several components \cite{dooley2016geo,aasi2015advanced,acernese2014advanced,aso2013interferometer}, and seems to find further improvements only in squeezing enhancement.  

\begin{figure}[ht!]
\centering
\includegraphics[width=.5\textwidth]{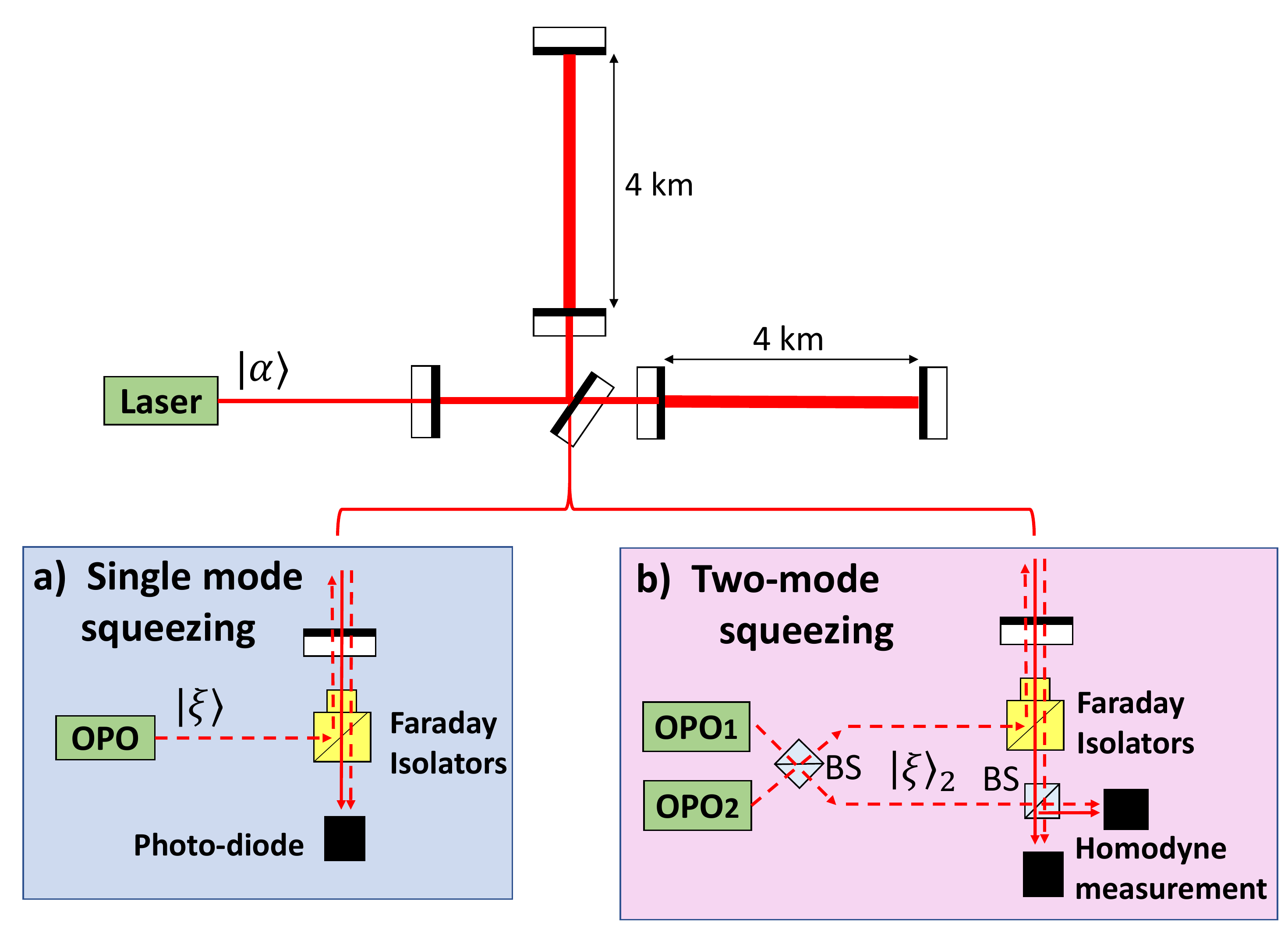}\caption{{\bf Conceptual scheme of LIGO interferometer seeded by squeezed states.} Squeezed states are injected in the interferometer by means of Faraday isolators, allowing light to pass along only specific directions. {\bf a} A single mode squeezed state $\ket{\xi}$, generated by an OPO, is injected in the Michelson interferometer. {\bf b} A two-mode squeezed state  $\ket{\xi}_2$ is generated by two squeezed states interfering in a BS. Then, one mode of $\ket{\xi}_2$ is injected in the interferometer, while the other mode interferes with the output of the interferometer in a beam splitter (BS) \cite{schnabel2017squeezed}.}
\label{fig:ligo}
\end{figure}

The conceptual scheme of the LIGO Michelson interferometer, seeded by squeezed states, is sketched in Fig. \ref{fig:ligo}. The 4 km-long arms of the interferometer contain Fabry-Perot cavities composed of two mirrors that reflect the light forcing it to travel across the arms multiple times, so enlarging the effective length of the arms and, consequently, the sensitivity of the interferometer. Recycling power mirrors are used to increase the optical intensity inside the interferometer. When single squeezed states are used to enhance the sensitivity, they are injected in the input of the interferometer (Fig. \ref{fig:ligo} a). Recently a squeezed vacuum state has been used in 
the Advanced LIGO detectors \cite{tse2019quantum,acernese2019increasing}. A two-mode squeezed state can be also exploited to enhance the detection. In this case, the two squeezed fields, with a fixed relative angle, interfere in a beam splitter generating a two-mode entangled state. One mode is injected in the interferometer, while the other interferes with the output of the interferometer in a second beam splitter. Finally, a homodyne measurement of the quadratures is performed (Fig. \ref{fig:ligo}b). Thanks to the simultaneous measurement of the quadratures, the possible disturbance signals can be recognized with respect to the signal to be analyzed \cite{steinlechner2013quantum}.

For a more in depth recent analysis on squeezed states we refer to review in Ref. \onlinecite{schnabel2017squeezed}.

\subsubsection{Platforms for other states}

Here we discuss other states, different from N00N and squeezed ones, that can be employed for quantum enhanced phase estimations. 

Let us consider the scenario where noises, losses, and low detection efficiency are present \cite{thomas2011real,datta2011quantum}, or when detection is restricted to projections onto states containing definite photon numbers along each output of interferometer (e.g. with a four-photon state along two modes, the detection could be restricted to $\ket{3,1}$ and $\ket{1,3}$ terms). In this configuration, the Holland-Burnett states can outperform N00N states for $N>4$ \cite{xiang2013optimal}. HB states with 4 \cite{xiang2011entanglement} and 6 \cite{xiang2013optimal,jin2016detection} photons were realized in polarization degree of freedom, through SPDC process. 

Photonic symmetric Dicke's states [Eq. \eqref{eq:dicke}] were generated in the polarization degree of freedom through bulk optics schemes with up to $N=6$ photons \cite{kiesel2007experimental,wieczorek2009experimental,prevedel2009experimental}. Furthermore, such states were generated by exploiting path-polarization hyper-entanglement, with two photons carrying $N=4$ qubits \cite{chiuri2012experimental}.

Other classes of states can be deterministically generated by exploiting the generalized HOM effect, implemented by a quantum Fourier transformation (QFT) acting on $N$ indistinguishable photons, one along each input mode of a $N$-mode interferometer. The QFT transformation is described by a unitary matrix $U_{j,k} = 1/\sqrt{N} \exp[\imath (j-1)(k-1)/N]$. QFT states with $N=2$ coincide with N00N state, while they are different for $N>2$. For instance, states at the output of a QFT with $N=3$ has the following expression in Fock basis: $\sqrt{2}/3 (\ket{3,0,0}+\ket{0,3,0}+\ket{0,0,3})+1/\sqrt{3} \ket{1,1,1}$. A typical interferometer in this framework is composed of a QFT that entangles the photons through the generalized HOM effect, and a QFT$^\dagger$ trasnformation that disentangles the photons after their phase-dependent evolution  \cite{motes2015linear,olson2017linear,su2017multiphoton}. In this way, phase super-sensitivity, beating SQL, can be observed if the unknown phase is put along one of the interferometer modes. However, this is true only for $N<7$ \cite{olson2017linear,su2017multiphoton}. An experimental realization of such scheme was performed with $N=2,3,4$, exploiting both path and polarization degrees of freedom in a bulk optical multimode interferometer \cite{su2017multiphoton}.

Approximated entangled coherent states [Eq. \eqref{eq:ecs}] were  generated by mixing squeezed vacuum states and coherent light inside a beam splitter in Ref. \onlinecite{israel2019entangled}. The  measurement was realized by performing photon-number resolving detection.

\subsection{ Other schemes and platforms}

OAM states can be exploited to perform ultra-sensitive measurements of rotations \cite{courtial1998measurement,lavery2013detection,jha2011supersensitive,fickler2012quantum, d2013photonic,zhang2019quantumr}. In this scenario, the value of angle of rotations can be embedded in relative phase shifts between OAM components. For instance, let us consider the state of a photon in a superposition of opposite OAM modes, with modulus of  OAM number $|m|$ as  $(\ket{m}+\ket{-m})/\sqrt{2}$. If a rotation of an angle $\theta$ is performed, whose sign is based on the value of the ancillary mode (say $+\theta$ for mode 1 and $-\theta$ for mode 2),  the state will evolve to $(e^{\imath m \theta}\ket{m}+e^{-\imath m \theta}\ket{-m})/\sqrt{2}$. Such state shows a rotation amplified by a factor $m$. Hence, the two angular momentum orientations $\pm m$ will show N00N-like interference fringes and are able to reach a sensitivity with an improved factor of $m$ as $\Delta \theta= 1/ (m \sqrt{\nu})$, where $\nu$ is the number of single photon probes. Such enhanced sensitivity arises from the superposition of $m$-quanta of OAM. Since single photon probes are exploited, advantages can be obtained both for generation, detection and robustness to losses with respect to N00N states. 

Other classes of useful sensitive high dimensional states are the so-called Kings of Quantumness states \cite{bjork2015extremal}. Those classes of states have been experimentally realized up to dimension 21 through OAM states \cite{bouchard2017quantum}.

If entangled photons carrying OAM are employed in interferometric setups, an amplification of the sensitivity due to both the carried OAM and the entanglement between probes can be obtained. This has been experimentally demonstrated with photonic platforms in Refs.  \onlinecite{jha2011supersensitive,fickler2012quantum, d2013photonic}.

Quantum enhanced sensitivities can be also obtained in SU(1,1) interferometers, in which the linear optical beam splitters are replaced by non-linear active optical interactions \cite{yurke19862,ou2012enhancement,li2014phase,chekhova2016nonlinear,sparaciari2016gaussian,giese2017phase}. These include parametric amplifiers based on FWM processes \cite{hudelist2014quantum} or parametric down conversion \cite{frascella2019wide}, that generate useful entanglement inside the interferometer. These platforms are robust against noises, thus guaranteeing quantum enhanced performances also in presence of losses \cite{plick2010coherent,marino2012effect,manceau2017detection,liu2018loss,gupta2018optimized}. General non linear effects in estimation strategies can lead to scalings beyond Heisenberg limit \cite{boixo2007generalized,boixo2008quantum,roy2008exponentially,napolitano2011interaction}.

Different photonic techniques exploiting quantum states for imaging have been reported \cite{moreau2019imaging}. For instance, spatial correlations between photons can be exploited for \emph{ghost imaging} \cite{gatti2004correlated,gatti2008quantum,erkmen2010ghost,shapiro2012physics,genovese2016real,padgett2017introduction,schori2018ghost,moreau2018resolution}. In this case two correlated photons, such us those belonging to pairs generated by a SPDC process, are employed to probe a sample with the goal of reconstructing its image. One of the photons propagates through the sample, and is measured by a photodetector without spatial resolution. Conversely, the other photon does not interact with the sample and is measured by spatial resolving detector. The image of the object is reconstructed by the combined information from the two correlated photons: the spatial information comes from detecting the photon that did not interact with the sample, and is triggered by detection of the photon that interacted with it. Hence, by exploiting the spatial correlation of the photons one can recover spatial information on the object. This is obtained by spatially resolving only the photon that did not interact with the object, and discarding the spatial information obtained from detection of the interacting photon. Quantum ghost imaging has been realized exploiting different techniques for the spatial resolving measurement stage \cite{pittman1995optical,strekalov1995observation,aspden2013epr,morris2015imaging,aspden2015photon}. 
The key ingredient of ghost imaging is the correlation between the two photons (or beams).  Hence, also classical correlated light can be used for this purpose  \cite{baleine2006correlated,shapiro2008computational,pepe2017diffraction,altmann2018quantum}. However, in the low probe intensity regime quantum probes can provide better performances \cite{gatti2008quantum,erkmen2009signal,brida2011systematic,meda2017photon}. Quantum ghost imaging has also been realized to perform a 3D tomography \cite{kingston2018ghost}, and in a configuration exploiting entanglement swapping \cite{bornman2019ghost}.

Imaging on samples with pair of photons can be performed even without detecting the photons interacting with the sample \cite{lemos2014quantum,lahiri2015theory}. This technique exploits two identical nonlinear crystals pumped by two laser beams coming from the same split laser. The pumped crystals emit pairs of non-degenerate photons, signal and idler. The idler photon from one crystal, once separated from the signal one, interacts with the sample and is subsequently sent along the second nonlinear crystal. The latter element is also pumped by the laser, and then can emit pairs of photons as well. In this way, there are two different paths for the signal photons, which are then sent to interfere in a beam splitter. Information is acquired from quantum interference between the two signal beams that did not interact with the object. Hence, one can perform phase and intensity imaging without measuring those photons which have interacted with the object. An experimental demonstration of such protocol has been demonstrated in Ref. \onlinecite{lemos2014quantum}. Finally, quantum mechanics allows to acquire information about an object without any system directly interacting with it. This is possible through interaction-free measurements \cite{elitzur1993quantum,vaidman1994realization,kwiat1995interaction,kwiat1998experimental,vaidman2003meaning}.  For a recent more detailed review on quantum imaging techniques we refer to Ref. \onlinecite{moreau2019imaging}.

Another approach which can be used for quantum phase estimation exploits weak measurements. In particular, this approach leads to interesting quantum phenomena including weak values \cite{aharonov1988result,dressel2014colloquium}, that can be employed for practical tasks in quantum metrology \cite{xu2013phase,zhang2015precision,combes2014quantum,viza2015experimentally,martinez2017ultrasensitive,sinclair2017weak,harris2017weak,vaidman2017weak,cimini2020anomalous}. 

The contrast of an object image can be enhanced by entangled photons through a scheme named quantum illumination \cite{lloyd2008enhanced,tan2008quantum,shapiro2009quantum,sanz2017quantum}. In this protocol, two entangled beams are generated, and one of the two beams is sent through a partially reflective object (sample). The reflected beam is finally jointly measured with the other beam who did not interact with the sample. Hence, exploiting the correlation between the beams one can recover the reflected light from the object, discriminating it from background noise. Experimental realizations of quantum illumination were performed using parametric down conversion beams detected by CCD camera \cite{lopaeva2013experimental} and using microwave frequency beams \cite{barzanjeh2015microwave}. The quantum enhancement of such scheme resists also in strongly noisy environments \cite{lloyd2008enhanced,tan2008quantum,zhang2015entanglement,zhang2013entanglement,gregory2020imaging}.

Finally, quantum interferometry can be realized also in the time-energy domain of photons, exploiting Franson interferometers \cite{franson1989bell,kwiat1993high,pe2005temporal,maclean2018ultrafast}. Furthermore, estimation of frequency as well as temporal separations between incoherent signals can be performed through mode-selective photon measurements \cite{donohue2018quantum}.

\section{Adaptive estimation protocols}
\label{qmsec5}

\begin{figure}[ht!]
\centering
\includegraphics[width=.5\textwidth]{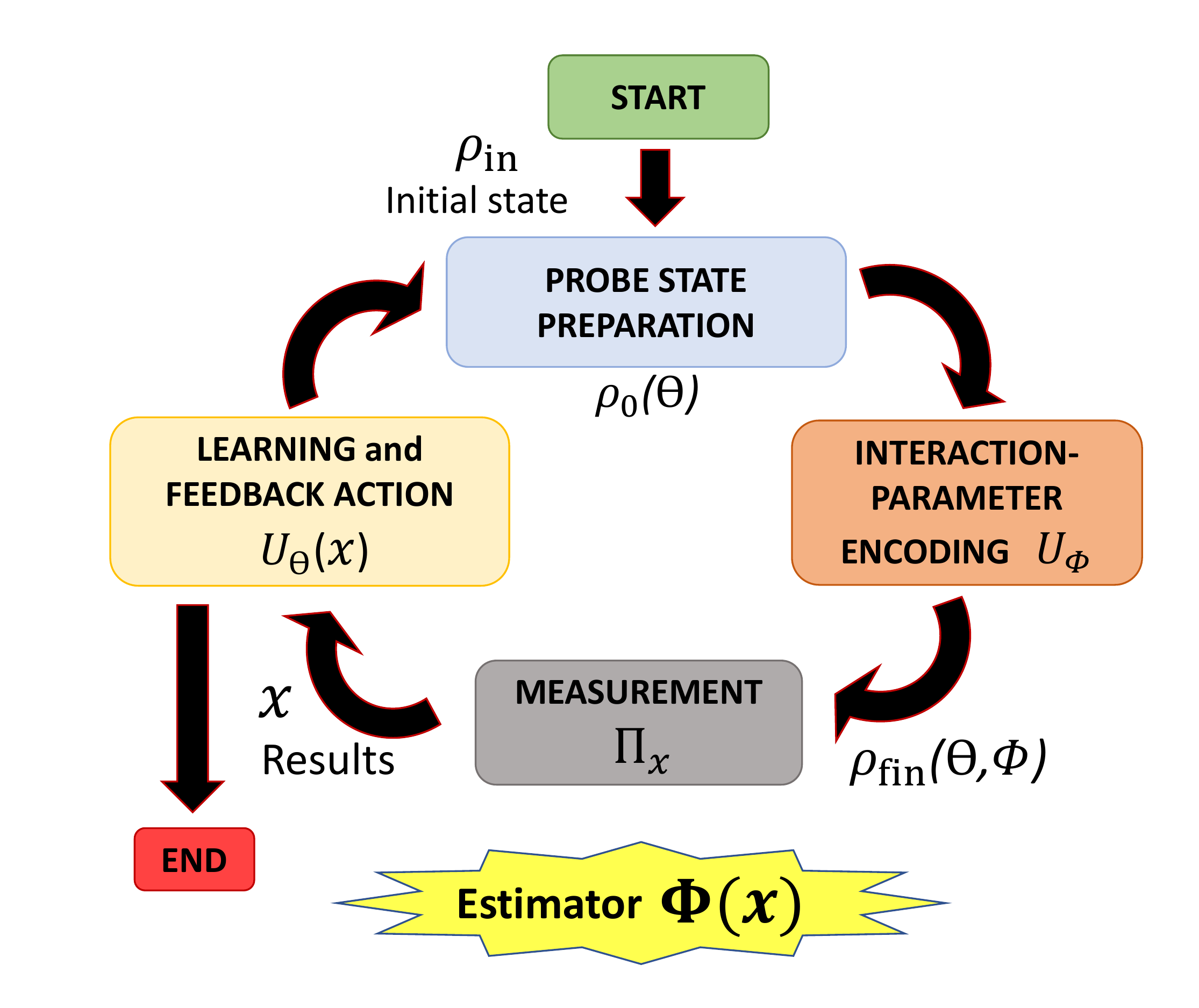}
\caption{{\bf Conceptual scheme of an adaptive estimation protocol.} The cycle of a general adaptive estimation protocol starts from an initial state $\rho_{\text{in}}$, that is prepared (blue box) in a state $\rho_0(\theta)$ through the action of $U_\theta$. Such state interacts with the unknown parameter $\phi$ (brown box), and then the output state $\rho_{\text{fin}}(\theta,\phi)$ undergoes an appropriately chosen measurement $\Pi_x$ (gray box). After such measurement, the results $X$ are exploited to define a suitable action $U_\theta(X)$ (orange box), employed to prepare the initial state of the next probe. In this way, the cycle is repeated for all the probes. At the end of the process, an estimator provides the final estimate of $\theta$.}
\label{adaptive}
\end{figure}

Different estimation protocols have been defined \cite{giovannetti2006quantum,demkowicz2014using}, and can be included in a few fundamental categories. A first example is provided by parallel protocols (Fig. \ref{parallel}) in which all the probes, entangled or not, interact in parallel with the system \cite{braunstein1996generalized,helstrom1976quantum}. Sequential (or multiround) protocols \cite{luis2002phase,rudolph2003quantum,higgins2007entanglement} are those where single probes interacts multiple times with the system, while ancilla-assisted ones \cite{van2007optimal,hotta2005ancilla,huang2016usefulness,huang2018noise,sbroscia2018experimental,wang2018entanglement} are those where a part of the probe, generally entangled with the other part, does not interact with the system and is directly measured. Such protocols can be non-adaptive \cite{giovannetti2011advances} or adaptive \cite{wiseman1995adaptive, berry2000optimal}. Here we focus on adaptive techniques, that represent a powerful tool to enhance the performances of estimation processes \cite{wiseman1995adaptive,berry2000optimal,doherty2000quantum,wiseman2009adaptive,serafini2012feedback,zhang2017quantumfeedback}. In non-adaptive estimation protocols, the available $m$ probes are sent through a fixed apparatus and, after collecting the full data set, a final estimate of the unknown parameter $\phi$ is obtained. Conversely, adaptive techniques make use of suitable controls on the experimental setup, namely some physical parameters $\bm{\theta}$, such as additional feedback phase shifts, that can be adjusted during the estimation.  Adaptive and entangled protocols can enhance metrology tasks, especially in presence of noise \cite{demkowicz2014using,demkowicz2017adaptive}. A discrete-time class of adaptive protocols can be schematically represented through the repetition, for each probe, of the four-step cycle as shown in Fig. \ref{adaptive}. 

$(i)$ The first step is dedicated to the preparation of an initial probe $\rho_{\text{in}}$, through a process $U_{\bm{\theta}} (\bm{x})$ that depends on certain parameters $\bm{\theta}$ and, if available, on the results $\bm{x}$ of previous measurements. 

$(ii)$ At a second stage, the prepared probe $\rho_0(\bm{\theta})$ interacts with the studied system and evolves under a unitary $U_\phi$ (for simplicity we are assuming unitary evolution) in $\rho_{\text{fin}}(\bm{\theta},\phi)$. 

$(iii)$ Then, a measurement $\Pi_{x}$ is performed and its outcome $x$ is recorded. 

$(iv)$ The final step of the cycle is post-processing of the measurement results. This step includes the choice of the parameters $\bm{\theta}$ determining the action $U_{\bm{\theta}}(\bm{x})$ to apply to the initial probe of the successive cycle.

This cycle is repeated for all the probes. Finally, an estimator $\Phi(\bm{x})$ based on all measurement results $\bm{x}$ provides an estimation of the unknown parameter $\phi$.

Exploiting adaptive protocols for quantum metrology was proposed in 1995 by Wiseman  \cite{wiseman1995adaptive}. Such protocols are necessary in order to overcome different issues. For instance, they can be used for the realization of the optimal POVM to saturate the QCRB. In certain scenarios, such POVMs can be hard or impossible to implement. In this case, approximation of such measurements can be achieved by adaptive techniques \cite{wiseman2009adaptive,gill2000state}. In particular, to approach the QCRB one has to maximize the Fisher Information of a given setup. However, the latter quantity depends in general on the unknown parameter. More specifically, given an initial probe, the QCRB is attainable only when the unknown parameter takes a value which maximizes the Fisher Information. Nevertheless, it can be demonstrated that, even with no prior knowledge on the unknown parameter, the QCRB can be asymptotically saturated by exploiting adaptive techniques \cite{barndorff2000fisher}.

A second scenario where an adaptive approach represents a useful resource is found for those systems where the output probabilities, calculated at different values of unknown parameter, take the same value. For instance, in a Mach-Zehnder interferometer seeded by single photons, the output probability ($P_1=1-P_0=\cos^2\frac{\phi}{2}$) is such that, in the range $\phi \in [0,2\pi]$, two different values of the phases lead to the same probability $P_1 \ne 0$. Indeed, the latter is not a monotonic function of the phase. Hence, without changing the relative phase shift during the experiment, it is impossible to discern the two equivalent phases leading to the same probability. Conversely, by changing the total phase shift during the estimation process, for instance through another known control phase, it is possible to solve such issue. In any case, when the output probabilities are periodic with a period less then $2\pi$, it is impossible to distinguish some phases. In such cases, one could employ an adaptive protocol where the probe state can change at each iteration, thus changing the likelihood function and its periodicity. For instance, during the first steps one can employ probes whose likelihood has no periodicity, in order to restrict the range of possible unknown phase values. When the range is sufficiently small, more sensible states with smaller periodicity can be used \cite{mitchell2005metrology}. However, the validity of such recipe depends on the problem symmetries.

Furthermore, an important task where adaptive protocols can be helpful is the convergence to the ultimate precision bounds in the limited data scenario \cite{rubio2019non, rubio2019quantum}. The latter regime characterizes different realistic conditions where the amount of resources that can be employed is restricted. In the single-parameter case, theorems guarantee that it is always possible to define suitable measurements and estimators, allowing to reach the minimum error achievable with a given probe state (see Sec.\ref{subsecQCRB}). However, this capability of reaching the ultimate bounds is guaranteed only in the asymptotic regime. Conversely, when only a limited number of probes is available, identifying the optimal strategies is a difficult task. To this end, one can employ adaptive protocols, leading to a boost in the convergence to the asymptotic limits.

Finally, adaptive protocols have to be taken into account to achieve the true quantum limits \cite{gorecki2020pi}. Importantly, feedback and error-correction schemes can be exploited to face noises and/or time-varying parameters  \cite{d1996feedback,berry2002adaptive,yonezawa2012quantum,kessler2014quantum,dur2014improved,roy2015robust,plenio2016sensing,pang2017optimal,sekatski2017quantum,demkowicz2017adaptive,layden2018spatial,zhou2018achieving,zhang2019quantum,zhuang2020distributed}.

There exist two prominent approaches to feedback-based phase estimation:  

- \emph{Online schemes:} 
At each step of the estimation protocol, the feedback is calculated according to the previous measurement result and a heuristic. An important class of these schemes is represented by Bayesian adaptive protocols. Here, at each step of the protocol, the posterior probability evolves based on measurement results. In this way, the posterior is used to calculate the optimal feedback action to be applied at next step. Note that optimality is defined depending on the particular problem and heuristic.

- \emph{Offline schemes:}   
The feedback values used during the experiments are computed before the estimation process. The goal is then the optimization of such sequence of feedback values. Different optimization techniques based on trial and error approaches can be exploited, such as those based on Particle Swarm Optimization (PSO) \cite{hentschel2010machine} and Differential Evolution (DE) \cite{lovett2013differential, palittapongarnpim2017learning,palittapongarnpim2019robustness}.

Finally, adaptive protocols can be also exploited to enhance state discrimination and more general in quantum tomography \cite{sugiyama2012adaptive,huszar2012adaptive,mahler2013adaptive,granade2016practical}. This has been experimentally demonstrated in the estimation of the photons polarization  \cite{okamoto2012experimental,kravtsov2013experimental,struchalin2016experimental,qi2017adaptive,okamoto2017experimental}. A detailed review on implementations of feedback controls in quantum systems can be found in Ref. \onlinecite{zhang2017quantumfeedback}. 

\subsubsection{Adaptive Bayesian protocols}
Bayesian estimation (Sec. \ref{sec:estimators}) naturally fits the requirements for adaptive protocols. In this framework, the posterior distribution is updated at each repetition of the estimation cycle [Eq.\eqref{bayes}]. The information encoded in this distribution can be exploited to choose the optimal feedback action according to the protocol heuristic. 

One of the first adaptive phase estimation, providing an experimental demonstration of the proposal in Ref. \onlinecite{wiseman1995adaptive}, was realized exploiting adaptive homodyne phase measurements on coherent states \cite{armen2002adaptive}. Coherent states with homodyne measurements were also employed for adaptive estimation of a continuously varying phase, beating SQL by exploiting smoothing  \cite{wheatley2010adaptive}. HL scaling, in this kind of schemes, cannot be achieved by employing coherent states as probes. However, an enhancement of a constant factor with respect to SQL can be obtained. Also coherent states discrimination can be performed through adaptive schemes \cite{cook2007optical}.

When employing quantum states, one can reach improved scaling in the estimation process. In this regime, when the phase to be estimated is completely unknown (flat prior distribution) adaptive techniques can be employed \cite{berry2000optimal,berry2001optimal}. This is the goal of an \emph{ab-initio} quantum phase estimation experiment that was experimentally realized by Ref. \onlinecite{xiang2011entanglement} using Holland-Burnett states. In such realization, a sequence of different states is used. In particular, single-, two- and four-photon HB states in the polarization degree of freedom were generated through a type-I SPDC process followed by interference in a polarizing beam splitter. These output states were detected by a probabilistic photon-number resolving detection.  The employed Bayesian protocol is composed of a first step with random feedback. Subsequently, after the measurement of a group of single-photon events, the posterior probability is updated and the next feedback is calculated by optimizing the expected sharpness function [Eq.\eqref{eq:sharpness}] over the possible results of the next measurements. Using suitable sequences of states (with photon numbers $N=1,2,4$), the SQL was surpassed \cite{xiang2011entanglement}. 

The SQL can be overcome by employing other classes of states, such as Gaussian squeezed states with squeezing parameter $r$ that reach a value for the variance \cite{monras2006optimal,olivares2009bayesian} equal to $V=1/[2N \sinh(2r)]$. Since for this resource state the optimal Fisher Information depends on the unknown phase, an adaptive protocol has to be employed, and Bayesian estimation can be exploited for this purpose. Given this class of input states, a Bayesian protocol for \emph{ab-initio} phase estimation has been experimentally realized using squeezed states and homodyne detection, together with real-time feedback \cite{berni2015ab}. The phase of a squeezed state is measured with respect to a local oscillator through homodyne detection. More specifically, a first set of data is exploited to perform a rough estimation of the phase. Then, the local oscillator phase is adjusted to the value that lead to the minimum error in the estimation process \cite{berni2015ab}. Finally, also two mode squeezed states can be exploited in adaptive protocols \cite{huang2017adaptive}. 

Bayesian adaptive estimation can be used to reach the HL with single photons in multipass configuration without the need of entanglement as demonstrated in Ref. \onlinecite{higgins2007entanglement}. In this case, single photons are employed for a multipass polarization interferometer estimating phases through a generalized Kitaev's algorithm \cite{kitaev1996quantum}. An adaptive hybrid approach, exploiting simultaneously polarization entangled 2-photon states and a multipass configuration (with $N=3$ passes per state, two for one photon and one for the other), achieved within $4\%$ the exact value of HL at finite number of resources \cite{daryanoosh2018experimental}. This implementation demonstrated the theoretical proposal of Ref. \onlinecite{wiseman2009adaptive}. The optimal state for this protocol is \cite{wiseman2009adaptive,daryanoosh2018experimental}: $\ket{\psi_{\text{opt}}}=c_0\ket{\Phi^+}+c_1 \ket{\Psi^+}$, with $\ket{\Phi^+}=(\ket{0,0}+\ket{1,1})/\sqrt{2}$, $\ket{\Psi^+}=(\ket{1,0}+\ket{0,1})/\sqrt{2}$ and $c_j=\sin[(j+1) \pi/5]/\sqrt{\sum_{k=0}^1 \sin[(k+1) \pi/5]^2}$ and was realized through a probabilistic control-Z gate \cite{ralph2010optical} between two SPDC photons. 

An efficient and robust adaptive Bayesian phase estimation protocol, called rejection filtering, \cite{wiebe2016efficient} was realized exploiting the evolution of pairs of photons in a silicon circuit. The latter implemented adaptive unitaries that depend on single events, extracted from collections of photon statistics \cite{paesani2017experimental}. 

An adaptive estimation experiment based on single-photon inputs was realized in a bulk Mach-Zehnder interferometer in the path degree of freedom, implementing two different Bayesian techniques \cite{piccoloLume}: (i) particle guess heuristic, in which at each step the feedback phase is randomly drawn from the posterior distribution\cite{wiebe2016efficient} and (ii) an optimal heuristic, which is derived analytically by optimizing the Bayesian mean square error of the future events over the feedback, under the assumption of narrow Gaussian prior \cite{piccoloLume}. In particular, the last optimized technique shows better performances than the PSO (discussed in details below) and the particle guess heuristics. Furthermore, such optimized technique has been experimentally demonstrated to be robust against different classes of noise.

\subsubsection{Machine Learning offline estimation techniques}

Offline machine learning techniques can be exploited to enhance quantum phase estimations. Machine learning techniques \cite{murphy2012machine, simon2013too} applied to physical problems represent a new, rich and continuously growing research area in which learning tools are used to enhance quantum information tasks \cite{carrasquilla2019reconstructing,torlai2019integrating,palmieri2020experimental,brajato2019optical,xu2018neural,rem2019identifying,cimini2019calibration,nichols2019designing,lian2019machine,agresti2019pattern,piccoloLume,rocchetto2019,magesan2015machine,butler2018,fischer2006,tairelly,melnikov,banchi2016quantum,wang2019quantum,Schuld2015,rupp2015,carleo2017solving, Biamonte2017,pilozzi2018machine,michimura2018particle,fosel2018reinforcement,bukov2018reinforcement,marcucci2019programming,o2019hybrid,you2019identification,xiao2019continuous,arrazola2019machine,nautrup2019optimizing,sabapathy2019production,kalantre2019machine,liu2020repetitive,krenn2016automated,peng2020feedback,schuff2020improving,iten2020discovering,gentile2020learning,cimini2020neural}. Such techniques can be also used to calibrate quantum sensors \cite{cimini2019calibration}. Note that also the opposite case is possible, namely quantum dynamics can enhance machine learning protocols \cite{rebentrost2014quantum,cai2015entanglement,Schuld2015, lloyd2014quantum,Biamonte2017,benedetti2019parameterized,lau2017quantum, dunjko2018machine,huggins2018towards,perdomo2018opportunities, wiebe2018quantum,steinbrecher2019quantum,romero2019variational,schuld2019quantum,cong2019quantum,killoran2019continuous,havlivcek2019supervised}. Remarkably, machine learning-based protocols have been developed also for adaptive quantum metrology \cite{hentschel2010machine, hentschel2011efficient, wiebe2014hamiltonian,wiebe2016efficient,paesani2017experimental,piccoloLume,palittapongarnpim2019robustness,lovett2013differential,peng2020feedback,schuff2020improving,Fiderer2020neural}.

Two significant machine learning techniques employed for quantum metrology with an offline approach are PSO \cite{hentschel2010machine,hentschel2011efficient} and DE \cite{palittapongarnpim2017learning,palittapongarnpim2019robustness}. Such techniques are able to self-learn the optimal feedback strategy to reach the ultimate limits on the scaling of the phase estimation uncertainty, with limited number of measurements. They are both based on reinforcement learning that is model-free, since it does not necessarily rely on the explicit model of the problem, but mainly on experience acquired from data. Even if a mathematical model is available, reinforcement learning techniques can surpass gradient-based greedy algorithms for non-convex optimizations in high-dimensional problems. In particular, PSO and DE are evolutionary algorithms \cite{eiben2003introduction,vikhar2016evolutionary}. Such algorithms often resemble biological evolution mechanisms and are characterized by the following features: the presence of a population of points  in the search space, the existence of a figure of merit called \emph{fitness} to be maximized and, finally, stochastic evolution of the solutions. One of the biggest advantage of evolutionary computation is the low probability of getting stuck at local optima of the function, since the space is explored by many candidate solutions and the optimization of the searching process happens in a quasi-random way. 

For phase estimation tasks, such approaches are applied to calculate, prior to the experiment, the sequence of optimal feedback phases shifts to be used during the adaptive experiments with $N$ probes. Considering a Mach-Zehnder interferometer, at each step $k$ of the experiment, the optimal feedback phase $\Phi_{k}$ can be updated according to the following Markovian rule with a logarithmic-search heuristic: 
\begin{equation}\label{eq:psofeedbackupdate}
    \Phi_{k} = \Phi_{k-1} - (-1)^{x_{k-1}} \Delta \Phi_{k},
\end{equation}
where $\Phi_{k-1}$ is the feedback phase at previous step, and $x_{k-1}=\{0,1\}$ is the result of the measurement at step $k-1$. The list of optimal phase shifts $\{\Delta \Phi_{k}\}$ for $k=1,\ldots N$ is called \emph{policy}. The final estimate for the unknown phase $\phi$ coincides with the last value $\Phi_{N}$ of the adaptive feedback phase at the end of the process according to $\Phi_{\text{est}}=\Phi_{N}$.

PSO is part of a class of unsupervised reinforcement learning algorithms for optimization problems \cite{eberhart1995new,blum2008swarm}, and can be exploited to compute the list of phase shifts $\{\Delta \Phi_{k}\}$ discussed above. The goodness of a policy is quantified by the sharpness of Eq.\eqref{eq:sharpness} relative to the estimation errors. Hence, the average of the sharpness is calculated over $P( \theta |\rho)$, that is the probability distribution of the error $\theta$ on the estimate given a policy $\rho$. In this way, the sharpness in Eq.\eqref{eq:sharpness} is the objective function that is maximized by PSO over the policies, and is related to the Holevo variance. When the sharpness is maximized, the Holevo variance is minimized. Given the number $N$ of employed photons in the estimation process, the goal of the PSO algorithm is to find the optimal policy by maximizing the associated sharpness. At each iterative step of the algorithm, every policy is mapped to a vector and compares its fitness with those relative to its neighborhood and to its past history. Then, the policies are updated according to a stochastic evolution rule depending on global and local optima. After a certain number of iterations, the last global optimum represents the solution of PSO. In Ref. \onlinecite{piccoloLume}, an adaptive scheme using PSO policies was realized using single photons in a path Mach-Zehnder interferometer, and SQL was approached after few photons ($\sim 20$).

However, it has been observed that PSO algorithm converges to optimal solutions only when the number of probes is small, and this limitation can be overcome by other techniques like Differential Evolution  \cite{lovett2013differential,palittapongarnpim2017learning}. DE is an evolutionary algorithm that performs a global optimization in the policies space by selecting and rejecting candidate policies according to their sharpness value. In particular, after a random initialization of candidate policies, at each iteration of the algorithm new polices are generated by combining randomly chosen policies. The policies with highest fitness values are then selected for the next step. This procedure is iterated until a halting condition for the fitness of the best policy is reached. These techniques are also resilient to different models of noise \cite{palittapongarnpim2019robustness}.

\section{Multiparameter quantum metrology} \label{Sec:multi}

In general, a physical process can involve more than one parameter. Analogously to the single parameter case, the estimation of multiple parameters can be enhanced by using quantum resources, giving rise to the emergent field of \emph{multiparameter quantum metrology} \cite{szczykulska2016multi,albarelli2020perspective}. A large effort has been devoted to such generalization of the single parameter case \cite{helstrom1976quantum,yuen1973multiple,young2009optimal,paris2009quantum,watanabe2010optimal,belavkin1976generalized,szczykulska2016multi,chiribella2012optimal,ballester2004estimation,monras2010information,monras2011measurement,ang2013optomechanical,albarelli2020perspective,helstrom1974noncommuting,ragy2016compatibility,zhang2014quantum,albarelli2019evaluating,chen2019optimal,chiribella2006joint,zhuang2018multiparameter,nichols2018multiparameter,bradshaw2017tight,matsubara2019optimal,yuan2016sequential,eldredge2018optimal,proctor2018multiparameter,li2018joint,berry2015quantum,kura2018finite,zhuang2017entanglement,kok2017role,liu2017control,tsang2011fundamental,liu2015quantum,knott2016local,baumgratz2016quantum,you2017multiparameter,yousefjani2017estimating,sidhu2017quantum,sidhu2018quantum,chen2017maximal,pham2011magnetic,shin2011quantum,tsang2009quantum,yang2019optimal,vaneph2013quantum,goldberg2018quantum,baumgratz2016quantum,suzuki2016explicit,komar2014quantum,matsumoto2002new,fujiwara2001estimation,nair2018quantum,zhang2017quantum,gorecki2019quantum,altorio2015weak,rubio2019bayesian,altenburg2018multi,liu2019quantum,tsang2019holevo,suzuki2019quantum,gessner2019metrologicalmulti,lu2020generalized,sidhu2019tight,carollo2019quantumness,sekatski2019optimal,suzuki2019information,gao2014bounds,macchiavello2003optimal,humphreys2013quantum,ballester2004entanglement,liu2016quantum,gagatsos2016gaussian,ge2018distributed,ciampini2016quantum,takeoka2017fundamental,you2019conclusive,pezze2017optimal,zhang2018scalable,gessner2018sensitivity,gatto2019distributed,guo2019distributed,li2019multi,polino2019experimental,valeri2020experimental,knysh2013estimation,vidrighin2014joint,yue2014quantum,genoni2011optical,crowley2014tradeoff,szczykulska2017reaching,roccia2017monitoring,roy2019fundamental, roccia2017entangling,yao2014multiple,roccia2018multiparameter,cimini2019quantum,acin2001optimal,fujiwara2003quantum,hayashi2006parallel,bagan2006optimal,kahn2007fast,hayashi2008asymptotic,zhou2015quantum, yang2019attaining,genoni2013optimal,steinlechner2013quantum,ast2016reduction,zhuang2018distributed,bradshaw2018ultimate,ivanov2018quantum,tsang2016quantum,vrehavcek2017multiparameter,bisketzi2019quantum,ang2017quantum,chrostowski2017super,vrehavcek2018optimal,backlund2018fundamental,yu2018quantum,genoni2019non,napoli2019towards,zhou2019quantum,grace2019approaching,hou2020ultimate,hall2018entropic,albarelli2019upper,gross2020one,rubio2020quantum,demkowicz2020multi,kull2020uncertainty}. In this scenario, while it may be possible to estimate separately the single parameters, in most of the cases simultaneous approach has to be adopted. Two main motivations can be identified in this direction: \textit{(i)} simultaneous estimation of the parameters can be more efficient, in terms of employed resources, with respect to the separate estimation; \textit{(ii)} in certain conditions, even if the parameter of interest is a single one, the estimation process unavoidably involves other parameters, such as noises, which have to be estimated simultaneously.

A large class of problems involves the estimation of multiple parameters and can then benefit of a quantum enhancement. Different examples are parameters estimation for gravitational waves detection \cite{freise2009triple,schnabel2010quantum}, multiple phases  \cite{macchiavello2003optimal,humphreys2013quantum,ballester2004entanglement,liu2016quantum,gagatsos2016gaussian,ge2018distributed,ciampini2016quantum,pezze2017optimal,takeoka2017fundamental,zhang2018scalable,gessner2018sensitivity,you2019conclusive,gatto2019distributed,guo2019distributed,li2019multi,polino2019experimental,valeri2020experimental}, phases and noises \cite{knysh2013estimation,vidrighin2014joint,yue2014quantum,genoni2011optical,crowley2014tradeoff,szczykulska2017reaching,roccia2017monitoring,roy2019fundamental, roccia2017entangling,yao2014multiple,roccia2018multiparameter,cimini2019quantum}, estimation of qubit mixed
states and quantum tomography \cite{acin2001optimal,fujiwara2003quantum,hayashi2006parallel,bagan2006optimal,kahn2007fast,hayashi2008asymptotic,zhou2015quantum,   yang2019attaining}, multidimensional fields \cite{baumgratz2016quantum}, force sensing \cite{tsang2011fundamental,ang2013optomechanical}, spin rotations \cite{vaneph2013quantum}, rotations about unknown axes \cite{goldberg2018quantum}, general functions of unknown parameters \cite{gross2020one,rubio2020quantum}, displacements in phase space \cite{genoni2013optimal,steinlechner2013quantum,ast2016reduction,bradshaw2017tight,zhuang2018distributed,bradshaw2018ultimate,ivanov2018quantum}, squeezing and displacement of a radiation \cite{chiribella2006joint}, imaging \cite{tsang2009quantum,giovannetti2011advances,shin2011quantum}, localization of incoherent point sources \cite{tsang2016quantum,vrehavcek2017multiparameter,bisketzi2019quantum,ang2017quantum,chrostowski2017super,vrehavcek2018optimal,backlund2018fundamental,yu2018quantum,napoli2019towards,zhou2019quantum,grace2019approaching}, spatial deformation of sources \cite{sidhu2017quantum,sidhu2018quantum}, local beam tracking \cite{qi2018ultimate}, sensing on biological systems \cite{taylor2016quantum}, range and velocity measurements \cite{zhuang2017entanglement}, atomic clocks networks \cite{komar2014quantum}, quantum sensing networks \cite{proctor2018multiparameter,eldredge2018optimal,qian2019heisenberg}, magnetic field imaging \cite{steinert2010high,pham2011magnetic,rondin2014magnetometry,hall2012high,arai2015fourier,zhuang2019simultaneous,hou2020ultimate}, Hamiltonian parameters \cite{zhang2014quantumh,kura2018finite} and general sensing technologies \cite{yousefjani2017estimating}.

Despite the large number of applications, multiparameter quantum metrology is characterized by several open questions with respect to the single parameter scenario. For instance, the possibility of saturating the ultimate quantum bound is not always guaranteed  \cite{helstrom1976quantum,helstrom1974noncommuting,yuen1973multiple,matsumoto2002new,fujiwara2001estimation,pezze2017optimal}. In parallel to investigations on the theoretical framework, growing interest is devoted in experimental implementations where vast unexplored areas still remain \cite{szczykulska2016multi,albarelli2020perspective}.

In the next sections we will briefly introduce the theoretical framework of multiparameter quantum metrology. Then, we will describe some specific multiparameter problems that have been studied with photonic platforms. Note that recent in-depth reviews on multiparameter quantum metrology can be found in Refs. \onlinecite{szczykulska2016multi,albarelli2020perspective}.

\subsection{Generalized theoretical framework for multiparameter quantum metrology}

The general scheme of a multiparameter estimation follows the same steps of the single parameter case (Fig. \ref{fig:figure1}): preparation of the probe state, interaction and parameters encoding, probe measurement and the estimator function.

Consider a multiparameter estimation task where $d$ unknown parameters $\bm{\lambda}=(\lambda_1,\lambda_2,...,\lambda_d)$ are obtained through a set of estimators  $\bm{\Lambda}(\bm{x})=(\Lambda_1(\bm{x}),\Lambda_2(\bm{x}),...,\Lambda_d(\bm{x}))$, after $\bm{x}$ measurement results. Each parameter $\lambda_i$, with $i=1,...,d$, can represent a physical quantity. When more than one parameter is involved in the process, the Fisher Information is generalized to the real-valued symmetric Fisher Information matrix ($\mathbb{F}$):

\begin{equation}
\mathbb{F}(\bm{\lambda})_{ij}=\sum_x\left[\frac{1}{ P(x|\bm{\lambda})}\frac{\partial P(x|\bm{\lambda})}{\partial\lambda_i}\frac{\partial P(x|\bm{\lambda})}{\partial\lambda_j}\right].
\end{equation}

The sensitivity of an estimator is quantified by its covariance matrix, which is defined as:
\begin{equation}
\label{CovMat}
C(\bm{\lambda})_{ij}=\sum_{\bm{x}}[\bm{\Lambda}(\bm{x})-\bm{\lambda}]_{i}\:[\bm{\Lambda}(\bm{x})-\bm{\lambda}]_{j} \; P(\bm{x}|\bm{\lambda}),
\end{equation}
with $i,j=1,...,d$. The covariance matrix provides a measure of the sensitivity relative to each parameter, while taking also into account the possible correlations between them.

In analogy with the single parameter case, a vector of estimators $\bm{\Lambda}(\bm{x})$ are said to be unbiased if the following relation holds: 
\begin{equation}
\sum_{\bm{x}}[\bm{\Lambda}(\bm{x})-\bm{\lambda}] \; P(\bm{x}|\bm{\lambda})=0.
\end{equation}
A locally unbiased estimator is an unbiased estimator that satisfies the following constraint: 
\begin{equation}
\sum_{\bm{x}}\Lambda_k(\bm{x}) \; \frac{\partial P(\bm{x}|\bm{\lambda})}{\partial \lambda_i}=\delta_{ik}.
\end{equation}

Note that, in the case of continuous-valued measurement outcomes $x$, an integral over $x$ will replace the sums in all these expressions.

For an unbiased estimator, the Cramer-Rao bound (CRB) \cite{helstrom1974noncommuting,lehmann2006theory} in the multiparameter case is generalized to the following matrix inequality: 
\begin{equation}
\label{eq:crbmatrix}
C(\bm{\lambda})\ge \mathbb{F}^{-1}(\bm{\lambda})/\nu,
\end{equation}
where $\nu$ is the number of independent measured probes. The CRB is well defined only when $\mathbb{F}$ is strictly positive, and thus invertible. In this case, the inequality in Eq.\eqref{eq:crbmatrix} can always be saturated by maximum likelihood estimator \cite{helstrom1974noncommuting} in the limit of large $\nu$. Conversely, local unbiased estimators can reach the CRB for any number of measurements $\nu$.

In complete analogy to the single parameter scenario, it is possible to define a Quantum Fisher Information matrix \cite{liu2019quantum,vsafranek2018simple} $\mathbb{F}_{\mathrm{Q}}$, that only depends on the initial probe state $\rho_0$ and on the transformation $U_{\bm{\lambda}}$. It is defined as:
\begin{equation}
\mathbb{F}_{\mathrm{Q}}(\bm{\lambda})_{ij}=\text{Tr}\left[\rho_{\bm{\lambda}}\frac{L_iL_j+L_jL_i}{2}\right],
\end{equation}
where $L_i$ is the symmetric logarithmic derivative of $\rho_{\bm{\lambda}}$ with respect to the parameter $\lambda_i$, defined as $\partial_{\lambda_i}\rho_{\bm{\lambda}}=(L_i\rho_{\bm{\lambda}}+\rho_{\bm{\lambda}} L_i)/2$. 
$\mathbb{F}_{\mathrm{Q}}$ has the following properties \cite{liu2019quantum}:

\emph{(i) Semi-definite positivity}: $\mathbb{F}_{\mathrm{Q}}\ge 0$;

\emph{(ii) Convexity}: $\mathbb{F}_{\mathrm{Q}}(p\,\rho_1+(1-p)\,\rho_2 )\le p\, \mathbb{F}_{\mathrm{Q}}(\rho_1)+ (1-p)\,\mathbb{F}_{\mathrm{Q}}(\rho_2 )\;$ for any $\rho_1, \rho_2$ and $p\in [0,1]$;

\emph{(iii) Additivity}: given $\nu$ independent probes $\rho_i$ ($i=1,...,\nu$), the Quantum Fisher Information matrix of the total product state $\rho^{\text{tot}}=\bigotimes_{i=1}^\nu \rho_i$ is: $\mathbb{F}_{\mathrm{Q}}(\rho^{\text{tot}})= \sum_{i=1}^\nu\mathbb{F}(\rho_i)$;

\emph{(iv)} $\mathbb{F}_{\mathrm{Q}}(\rho)= \mathbb{F}_{\mathrm{Q}}(U \rho U^\dagger)$, for any unitary $U$ independent from the unknown parameters $\bm{\lambda}$.

A review on the properties and applications and calculation techniques of Quantum Fisher Information matrix can be found in Ref. \onlinecite{liu2019quantum}, including a discussion on the infinitesimal generators of the parameters.

The Quantum Cramer-Rao bound (QCRB) in the multiparameter case is the following matrix inequality:
\begin{equation}
\label{QCRBM}
C(\bm{\lambda})\ge \frac{\mathbb{F}^{-1}(\bm{\lambda})}{\nu} \ge \frac{\mathbb{F}_{\mathrm{Q}}^{-1}(\bm{\lambda})}{\nu}.
\end{equation}
In particular, by summing over the diagonal elements of the matrix inequality \eqref{QCRBM}, one can estimate the precision of a multiparameter estimator as the trace of the covariance matrix in Eq. (\ref{CovMat}), that obeys the scalar bound: 
\begin{equation}
\label{MQCRBSvar}
\sum_{i=1}^{d}(\Delta \Lambda_i) ^2 \geq\;\frac{\text{Tr}\left[ \mathbb{F}^{-1}(\bm{\lambda}))\right]}{\nu}\geq\;\frac{\text{Tr}\left[ \mathbb{F}_{\mathrm{Q}}^{-1}(\bm{\lambda}))\right]}{\nu}.
\end{equation}
The QCRB is saturated when the equality in the second part of Eq. (\ref{QCRBM}) is reached. Note that it is also possible to define other quantities and bounds, and classify different multiparameter problems \cite{albarelli2020perspective, suzuki2019information}. For instance, in the multiparameter scenario, different perspectives can be provided by other metrics such as the right logarithmic derivative (RLD) $R_i$ relative to a state $\rho_{\bm{\lambda}}$ \cite{yuen1973multiple,helstrom1974noncommuting} defined as: $\partial_i \rho_{\bm{\lambda}}=\rho_{\bm{\lambda}}\, R_i$. In this case, a matrix $\mathbb{I_R}(\bm{\lambda})_{ij}$ can be defined as $\mathbb{I_R}(\bm{\lambda})_{ij}= \text{Tr}[R^\dagger_i\, \rho_{\bm{\lambda}}\,R_j]$. The following bound can be demonstrated \cite{sidhu2020geometric}: $C(\bm{\lambda})\ge \mathbb{I_R}^{-1}(\bm{\lambda})$. In some multiparameter cases this bound can be tighter, that is more accurate, than the QCRB in Eq.\eqref{QCRBM}. However, hereafter we will consider only the QCRB with the metric defined by symmetric logarithmic derivative.

Considering pure probe states ($\rho_{\bm{ \lambda}}\rightarrow |\Psi_{\bm{\lambda}}\rangle$),   $\mathbb{F}_{\mathrm{Q}}$ can be expressed according to the following relation:
\begin{equation}
\label{qfimpure}
\mathbb{F}_{\mathrm{Q}}(\bm{\lambda})_{ij}= 
4 \text{Re}[\langle \partial_{\lambda_i} \Psi_{\bm{\lambda}}|\partial_{\lambda_j} \Psi_{\bm{\lambda}}\rangle]+ 4 \langle \partial_{\lambda_i} \Psi_{\bm{\lambda}}|\Psi_{\bm{\lambda}}\rangle \langle \partial_{\lambda_j} \Psi_{\bm{\lambda}}|\Psi_{\bm{\lambda}}\rangle.
\end{equation}
where $|\partial_{\lambda_i}\Psi_{\bm{\lambda}}\rangle \equiv \partial|\Psi_{\bm{\lambda}}\rangle/ \partial \lambda_i$.

In order to find the best possible accuracy on the estimation, it is fundamental to find necessary and sufficient conditions to saturate the QCRB. As previously anticipated, the possibility of achieving the ultimate quantum bounds in multiparameter estimations is not guaranteed \cite{helstrom1976quantum,helstrom1974noncommuting,yuen1973multiple,matsumoto2002new,fujiwara2001estimation,pezze2017optimal}, at variance with the single parameter case \cite{paris2009quantum}. Indeed, when different parameters have to be estimated, the corresponding optimal measurements may not commute, thus making impossible their implementation in a single experiment \cite{barndorff2000fisher}. In this way, the capability of achieving the ultimate bounds is forbidden.

Conversely, $d$ parameters are said \emph{compatible} if they can be simultaneously estimated by some proper probe state, measurement and estimator, in such a way that the precision on each parameter is equal to the optimal precision achievable estimating each individual parameter separately \cite{ragy2016compatibility}. In this case, a simultaneous estimation process achieves a reduction of the resources by a factor $d$ compared to any separate strategy. Hence, there is an advantage in simultaneously estimating the parameters with respect to performing a separate estimation process.

A necessary condition for the attainability of the multiparameter QCRB inequality is provided by the following constraint \cite{ragy2016compatibility,matsumoto2002new}:
\begin{equation} \label{eq:saturability}
\text{Tr}[\rho_{\bm{\lambda}}\,\,[L_i,L_j]]=0.
\end{equation}
The latter equality corresponds to requiring that the optimal measurements for the estimation of the single parameters are compatible observables, which in general may not be satisfied. Importantly, for pure states there exists a necessary and sufficient condition for the saturation of the QCRB. If $\mathbb{F}_{\mathrm{Q}}$ corresponding to state $|\Psi_{\bm{\lambda}}\rangle$ is invertible, the QCRB can be saturated if and only if \cite{matsumoto2002new}:
\begin{equation}
\label{conns}
\text{Im}[\langle  \Psi_{\bm{\lambda}}|L_i L_j|\Psi_{\bm{\lambda}}\rangle]=0 \; \;\;\;\; \;\;\;\;\;\;\forall\;\;\; i,\; j.
\end{equation} 
Here, $L_i(\bm{\lambda})$ has the following expression for pure states: $\label{symmder} L_i(\bm{\lambda})=2(|\partial_{\lambda_i} \Psi_{\bm{\lambda}}\rangle \langle\Psi_{\bm{\lambda}}|+| \Psi_{\bm{\lambda}}\rangle \langle \partial_{\lambda_i}\Psi_{\bm{\lambda}}|)$. In Ref. \onlinecite{pezze2017optimal} the authors generalize such results. In particular, in the case of pure states necessary and sufficient conditions on projective measurements are derived such that the Fisher Information matrix $\mathbb{F}$ is equal to  $\mathbb{F}_{\mathrm{Q}}$ even if $\mathbb{F}_{\mathrm{Q}}$ is not invertible. If $\mathbb{F}_{\mathrm{Q}}$ is invertible, such conditions are necessary and sufficient also for the saturation of QCRB. When the generators of the parameters commute and the probe state is pure the QCRB can be saturated \cite{ragy2016compatibility,pezze2017optimal}.

In parallel to the single parameter case, $\mathbb{F}_{\mathrm{Q}}$ is related to the geometric distance between states, generalizing relation \eqref{qfibures}. Let us consider an infinitesimal variation $\delta \bm{\lambda_i}$ of the parameter vector $\bm{\lambda_i}$. The following equality holds \cite{paris2009quantum}:  
\begin{equation}
\label{qfimbures}
\tilde{D}_{\mathrm{B}}(\rho_{\bm{\lambda}},\rho_{\bm{\lambda}+\delta \bm{\lambda}})^2 = \frac{1}{8}\sum_{ij}\mathbb{F}_{\mathrm{Q}\,\,ij}(\rho_{\bm{\lambda}})\;\delta  \lambda_i \;\delta \lambda_j,
\end{equation}
where $\tilde{D}_{\mathrm{B}}$ is the Bures distance.
A technique able to optimize Bayesian multiparameter estimation in presence of limited data has been proposed in Ref. \onlinecite{rubio2019bayesian}.

Despite the broad range of applications, the number of experimental implementation of quantum multiparameter estimation tasks are surprisingly few. In this scenario, photons can be employed with different schemes and approaches \cite{szczykulska2016multi,albarelli2020perspective,you2017multiparameter,gessner2019metrologicalmulti}. In the next sections we list some of the problems that have been approached through photonic platforms.

\subsection{Multiphase estimation}

An important task in quantum multiparameter estimation is provided by those problems where the physical quantities to be estimated are multiple phases. This scenario has been intensively studied in the last years \cite{proctor2018multiparameter,you2017multiparameter,zhang2017quantum,macchiavello2003optimal,humphreys2013quantum,ballester2004entanglement,liu2016quantum,gagatsos2016gaussian,takeoka2017fundamental,ge2018distributed,ciampini2016quantum,pezze2017optimal,zhang2018scalable,gessner2018sensitivity,gatto2019distributed,guo2019distributed,li2019multi,polino2019experimental,valeri2020experimental}. More specifically, the unknown parameters are relative phases corresponding to different paths in an interferometer with respect to a common reference. Besides direct mapping of this problem to quantum imaging, multiphase estimation can represent a benchmark suitable for tests of quantum multiparameter protocols. Its importance and generality derives also from the fact that unitary evolutions generally introduce a phase in the evolved states. 

\begin{figure*}[ht!]
\centering
\includegraphics[width=1.\textwidth]{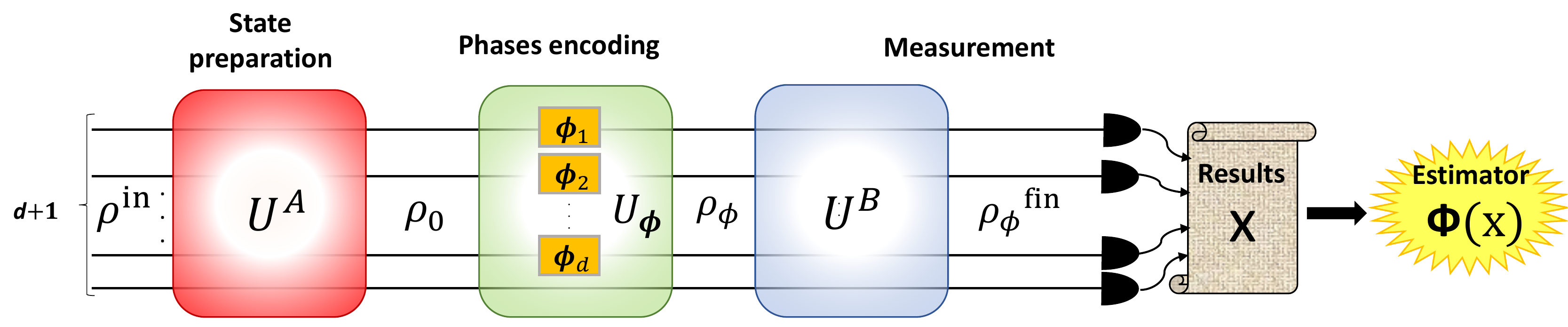}
\caption{{\bf Multiphase estimation scheme.} An initial probe $\rho^{\text{in}}$, living in the space of the ($d+1$) paths, is prepared in a state $\rho_{0}$ through a unitary evolution $U^A$. Then, the probe interacts with the phases $\phi_1,...,\phi_d$ according to an evolution $U_{\bm{\phi}}$. The state is measured by means of a unitary $U^B$ followed by a projective measurement, giving outcome $\bm{x}$. Finally, an estimate of the unknown phases is given by a suitable estimator $\bm{\Phi}(\bm{x})$.}
\label{fig:multiphasescheme}
\end{figure*}

Let us now consider multiphase estimation in a multiarm interferometer. Here, the unknown parameters are a set of phases (relative to a reference) along $d$ arms of an interferometer:  $\bm{\phi}=(\phi_1, \phi_2,...,\phi_{d})$.  The general scheme of a multiphase estimation is sketched in Fig. \ref{fig:multiphasescheme}.
Preparation of the probe along the $(d+1)$ paths is realized by an operation $U^A$, considered to be unitary for simplicity. After the evolution $U_{\bm{\phi}}$, that depends on the unknown phases $\phi_1, ... ,\phi_d$, the state is measured through a second unitary $U^B$ and projective measurements performed on the output paths. Finally, a suitable estimator $\bm{\Phi}(\bm{x})=[ \Phi_1(\bm{x}),  \Phi_2(\bm{x}),...,  \Phi_{d}(\bm{x})]$ provides an estimate of the phases by exploiting the $m$ measurement outcomes $\bm{x}=(x_1,...,x_m)$.

For pure input probes, prepared in $|\Psi_0 \rangle$, the state after the phase unitary evolution $U_{\bm{\phi}}$ reads $|\Psi_{\bm{\phi}}\rangle=U_{\bm{\phi}} |\Psi_0 \rangle$, where $U_{\bm{\phi}}=e^{(i\sum_{i=1}^{d}O_i\phi_i)}$. In this expression, each $O_i$ represents the generator of the phase shift $\phi_i$ along the mode $i$. When the operators $O_i$ mutually commute, and hence $[O_i, O_j]=0$ $\forall i,j$, the Quantum Fisher Information matrix $\mathbb{F}_{\mathrm{Q}}$ takes the following form: 
\begin{equation}
\label{qfimcova}
\mathbb{F}_{\mathrm{Q}}(\bm{\phi})_{ij}=4 [\langle O_i O_j\rangle-\langle O_i\rangle \langle O_j \rangle],
\end{equation}
where the average $\langle \cdot \rangle$ is calculated with respect to state $|\Psi_{\bm{\phi}}\rangle$. When the phases are those corresponding to independent modes, the generators are $O_i=n_i$ (see Sec. \ref{qmsec2}), where $n_i$ is the photon number operator for mode $i$. Since $[n_i, n_j]=0$  $\forall i,j$, from (\ref{qfimcova}) we find that $\mathbb{F}_{\mathrm{Q}}(\bm{\phi})_{ij}=4 [\langle n_i n_j\rangle-\langle n_i\rangle \langle n_j\rangle]$. Hence, the quantum Fisher Information $F_{\mathrm{Q}\phi_i}$ of a single phase $\phi_i$ corresponds to:
\begin{equation}
F_{\mathrm{Q} \phi_i}=\mathbb{F}_{\mathrm{Q}ii}=4\langle (\Delta n_i)^2 \rangle, 
\end{equation}
where $(\Delta n_i)^2$ is the variance of the photon number operator $n_i$. 

One of the first studies on simultaneous quantum enhanced estimation of multiple independent phases was performed in Ref. \onlinecite{humphreys2013quantum}. The authors considered probe states with a fixed number of photons, and a number $d$ of independent phase differences to be estimated for $d$ modes of an interferometer with respect to an additional reference mode. The simultaneous estimation of the phases can provide an advantage in the variance that scales as $O(d)$, with respect to the best quantum strategy that estimates such phases individually \cite{humphreys2013quantum}. In particular, this result is demonstrated using suitable optimized projective measurements on the optimal quantum probe states of the form:
\begin{equation}\label{eq:humpreysstates}
\begin{split}
  \ket{\Psi}_{\text{opt}}&=  \frac{1}{\sqrt{d+\sqrt{d}}}[|0,N, \ldots, 0, 0 \rangle+...+|0,0, \ldots, N, 0\rangle+\\
 & +|0,0, \ldots 0,N \rangle]+
  \sqrt{\frac{\sqrt{d}}{d+\sqrt{d}}}|N,0, \ldots 0,0\rangle,
  \end{split}
\end{equation} 
where $N$ is the number of photons contained in the probe state. The state is distributed along $d+1$ modes and the last term of the superposition indicates $N$ photons occupying the reference arm. Such optimal states lead to a total variance equal to:
\begin{equation}\label{eq:humpreysvariance}
 \sum_{i=1}^{d} (\Delta \phi_i)^2_{\text{opt}} \geq\;\text{Tr}\left[ \mathbb{F}_{\mathrm{Q}}^{\mathrm{opt}\;-1} \right]= \frac{(1+\sqrt{d})^2\, d}{4\,N^2},
\end{equation} 
This leads to an advantage (in the variance) of a factor $O(d)$ with respect to the optimal separate quantum single-phase estimation leading to $\text{Tr}\left[ \mathbb{F}_{\mathrm{Q}}^{\mathrm{sep}\;-1} \right] \ge d^3/N^2$. This enhancement achieved by performing simultaneous estimation can be found also with noncommuting unitary parameter generators \cite{baumgratz2016quantum} and in the presence of small amount of losses \cite{yuen1973multiple}. A simultaneous multiphase estimation can even provide a higher advantage by using entangled coherent states \cite{liu2016quantum}.

Multiphase estimation in multimode interferometers has been theoretically studied in Refs. \onlinecite{ciampini2016quantum,spagnolo2013three}. A  bound on the achievable sensitivity using separable probe states has been obtained\cite{ciampini2016quantum}, providing conditions of useful entanglement for the simultaneous estimation. A multimode interferometer is composed of two cascaded $(d+1)$-mode balanced multiport splitters (the $(d+1)$-mode extension of beam splitters), resembling the structure of a Mach-Zehnder interferomer. The internal modes include $d$ independent phase shifts between the different internal paths with respect to one of the modes acting as a reference. In Ref. \onlinecite{ciampini2016quantum} the authors study input multimode Fock states $|1\rangle_1 \otimes|1\rangle_2\cdot \cdot \otimes |1\rangle_{d+1} \equiv |11\cdot \cdot 1\rangle$, where $|1\rangle_i$ represent a single photon along the mode $i$.  The benchmark for the sensitivity in Eq.\eqref{MQCRBSvar} is given by the lower estimator variance, achievable by using $m$ separable photons to jointly estimate the $d$ phases \cite{pezze2009entanglement,ciampini2016quantum}: 
\begin{equation}
\label{sepbounf}
\sum_{i=1}^{d}\Delta \phi_i^2 \geq\;\frac{\text{Tr}\left[ \mathbb{F}^{-1}(\bm{\phi}))\right]}{m}\geq \;\frac{d}{m}.
\end{equation}
This limit is valid for each separable state transformed by the action of the phase generators, and for all possible POVMs. Hence, it represents the classical limit in this scenario. Useful entanglement is then present in the state when the variance of the estimator is lower than the bound (\ref{sepbounf}). Such bound can be surpassed by injecting indistinguishable photons into the multimode interferometer\cite{ciampini2016quantum}. To reach optimal and symmetric bounds for each value of the jointly estimated phases, an adaptive estimation protocol can be in principle exploited. This is obtained by employing additional control phases along the mode of the interferometer to perform adaptive measurements \cite{ciampini2016quantum}.

A deeper insight in multiphase estimation is obtained by using the CRB/QCRB inequality in its matrix formulation of Eq.\eqref{QCRBM}: $C(\bm{\phi})\ge \mathbb{F}^{-1}(\bm{\phi})/\nu \ge \mathbb{F}_{\mathrm{Q}}^{-1}(\bm{\phi})/\nu$, being $\nu$ the number of repeated independent measurements. The relevance of considering the covariance matrix $ C(\bm{\phi})$ to study the sensitivity bounds is highlighted by the possibility to compare any target scenario, with corresponding Fisher information matrix $\mathbb{F}_{\text{target}}$, with a benchmark state associated to a Fisher information $\mathbb{F}_{\text{bench}}$. As shown by Ref. \onlinecite{gessner2018sensitivity} such comparison can be studied through the matrix $\mathbb{F}_{\text{target}}$ - $\mathbb{F}_{\text{bench}}$. Indeed, the number of positive eigenvalues of this matrix corresponds to the number of independent combinations of the unknown parameters \cite{gessner2018sensitivity} for which the target state provides an enhancement compared to the benchmark state.

\subsubsection{Photonic platforms for multiphase estimation problems}

Photonic systems represent the most natural platform for multiphase estimation problems. Surprisingly, not many experimental realizations of quantum multiphase estimation have been reported. 

As previously discussed, a relevant benchmark problem is represented by the estimation of different optical phases along different spatial paths, with direct application in the vast area of imaging. Integrated circuits represent an ideal and scalable platform to investigate experimentally such scenario. Besides the quality of spatial mode interactions, integrated photonics provides the stability that is necessary to estimate relative phases along different paths, which is almost impossible to achieve in bulk optics platforms because of thermal fluctuations and mechanical vibrations.

In Ref. \onlinecite{polino2019experimental} the authors realized the first experimental implementation of multiphase estimation enhanced by quantum states.
The employed platform is an integrated three-mode interferometer realized through the femtosecond laser writing technique. Such device is composed by two cascaded tritters (the three-mode analogue of beam splitters) \cite{spagnolo2013three, spagnolo2013general} and includes six reconfigurable thermo-optic phase shifters (see Fig.\ref{fig:multiexp}a). The first tritter, described by an unitary $U^A$ prepares the input probe state starting from indistinguishable photons, through an HOM interference effect. The final tritter, described by $U^B$, is part of the measurement process together with single-photon detectors. After calibrating the device, the capability to achieve quantum advantage in multiphase estimation was experimentally demonstrated by performing two-photon measurements \cite{polino2019experimental}. In particular, the Fisher Information of the device $\mathbb{F}_{\text{exp}}$  was estimated from experimental data and compared with that relative to the optimal simultaneous strategy with separable probes ($\mathbb{F}_{\text{cl}}$). For some values of the unknown phases, the matrix $\mathbb{F}_{\text{exp}}$ - $\mathbb{F}_{\text{cl}}$ has two positive eigenvalues demonstrating a quantum advantage reached by the circuit. Such advantage can be in principle extended to all pair of phases through adaptive protocols. The sensitivity enhancement was achieved experimentally with respect to classical strategies, considering as resources the number of effectively detected coincidences \cite{polino2019experimental}. The same setup has also been exploited in Ref. \onlinecite{valeri2020experimental} for the implementation of a Bayesian adaptive multiphase estimation \cite{granade2012robust} using single photons inputs.

Recently, distributed quantum sensing of the linear combination (arithmetic average) of multiple small phases along four distant nodes was performed  \cite{guo2019distributed}. The scenario is a network of $M$ nodes along which independent relative phase shifts $\phi_i$, with $i=1,...,M$, one for each node, are experienced by the probes. The final goal is to estimate the arithmetic average of the phases: $\bar{\phi}=\sum_{i=1}^M \phi_i/M$.  The employed probe state is a squeezed coherent state of the form $D(\alpha)S(r)\ket{0}$, where $D(\alpha)$ is the displacement operator [Eq.\eqref{displacement}] with amplitude $\alpha$, and $S(r)$ is the squeezing single mode operator [Eq.\eqref{squeezingOp}] with squeezing parameter $r$. The output state is detected through homodyne detectors along each node, thus measuring the phase quadratures $P_i$ ($i=1,...,M$) representing the estimators for the phases. Given such kind of state, two classes of estimation experiments are possible: $(i)$ separable estimation in which $M$ independent and identical squeezed coherent probes are sent each along a single node, thus separately estimating the associated phases, and $(ii)$ entangled estimation in which a single initial squeezed coherent state is equally divided along the $M$ nodes by initial beam splitters, that generate mode entanglement in the probe state (Fig. \ref{fig:multiexp}b). The authors in Ref. \onlinecite{guo2019distributed} showed that, in the ideal case of unitary transmission, the optimal sensitivity for the entangled estimation shows a Heisenberg scaling $1/(MN)$ in both the number of modes $M$ and mean number of photon $N$. This is obtained by optimizing over the initial probe state. Conversely, a separable estimation leads to a SQL scaling in $M$ and Heisenberg scaling in $N$: $1/(\sqrt{M}N)$. The authors experimentally demonstrated this entangled advantage in a network of $M=4$ nodes and a probe state generated by an OPO at wavelength $1550$ nm. In particular, using optimal probes containing $N \approx 2.5$ photons per mode, the measured standard deviation of $\bar{\phi}$ estimated was found equal to $\sigma_{\text{ent}}=0.099 \pm 0.003$  for the entangled estimation strategy, while being equal to $\sigma_{\text{sep}}=0.118 \pm 0.002$ for the separable estimation one \cite{guo2019distributed}.

\subsection{Simultaneous quantum estimation of phase and noises}

Realistic scenarios involve the unavoidable presence of noisy channels. The effects of noisy parameters inside optical interferometers were considered in several theoretical and experimental investigations \cite{degen2017quantum,escher2012quantum,genoni2012optical,genoni2011optical,datta2011quantum,kacprowicz2010experimental,brivio2010experimental,kolodynski2010phase,knysh2011scaling,ma2011quantumF,demkowicz2009quantum,dorner2009optimal,escher2011general,chin2012quantum,chaves2013noisy,knott2014effect,chabuda2020tensor}. In practical applications of phase estimation problems, the theoretical achievable quantum-enhanced precision is limited by photon losses \cite{datta2011quantum,demkowicz2009quantum,demkowicz2009quantum,dorner2009optimal,kacprowicz2010experimental} and phase diffusion \cite{demkowicz2012elusive,genoni2011optical,brivio2010experimental,genoni2012optical,escher2012quantum}, and vanishes when significant noise occurs\cite{yue2014quantum,demkowicz2012elusive}. In order to achieve effective quantum enhancement, optical quantum sensors require to take into account all these imperfections, often resulting in trade-off conditions on the achievable sensitivities. In this scenario, multiparameter estimation of both phase and noise represents a valid solution. A possible approach can be performing an a-priori characterization of noise before the estimation process. However, in many cases  time-varying systematical errors cannot be characterized in advance, such as phase oscillations due to thermal or mechanical fluctuation of optical systems \cite{musha1982optical,du2013sensitivity}. In these cases simultaneous estimation of phase and noise is necessary \cite{vidrighin2014joint}. All these studies generally require calculation of multiparameter bounds in which noise is considered as a non-unitary evolution.

We describe below different classes of multiparameter  phase and noise estimation implemented in photonic platforms: phase and phase diffusion, and phase and visibility.

\subsubsection{Phase and phase diffusion estimation}
Characterization of phase diffusion mechanisms can provide a more complete information on the quantum sensor. The problem of estimating a single phase inside an interferometer in the presence of phase diffusion can be modeled by an out-of control random phase shift, according to a Gaussian distribution of standard deviation $\Delta$ \cite{vidrighin2014joint,knysh2013estimation}. Such width represents the noise strength, and the associated non-unitary evolution can be described by the following action in the Fock basis \cite{vidrighin2014joint}:
\begin{equation}
    \mathcal{C}_{\Delta}=  e^{-\Delta^2 (m-n)^2} \ket{m} \bra{n},
\end{equation}
which causes an exponential damping of the coherence terms. Hence, starting from a 2-dimensional pure state $\ket{\Psi}_0=\cos \frac{\theta}{2} \ket{0}+\sin \frac{\theta}{2} \ket{1}$, after the a phase shift evolution $e^{i \phi}$ along mode 1 and a dephasing process, the final mixed state will be:
\begin{equation} \label{eq:statephasediff}
  \rho_{\text{dep}}=  \begin{pmatrix}
\cos^2(\frac{\theta}{2})&\sin(\frac{\theta}{2})\cos(\frac{\theta}{2})  e^{-i \phi-\Delta^2}\\
\cos(\frac{\theta}{2})\sin(\frac{\theta}{2})e^{i \phi-\Delta^2}&\sin^2(\frac{\theta}{2})\\
\end{pmatrix}.
\end{equation}

Multiparameter estimation of phase and dephasing has been investigated in various scenarios and platforms, from single qubit systems \cite{altorio2015weak,ragy2016compatibility} to larger number of qubits \cite{knysh2013estimation}, as well as considering both independent \cite{ragy2016compatibility} and collective dephasing \cite{szczykulska2017reaching}. The quantum Fisher Information matrix $\mathbb{F}_{\mathrm{Q}}$ relative to the considered qubit $\ket{\Psi}_0$ reads \cite{vidrighin2014joint}: 
\begin{equation}
   \mathbb{F}_{\mathrm{Q}}(\theta, \Delta)= \begin{pmatrix}
e^{-2 \Delta^2}&0\\
0&\frac{4 \Delta^2}{e^{2 \Delta^2}-1}\\
\end{pmatrix}.
\end{equation}

In this case the necessary and sufficient condition to saturate the QCRB [Eq.\eqref{eq:saturability}] is satisfied. Hence, there exists an optimal measurement such that the errors on both the two parameters reach the ultimate limits predicted by the QCRB. It is then necessary to define the optimal measurement. Such task in this scenario highlights a fundamental difference between single and multiparameter estimation. In the former case, quantum resources in the measurement stage do not improve the achievable sensitivity \cite{giovannetti2006quantum}. In the latter case, entangled measurements of probes can lead to an advantage in certain conditions \cite{vidrighin2014joint,roccia2017entangling}. 
The estimation of phase and phase diffusion is one of these scenarios. To formalize the problem, let us consider the following quantity $\kappa$ (for an estimation of $d$ parameters) \cite{ballester2004estimation,vidrighin2014joint,roccia2017entangling}:
\begin{equation}
     \kappa = \frac{1}{\nu}\sum_{i=1}^d \frac{1}{\mathbb{F}_{\mathrm{Q}ii}\;\,\mathbb{F}^{-1}_{ii}},
\end{equation}
being $\nu$ the number of independent probes. This figure of merit allows to quantify how much the chosen measurement is close to saturate the quantum limits relative to the employed probes and described by $\mathbb{F}_{\mathrm{Q}}$. The question is when the QCRB, $\kappa \le d$, is saturated.

In the qubit case (comprising N00N states), it has been demonstrated that $\kappa \le 1$ for any estimation involving two parameters, if the separate probes are independently measured \cite{gill2000state,vidrighin2014joint}. Then, in this case the QCRB cannot be saturated. On the contrary, for the estimation of phase and dephasing, we find $\kappa \le 1.5$ when two qubit probes are collectively measured \cite{vidrighin2014joint, roccia2017entangling}. Hence, in this case, even if the probes are separable, an advantage can be obtained with entangling measurements. Note that collective measurements do not always provide an enhancement for any multiparameter estimation. For instance, entangling measurements do not improve multiphase estimation tasks \cite{roccia2017entangling}.

\begin{figure*}[ht!]
\centering
\includegraphics[width=1.\textwidth]{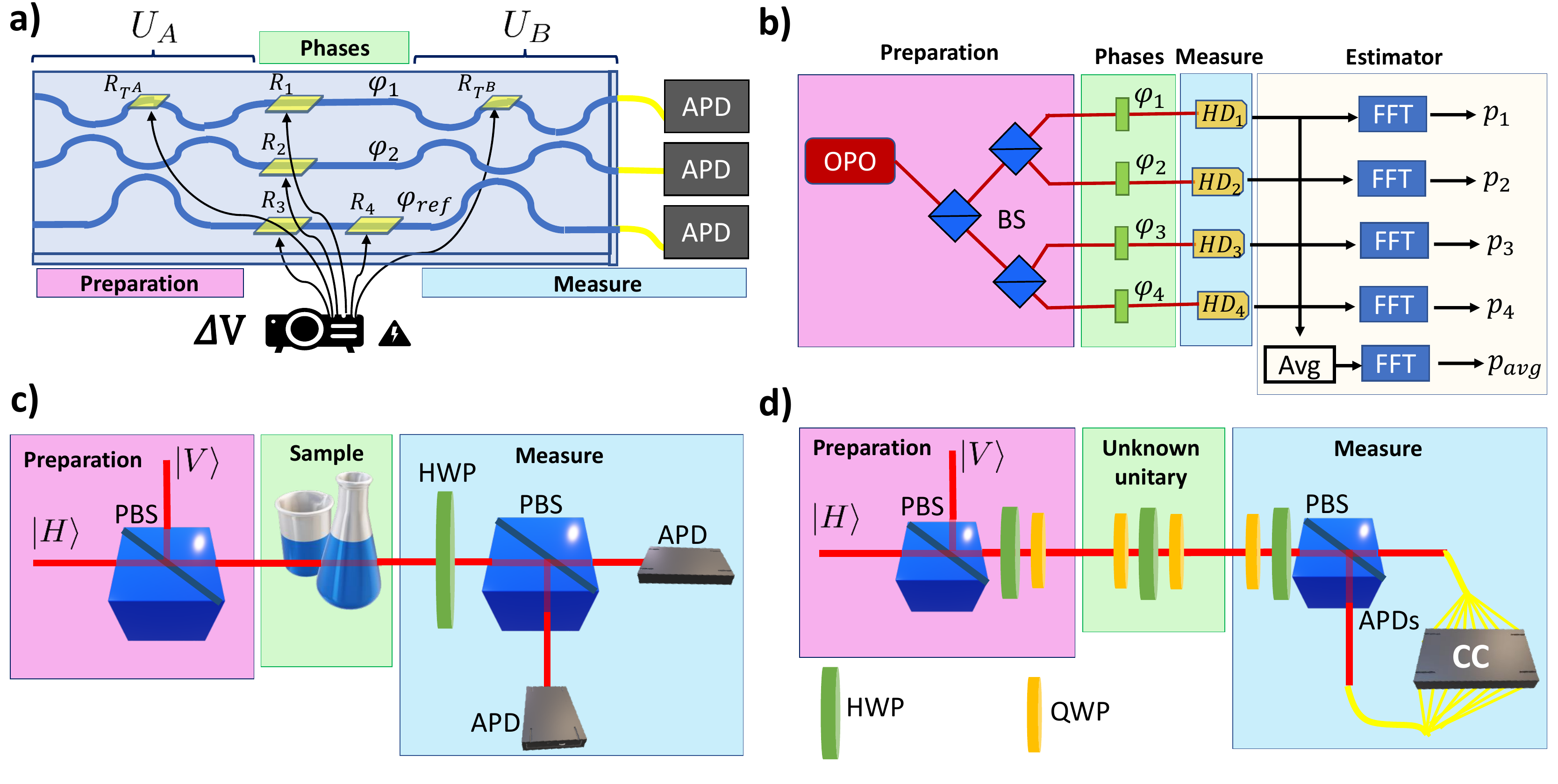}
\caption{{\bf Photonic platforms for multiparameter problems.}  {\bf a} Integrated platform for the simultaneous estimation of two phases $\Delta \phi_1 =\varphi_1-\varphi_{\text{ref}}$ and  $\Delta \phi_2=\varphi_2-\varphi_{\text{ref}}$, realized in Ref. \onlinecite{polino2019experimental}. The probe states are two indistinguishable photons. The unitaries $U_{A/B}$ represent the 2-D decomposition of tritters, while $R_i$ are the resistors used to tune the phases \cite{polino2019experimental}.  {\bf b} Scheme of the apparatus for the distributed sensing of the average of four phases $\varphi_1,...,\varphi_4$, realized in Ref. \onlinecite{guo2019distributed}. In the entangled estimation, the probes are squeezed coherent states generated in the  optical parametric oscillator (OPO) and  distributed along the four nodes through $50:50$ beam splitters (BS) \cite{guo2019distributed}. The measurement of the phase quadrature $p_i$, along each node $i$,  is performed by the homodine detection $HD_i$. Finally the average estimation is performed. {\bf c} Scheme of the Mach-Zehnder interferometer in polarization, realized to perform measurements of phase and visibility of different samples in Refs.  \onlinecite{roccia2018multiparameter,cimini2019quantum,cimini2019adaptive}. The probes are two-photon N00N state in polarization.  {\bf d} Experimental apparatus for the estimation of the parameters describing unknown processes in the polarization space,  realized in Ref. \onlinecite{zhou2015quantum}. The probes are four-photon HB states in polarization and the final measurement of the coincidence events (CC) is realized by a probabilistic photo-number detection.}
\label{fig:multiexp}
\end{figure*}

An experimental implementation of such improvement on simultaneous phase and dephasing estimation was demonstrated in Ref. \onlinecite{roccia2017entangling}. In this work, the probes are polarization qubits and the two-qubit entangled measurements are Bell measurements realized through a probabilistic controlled-sign gate. Given an arbitrary two-qubit state, such gate introduces a phase sign "-1" on all terms $\ket{V}\ket{V}$ in which both qubits are vertically polarized. The controlled-sign gate can be realized in the polarization degree of freedom through partially polarizing beam splitter whose transmission coefficients are appropriately chosen \cite{langford2005demonstration,kiesel2005linear,okamoto2005demonstration}. In this way, the authors performed simultaneous estimation of phase and dephasing, and an optimal value $\kappa^{\mathrm{opt}} = 1.18\pm 0.02$ was obtained \cite{roccia2017entangling}.

In a different work\cite{altorio2015weak}, weak measurements were exploited to experimentally perform multiparameter estimation of a phase shift and its phase diffusion with classical probes.

\subsubsection{Phase and visibility estimation}

In certain scenarios, noise can be detrimental to the estimation of a parameter. In these cases, even if the noise value is not of interest, a multiparameter approach can be used to estimate the desired unknown parameter. A certain class of noise processes in interferometers can be modeled by the interference visibility $v \le 1$. This parameter quantifies the quality of probes and apparatus, such as the visibility of HOM interference of two photons in a beam splitter. 

In Ref. \onlinecite{roccia2018multiparameter} an optical phase shift $\phi$ and noise over the probe state, measured in terms of visibility $v$ of the interference fringes, have been simultaneously estimated. The output probability distributions of the system depend on both $\phi$ and $v$. and this implies the necessity of a multiparameter estimation. If the visibility is not properly, estimated, the estimation of the phase shift would suffer of a bias. In the employed apparatus, that is a Mach-Zehnder, a $N=2$ N00N state in the polarization degree of freedom interferes with the optical phase (Fig. \ref{fig:multiexp}c). Coincidence measurements are performed to estimate the two unknown parameters. First, two-parameter estimation is made for the pre-calibration of the apparatus. Then, the scheme is exploited to study the optical activity of two different biological samples, fructose and sucrose aqueous solutions, through a Bayesian learning approach. The quality of the experimental estimation with respect to the CRB has been verified through a likelihood ratio test, defined as:  $l = m^2 \text{Tr}(\mathbb{F}\cdot C(\phi,v))-m(\ln{\text{det}(C(\phi,v))}+\ln{\text{det}(m \mathbb{F})})-2$. More specifically, the null hypothesis corresponds to the covariance matrix $C(\phi,v)$ of the two parameters saturating the CRB. This quantity is distributed as a $\chi^2$ variable with 3 degrees of freedom. For the fructose solution the authors obtained a sensitivity $l_f=2.63$, while for sucrose $l_s=0.10$. Both these values are compatible with the null hypothesis in a $95\%$ confidence interval.

The same scheme has been used also to perform tracking of a chemical process relative to the acid hydrolysis of sucrose \cite{cimini2019quantum}. In this case, the change of the sample optical activity from dexorotatory to levorotatory, due to the chemical reaction, turns in a phase variation between the polarizations and is measured with quantum super-resolution. More specifically, the real-time visibility $v(t)$ and phase $\phi(t)$ are monitored as function of the time $t$. 

Finally, this setup was used for estimation of real-time invertase enzymatic activity, through a Bayesian adaptive technique \cite{cimini2019adaptive}. At each step, an additional known and controlled phase is chosen to maximize the Fisher Information depending on the current knowledge of the unknown phase.

\begin{table*}[t!]
\centering
\begin{tabular}{|l|l|l|l|l|}
\hline
{\bf Estimation}&{\bf Refs.}&{\bf Photonic platform } & {\bf Mean number }&{\bf Probe} \\
{\bf problems}& &{\bf }& {\bf of photons}&{\bf } \\
\hline
Two phases& [\onlinecite{polino2019experimental}]&Integrated three-arm interferometer& $N=2$ &indistinguishable \\
 && &&photons\\
\hline
Average of  & [\onlinecite{guo2019distributed}] &4-node network&$N\approx 4 \times 2.5$&  squeezed coherent states
\\
four phases  &&&&
\\
\hline
Phase and& [\onlinecite{roccia2017entangling}]&Bulk two-mode polarization interferometer & $N=2$ & separable two-qubit state
\\
phase diffusion&&with entangling measurement&&\\
 \hline
Phase and& [\onlinecite{roccia2018multiparameter}]&Bulk two-mode polarization MZI interferometer& $N=2$ &  N00N state
\\
visibility&&measurements on fructose and sucrose&&
\\
  \hline
Phase and&[\onlinecite{cimini2019quantum}]&Bulk two-mode polarization MZI interferometer& $N=2$ & N00N state
\\
visibility&&real-time measurement of sucrose acid hydrolysis&&
\\
 \hline
Phase and&[\onlinecite{cimini2019adaptive}] &Bulk two-mode polarization MZI interferometer&$N=2$ &  N00N state
\\
visibility&&adaptive real-time measurement of invertase enzymatic activity& &
 \\
\hline
Centroid and separation& [\onlinecite{parniak2018beating}]&HOM interference in a BS and spatial resolution detection& $N=2$ & indistinguishable\\
 of two incoherent sources&&& &photons\\
\hline
Unitary process in& [\onlinecite{zhou2015quantum}]&Bulk polarization interferometer& $N=4$ &HB states\\
 polarization space&& &&\\
 \hline
\end{tabular}
\caption{Table of some photonic platforms exploiting quantum schemes for multiparameter estimation problems.} 
\label{tab:multi}
\end{table*}

\subsection{Other scenarios}
Photonic sensors are exploited in other multiparameter scenarios. Here we briefly list some examples. 
 
A first scenario is the estimation of separations between incoherent point sources \cite{tsang2016quantum,lupo2016ultimate,nair2016far,tsang2017subdiffraction,yang2016far,tham2017beating,gefen2019overcoming,vrehavcek2017multiparameter,bisketzi2019quantum,ang2017quantum,tang2016fault,chrostowski2017super,donohue2018quantum,vrehavcek2018optimal,tsang2019quantum,paur2016achieving,grace2019approaching,backlund2018fundamental,yu2018quantum,bonsma2019realistic,napoli2019towards,prasad2019quantum,zhou2019quantum}.
In the case of two point sources, the two parameters are the difference of the two positions $\lambda_1=x_1-x_2$, and the corresponding centroid $\lambda_2=(x_1+x_2)/2$.
In Ref. \onlinecite{parniak2018beating}, the authors proposed and realized experimentally the simultaneous estimation of the centroid and separation of two point sources, exploiting the HOM effect. In this experiment, photons generated by the two sources are sent to the input ports of a beam splitter. The coincidences between outputs of the beam splitter and two-photon events along the same output are measured with a spatial resolution technique \cite{jachura2016mode}. 
 
A different multiparameter task is related to system characterization. A quantum-enhanced tomography of an unknown unitary process acting on the polarization degree of freedom was realized using multiphoton quantum states in Ref. \onlinecite{zhou2015quantum}. The task is also called Quantum Process Tomography  \cite{mitchell2003diagnosis,altepeter2003ancilla,o2004quantum,lobino2008complete,mohseni2008quantum,rozema2014optimizing}. In a multiparameter approach, the aim is to simultaneously estimate the parameters characterizing a quantum evolution, by exploiting opportunely chosen probes and measurements. In Ref. \onlinecite{zhou2015quantum} quantum process tomography was performed on unknown unitary evolutions in the 2-dimensional space of photonic polarization. Given a polarization single photon state $c_H \ket{H}+c_v \ket{V}$, where $H$ and $V$ are polarization states and $c_H$ and $c_V$ are the corresponding amplitudes, a unitary evolution $U$ acting on this state can be described by a matrix $U=\begin{pmatrix} a+i\,b&c+i\,d\\  -c+i\,d&a-i\,b\end{pmatrix}$, with $a,b,c,d \in \mathbb{R}$ and $a^2+b^2+c^2+d^2=1$. Hence the estimation regards three independent parameters and is performed by measuring three output probabilities $p_{HV}$, $p_{AD}$ and $p_{RL}$ along three different polarization bases. The authors employed HB 4-photon polarization states split along the two polarization modes $H$ and $V$: $\ket{2}_{H}\ket{2}_{V}$. The state is generated by a type-I non-collinear SPDC process in which the polarization of photons emitted along one mode is rotated by $45^\circ$. All photons along the two modes are recombined in a single spatial mode through a polarizing beam splitter. After its preparation, the 4-photon HB state passes through the unitary evolution, performed by three cascaded waveplates, and is finally measured by a polarization selection stage and probabilistic photon-number resolving detection (Fig. \ref{fig:multiexp}d). The probabilities are estimated from data through a maximum-likelihood technique and a quantum enhancement is observed in the estimation of random unitaries, considering the four-photon detected events as the employed resources \cite{zhou2015quantum}.

In measurements of squeezed light, scattering of photons from the meter can cause parastic signal, called parasitic interference \cite{vinet1996scattered,ottaway2012impact}. This affects also measurements of gravitational waves, limiting the possible quantum advantages \cite{aasi2015advanced}.
The authors in Ref. \onlinecite{steinlechner2013quantum} theoretically and experimentally demonstrated the concept of quantum dense metrology. In this scheme, a two-mode squeezed state is exploited in order to identify the parasitic noise and discard the corrupted data recovering quantum advantage. The setup is the same represented in Fig.\ref{fig:ligo}b. To this end, both the quadratures are simultaneously estimated beyond the SQL. Quantum dense metrology technique was applied also in modified protocols able to enhance the  reduction of noises \cite{ast2016reduction}.

\section{Conclusions and Perspectives}

In this Review we provided an overview of the current state of the art in photonic technologies for quantum metrology applications. In particular, starting from the theoretical fundamental ingredients, we have discussed the most commonly adopted strategies and platforms for photonic systems, with particular attention towards application of adaptive strategies for efficient extraction of information. Finally, we have provided an overview on the recently expanding field of multiparameter estimation, which has potential applications in a large variety of fields where the process inherently involves multiple physical quantities at once.

Several open points, both from a theoretical and experimental point of view, still has to be addressed towards development of photonic quantum sensors capable to provide quantum enhancement in realistic noisy conditions \cite{ono2010effects,chin2012quantum,jarzyna2013matrix,chaves2013noisy,demkowicz2014using,pirandola2017ultimate,chabuda2020tensor}. For photonic systems, the main challenges is represented by losses within the apparatus. Theoretical studies \cite{kolodynski2010phase,escher2011general,giovannetti2011advances,knysh2011scaling,demkowicz2012elusive} have indeed shown the detrimental effect of losses towards reaching sub-SQL performances, in particular when the number of involved photons $N$ in the prepared probes is large. Indeed, no quantum enhancement (in terms of scaling in $N$) can be achieved when losses are large enough, leaving space only for a constant improvement in this regime. All these results lead to a large effort towards development of appropriate metrological strategies capable of providing a more robust behavior in a noisy and lossy scenario \cite{dorner2009optimal,demkowicz2010multi,kacprowicz2010experimental,vitelli2010opa,spagnolo2011opa,ozeri2013heisenberg,kolodynski2013tools,nichols2016practical,unden2016quantum,jachura2016mode,cable2010parameter,matthews2016towards,wang2018entanglement,albarelli2018restoring,bai2019retrieving}. In parallel, experimental effort has been devoted to achieve technological advances in photonic systems. Such effort has enabled the first recent experimental demonstration of unconditional violation of the SQL in a two-photon experiment\cite{slussarenko2017unconditional}. In this direction, a significant amount of work still has to be done to improve the performances of photonic platforms to obtain quantum-enhancement in more complex scenarios. To this end, a significant intermediate step before reaching an unconditional violation of the SQL for larger sensors would be to obtain improved performances with respect to classical strategies, by comparing the achieved sensitivities in the presence of the same noise conditions. Note that, in the presence of noise, entangled estimation protocols employing external ancillas can lead to a higher sensitivity than the one achievable by using sequential unentangled estimations \cite{demkowicz2014using,huang2016usefulness,huang2018noise}. In this sense, noisy cases are those where quantum advantage becomes evident\cite{demkowicz2015quantum}.

A second promising research direction can be found in the multiparameter scenario, which has recently received growing attention for its wide range of applications \cite{szczykulska2016multi,albarelli2020perspective}. Additionally, recent studies have shown that a multiparameter approach can provide some advantages in the presence of noise \cite{vidrighin2014joint,roccia2018multiparameter}. While the last few years have reported significant advances in the field, both in terms of theoretical background and of technological platforms, there are still several open points. Indeed, general recipes for the definition of optimal probe states for a general multiparameter scenario are still lacking. A similar issue is present for the definition of the optimal measurement strategies, in particular in the presence of noise which inherently requires considering mixed probe states. Finally, minimally-invasive scenarios such as those involving biological system \cite{taylor2013biological} require the development of quantum strategies tailored to obtain optimal performances when only limited data are available \cite{rubio2019quantum,rubio2018non}. 

\begin{acknowledgments}
We thank Ilaria Gianani and Marco Barbieri for useful discussions. This work is supported by the Amaldi Research Center funded by the  Ministero dell'Istruzione dell'Universit\`a e della Ricerca (Ministry of Education, University and Research) program ``Dipartimento di Eccellenza'' (CUP:B81I18001170001), by MIUR via PRIN 2017 (Progetto di Ricerca di Interesse Nazionale): project QUSHIP (2017SRNBRK), by QUANTERA HiPhoP (High dimensional quantum Photonic Platform; grant agreement no. 731473), 
by the ID n. 61466 grant from the John Templeton Foundation, as part of the “The Quantum Information
Structure of Spacetime (QISS)” Project (qiss.fr) (the opinions expressed in this
publication are those of the author(s) and do not necessarily reflect the views of the John Templeton Foundation) and by the Regione Lazio programme “Progetti di Gruppi di ricerca” legge Regionale n. 13/2008 (SINFONIA project, prot. n. 85-2017-15200) via LazioInnova spa.
\end{acknowledgments}

\providecommand{\noopsort}[1]{}\providecommand{\singleletter}[1]{#1}%
%


\end{document}